\documentclass{article}
\usepackage[utf8]{inputenc}
\usepackage[left=3cm,right=3cm,top=3cm,bottom=3cm]{geometry}
\usepackage{amsmath}
\usepackage{bbold}
\usepackage{amssymb}
\usepackage{graphicx}
\usepackage{esint}
\usepackage{enumitem}
\usepackage{multirow}
\usepackage{graphicx}
\usepackage{hhline}
\usepackage{hyperref}
\usepackage{centernot}
\usepackage{amsthm}
\hypersetup{
  colorlinks   = true,    
  urlcolor     = blue,    
  linkcolor    = blue,    
  citecolor    = black     
}

\newcommand\Tr{\mathrm{Tr}}

\newcommand{\xdownarrow}[1]{%
  {\left\downarrow\vbox to #1{}\right.\kern-\nulldelimiterspace}
}

\title{Wilson line-based action for gluodynamics at the loop level}

\author{Hiren Kakkad$^a$,
Piotr Kotko$^a$,
Anna Stasto$^b$
\\ \\
$^a${\it AGH University}\\ 
{\it Faculty of Physics and Applied Computer Science,} \\ 
{\it al. Mickiewicza 30, 30-059 Krakow, Poland} \\ \\
$^b${\it The Pennsylvania State University, Physics Department}\\ 
{\it 104 Davey Lab, University Park, PA 16802, USA }
}

\date{}

\begin{document}
\maketitle

\begin{abstract}
We develop quantum corrections to 
the Wilson line-based action which we recently derived through a transformation that eliminates triple gluon vertices from the Yang-Mills action on the light-cone. The action efficiently computes high multiplicity tree-level split-helicity amplitudes with the number of diagrams following the Delannoy number series. However, the absence of the triple gluon vertices results in missing loop contributions. To remedy this, we develop two equivalent approaches using the one-loop effective action method to systematically incorporate loop contributions to our action. In one approach there are solely Yang-Mills vertices in the loop whereas the other uses the interaction vertices of our action along with the kernels of the solution of our transformation in the loop. In addition to demonstrating the equivalence of both approaches, we validated the quantum completeness of the former by computing all 4-point one-loop amplitudes which could not be previously computed. Both of our approaches are easily extendable to develop quantum corrections to other reformulations of the Yang-Mills theory obtained via non-linear classical field transformations eliminating interaction vertices.

\end{abstract}

\section{Introduction}
\label{sec:Intro}

The simplicity of the tree level pure gluonic amplitudes is obscured within the traditional Feynman diagram approach. It, however, becomes evident upon half-Fourier transform\footnote{The negative helicity spinor $\widetilde\lambda_i^{\dot\alpha}$ corresponding to the four momenta of  external gluons $p_i^{\dot\alpha \alpha}  =   \widetilde\lambda_i^{\dot\alpha}\lambda_i^\alpha\,$ is Fourier transformed to $\mu_i^{\dot\alpha}$ to obtain the  amplitude in terms of twistor coordinates $Z_i = (  \mu^{\dot\alpha}_i, \lambda^\alpha_i )$.} to the twistor space $\mathbb{PT}$, where these localize on algebraic curves of different degrees \cite{Witten2004}. In general, the tree level $\mathrm{N}^k\mathrm{MHV}$ pure gluonic amplitudes with $k+2$ negative helicities localize on algebraic curves of degree $k+1$ and genus zero in twistor space.  This implies the maximally helicity violating ($\mathrm{MHV}$) amplitudes live on the simplest algebraic curves of degree one $\mathbb{CP}^1 \subset \mathbb{PT}$ -- a Riemann sphere (sometimes dubbed as a complex line) -- in the twistor space. Apart from unraveling the simple structures underlying the gauge theory amplitudes, \cite{Witten2004} also laid the foundation for developing new approaches to computing pure gluonic amplitudes.

\begin{figure}
    \centering
    \includegraphics[width=14cm]{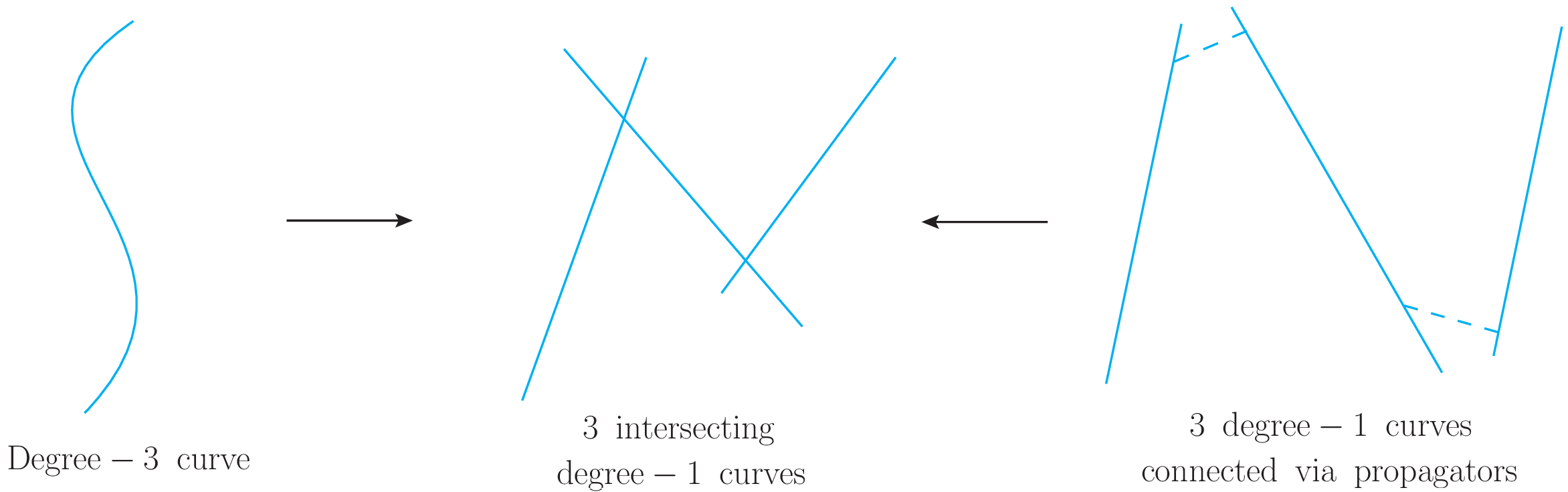}
    \caption{
    \small  
    The figure demonstrates the equivalence between the completely connected and the completely disconnected approach to computing amplitudes \cite{gukov2004equivalence}. On the left, we have a single degree-3 curve in the twistor space. Integrating over the moduli space of this allows one to compute an NNMHV amplitude. This is the completely connected approach. The same amplitude can be computed for a set of 3 degree-1 curves connected via twistor space propagators as shown on the right. The equivalence states that there exists a locus on which the degree-3 curve reduces to three intersecting degree-1 curves and on the same locus the propagators used in the disconnected approach shrink resulting in the same as shown in the middle.}
    \label{fig:TS_deg}
\end{figure}

The \textit{incidence relations} 
\begin{equation}
    \mu^{\dot\alpha}= x^{\dot\alpha \alpha} \lambda_\alpha 
    \label{eq:inc_rel}
\end{equation}
define the mapping between the points $x^{\dot\alpha \alpha}$ in complexified Minkowski space $\mathbb{M}_\mathbb{C}$ and the points $Z^A=(  \mu^{\dot\alpha}, \lambda_\alpha )$\footnote{$(  \mu^{\dot\alpha}, \lambda_\alpha ) \sim ( c \mu^{\dot\alpha},  c\lambda_\alpha )$ where $c \in \mathbb{C}^* = \mathbb{C}/\{0\}$.} in twistor space $\mathbb{PT}$. An important outcome of the incidence relation is that degree one curves $\mathbb{CP}^1 $ on twistor space correspond to points in Minkowski space. This correspondence implies that $\mathrm{MHV}$ amplitudes -- supposedly, non-local objects -- could be treated as local interaction vertices in Minkowski space. This realization gave birth to the Cachazo-Svrcek-Witten (CSW) rules \cite{Cachazo2004} for computing amplitudes. In this approach, the MHV amplitudes continued off-shell are used as interaction vertices that are local in light-cone time $x^+$ \cite{Mansfield2006}. Gluing these using scalar propagators allows one to compute tree amplitudes with any helicity configuration. In the twistor space, this means that the tree level $\mathrm{N}^k\mathrm{MHV}$ amplitudes could be obtained by connecting  $k+1$ degree-one curves via propagators given as $(0,2)$-form on $\mathbb{CP}^3 \times \mathbb{CP}^3$. However, in \cite{Roiban1,Roiban2,Roiban3}, the authors demonstrated that the $\overline{\mathrm{MHV}}$ $(- - \dots - + +)$ as well as the $\mathrm{N}\mathrm{MHV}$ $(- -  - + \dots +)$ amplitudes could be obtained from a single connected curve in twistor space. We, therefore, have two different approaches to compute the same amplitude: the \textit{completely disconnected} approach (or the CSW rules) where one use $k+1$ degree-one curves to compute $\mathrm{N}^k\mathrm{MHV}$ amplitude, and the \textit{completely connected} approach where one uses a single connected curve of degree $k+1$\footnote{The integral over the moduli space of these curves gives the $\mathrm{N}^k\mathrm{MHV}$ amplitude.}. These two approaches were later shown to be equivalent in \cite{gukov2004equivalence}. Simply put, the authors demonstrated that there exists a locus where a degree $d$ curve degenerates into $d$ intersecting degree-one curves, and on the same locus the propagators (joining degree-one curves in the CSW approach) shrink to zero size resulting in $d$ intersecting degree-one curves (see Figure~\ref{fig:TS_deg}). In the process, they conjectured the existence of intermediate approaches using curves of different degrees $d_i$ such that $\sum_i d_i = k+1$ for $\mathrm{N}^k\mathrm{MHV}$ amplitude. This then led to the formulation of non-MHV vertices in \cite{Bena_2005} for computing tree level amplitudes thus providing an explicit realization of such an intermediate prescription.

All the above ideas are inspired by the twistor space representation of the gauge theory amplitudes. But providing an action-based realization of these in Minkowski space starting with the Yang-Mills action is yet another interesting research direction. In \cite{Mansfield2006}, an action (the so-called 'MHV action') was derived via canonically transforming the light-cone Yang-Mills action such that the Feynman rules for computing the tree level amplitudes correspond to the CSW rules. However, no such action has been known, to the best of our knowledge, that could correspond to one of the intermediate approaches conjectured in \cite{gukov2004equivalence}. In \cite{Kakkad:2021uhv}, we derived a new action takes goes beyond the CSW rules. It has the following schematic representation (we provide further details later):
\begin{align}
S\left[Z^{\bullet},Z^{\star}\right] =\int dx^{+} \Bigg\{ & 
-\int d^{3}\mathbf{x}\,\mathrm{Tr}\,\hat{Z}^{\bullet}\square\hat{Z}^{\star} \nonumber \\
 & + \mathcal{L}_{--++}+ \mathcal{L}_{--+++}+\mathcal{L}_{--++++} + \dots \nonumber \\
& + \mathcal{L}_{---++}+ \mathcal{L}_{---+++}+\mathcal{L}_{---++++} + \dots \nonumber \\
& \,\, \vdots \nonumber \\
& + \mathcal{L}_{---\dots -++}+ \mathcal{L}_{---\dots -+++}+\mathcal{L}_{---\dots -++++}+ \dots
\Bigg\}
\,,\label{eq:Z_action1}
\end{align}
where $Z^{\bullet}$ and $Z^{\star}$ are the fields that in the on-shell limit correspond to two transversely polarized gluons.
It consists of a scalar propagator connecting opposite helicity fields and an infinite set of interaction vertices with the following generic form: 
\begin{equation}
    \mathcal{L}_{\underbrace{-\,\cdots\,-}_{m}\underbrace{+ \,\cdots\, +}_{n-m}}= 
   \int\!d^{3}\mathbf{y}_{1}\dots d^{3}\mathbf{y}_{n} \,\, \mathcal{U}^{b_1 \dots b_{n}}_{-\dots-+\dots+}\left(\mathbf{y}_{1},\cdots \mathbf{y}_{n}\right) 
   \prod_{i=1}^{m}Z^{\star}_{b_i} (x^+;\mathbf{y}_{i})
   \prod_{j=1}^{n-m}Z^{\bullet}_{b_j} (x^+;\mathbf{y}_{j}) \, .
   \label{eq:Z_vertex_lagr_pos}
\end{equation}

The first row of interaction vertices in Eq.~\eqref{eq:Z_action1} consists of MHV vertices $(--+\dots++)$ which give the corresponding MHV amplitudes in the on-shell limit. In the on-shell limit, these would therefore localize on degree-one curves in the twistor space as was demonstrated in \cite{Witten2004}. The first column, on the other hand, consists of $\overline{\mathrm{MHV}}$ vertices $(--\dots-++)$, and these too give the corresponding amplitudes in the on-shell limit. These would therefore localize on curves of degree $k$ ($k+1$ being the number of negative helicities in $(--\dots-++)$) in the twistor space as was demonstrated in \cite{Roiban2}. As for the other vertices\footnote{We note that these vertices in our action do not correspond to the non-MHV vertices of \cite{Bena_2005}. What they call non-MHV vertices are non-MHV amplitudes computed using the CSW rules and then the external legs are taken off-shell. Our vertices with some given helicity configuration, on the other hand, consist of only a subset of the contributions necessary to compute the amplitude with the same configuration.}, say $\mathrm{N}^k\mathrm{MHV}$ vertex, these correspond to a subset of all possible degenerations of a degree $k+1$ curve into $k+1$ degree-one curves i.e. these consist of a subset of all the contributions originating in the CSW rules necessary to compute the $\mathrm{N}^k\mathrm{MHV}$ amplitude. We shall discuss this point briefly in Section \ref{sec:Zaction} (see Section 3.4 of \cite{kakkad2023scattering} for more details). Note that these observations hold only in the on-shell limit. However, they imply that our action Eq.~\eqref{eq:Z_action1} should provide an off-shell Minkowski space realization of an approach that would use algebraic curves of different degrees to compute gauge theory tree level amplitudes in the twistor space as was conjectured in \cite{gukov2004equivalence}. 

 Our action Eq.~\eqref{eq:Z_action1}, just like the MHV action \cite{Mansfield2006}, was derived by canonically transforming the light-cone Yang-Mills action. The transformation eliminated both the triple point interaction vertices from the latter. Although beneficial at the tree level -- and also for providing Minkowski space realization to the above twistor space ideas -- the elimination of the triple point interaction vertices results in missing contributions at the loop level. For instance, the all-plus $(+ + \dots +)$ one-loop gluonic amplitude is zero in both the actions. But it is well known to be a rational function of spinor products \cite{Bern:1991aq,Kunszt_1994}. This amplitude is the sole quantum correction to the self-dual sector of the Yang-Mills and has also been attributed to quantum anomalies \cite{Bardeen1996,Chattopadhyay:2020oxe,Monteiro:2022nqt} (for further details on the amplitudes in the self-dual sector see \cite{Chalmers1996,Cangemi1997,Rosly1997,Monteiro2011}). However, when deriving the MHV action, the self-dual sector gets mapped to a free term via Mansfield's transformation  resulting in missing loop contributions necessary to compute $(+ + \dots +)$ one-loop amplitude. 

 Numerous attempts have been made to restore the missing loop contributions in the MHV action. For instance, in \cite{Ettle2007}, the authors used the violation of the S-matrix equivalence theorem to restore the contributions to the all-plus helicity one-loop amplitude. In \cite{Brandhuber2007} yet another approach, based on a four dimensional regularization scheme, the so-called world-sheet regularization of Chakrabarti, Qiu, and Thorn (CQT) \cite{CQT1,CQT2}, was used to introduce all-plus helicity one-loop vertices to the MHV action. These vertices originate from the canonical transformation of a $(++)$ gluon self-energy counterterm needed in this regularization scheme. Finally, in \cite{Boels_2008} the authors used the massive CSW rules (i.e. the CSW rules for Yang-Mills gauge theory coupled to a massive colored scalar). Following this, the $n$-point all plus helicity and single minus helicity amplitudes were developed in \cite{Elvang_2012}.

 Recently, in \cite{Kakkad_2022}, we used a yet different route to systematically develop quantum corrections to the MHV action. The starting point was the one-loop effective Yang-Mills action where the classical action is separated from the one-loop corrections. To this we applied Mansfield's transformation to obtain the classical MHV action plus one-loop contributions. We validated the one-loop effective MHV action by computing the four point $(++++)$ and $(+++-)$ one-loop amplitudes. Although our approach was successful in developing one-loop corrections to the MHV action, a major drawback was that the MHV vertices were not explicit in the loop. The aim of the current work is, therefore, two fold. Firstly, extend the one-loop effection action approach employed in \cite{Kakkad_2022} to systematically develop quantum corrections to the new action Eq.~\eqref{eq:Z_action1}. Secondly, do so in such a way that the interaction vertices of our action are explicit in the loop. In the process, we will also report our recent discovery about the number series followed by the number of diagrams required to compute tree-level amplitudes using our action Eq.~\eqref{eq:Z_action1}.

 The paper is organized as follows. In Section \ref{sec:Zaction}, we briefly review the derivation of our action Eq.~\eqref{eq:Z_action1}. After that, to make the correspondence between the vertices of our action and curves of different degrees in twistor space apparent, we show that the vertices in our action in the on-shell limit can be decomposed into a collection of MHV vertices connected via scalar propagators. In Section \ref{sec:DEL}, we discuss the \textit{Delannoy} numbers appearing in the computation of tree amplitudes using our action Eq.~\eqref{eq:Z_action1}. Section \ref{sec:GAOL} contains the main results of our work where we employ the one-loop effective action approach to develop quantum corrections to our action Eq.~\eqref{eq:Z_action1} such that the interaction vertices are explicit in the loop. In the process, we also demonstrate that our previous approach \cite{Kakkad_2022} can also be extended to \textit{equivalently} develop quantum corrections. Details of some of the calculations and derivations are in the Appendices.
\section{The Wilson line-based action}
\label{sec:Zaction}

In this section, we briefly recall the main result of \cite{Kakkad:2021uhv} where we derived a new classical action, which we shall dub the "Z-field action" hereafter. 

In \cite{Kakkad:2021uhv}, we demonstrated that the Z-field action Eq.~\eqref{eq:Z_action1} can be derived in two equivalent ways. First is via canonically transforming the MHV action \cite{Mansfield2006} such that it eliminates the $(+ - -)$ triple point MHV vertex. Second is via canonically transforming the light-cone Yang-Mills action \cite{Scherk1975} such that it eliminates both $(+ - -)$ and $(+ + -)$ triple point interaction vertices at once. In \cite{Kakkad:2021uhv}, we used the former to derive the action i.e. the explicit expressions for the interaction vertices. In this section, we will use the latter approach to rederive the Z-field action. 

The Z-field action reads
Eq.~\eqref{eq:Z_action1}
\begin{align}
S\left[Z^{\bullet},Z^{\star}\right] =\int dx^{+} \Bigg\{ & 
-\int d^{3}\mathbf{x}\,\mathrm{Tr}\,\hat{Z}^{\bullet}\square\hat{Z}^{\star} \nonumber \\
 & + \mathcal{L}_{--++}+ \mathcal{L}_{--+++}+\mathcal{L}_{--++++} + \dots \nonumber \\
& + \mathcal{L}_{---++}+ \mathcal{L}_{---+++}+\mathcal{L}_{---++++} + \dots \nonumber \\
& \,\, \vdots \nonumber \\
& + \mathcal{L}_{---\dots -++}+ \mathcal{L}_{---\dots -+++}+\mathcal{L}_{---\dots -++++}+ \dots
\Bigg\}
\,,\label{eq:Z_action}
\end{align}
where $x^{+}$ is the light-cone time and $\mathbf{x}\equiv\left(x^{-},x^{\bullet},x^{\star}\right)$. We use the so-called "double-null" coordinates, in which the contravariant components of a four vector read
\begin{gather}
v^{+}=v\cdot\eta\,,\,\,\,\,\,\,\,\, v^{-}=v\cdot\widetilde{\eta}\,, \,\,\,\,\,\,\,\,
v^{\bullet}=v\cdot\varepsilon_{\bot}^{+}\,,\,\,\,\,\,\,\,\, v^{\star}=v\cdot\varepsilon_{\bot}^{-}\,, \label{eq:DNC_cor}
\end{gather}
 where ${\eta}^\mu$, $\widetilde{\eta}^\mu$ are the two light-like and $\varepsilon_{\perp}^{\pm\,\mu}$ are the two space-like basis four vectors 
\begin{gather}
\eta^\mu=\frac{1}{\sqrt{2}}\left(1,0,0,-1\right)\,,\,\,\,\,\widetilde{\eta}^\mu=\frac{1}{\sqrt{2}}\left(1,0,0,1\right)\, ,\,\,\,\, \varepsilon_{\perp}^{\pm\,\mu}=\frac{1}{\sqrt{2}}\left(0,1,\pm i,0\right)\,.
\label{eq:DNC_basis}
\end{gather}
Lowering of the indices in Eq.~\eqref{eq:DNC_cor} results in the flipping $+\leftrightarrow -$ and $\star \leftrightarrow  \bullet$ where the latter is accompanied by a change of sign. In these coordinates, the dot product of two four vectors is $u\cdot w=u^{+}w^{-}+u^{-}w^{+}-u^{\bullet}w^{\star}-u^{\star}w^{\bullet}$. Thus, $\square=\partial^\mu \partial_\mu = 2(\partial_+\partial_- - \partial_{\bullet}\partial_{\star})$. For the fields, in Eq.~\eqref{eq:Z_action}, we use $\hat{Z}=Z_at^a$ where $t^a$ are the generators of the color in the fundamental representation\footnote{The generators $t^a$ satisfy, $\left[t^{a},t^{b}\right]=i\sqrt{2}f^{abc}t^{c}$ and $\mathrm{Tr}(t^{a}t^{b}) = \delta^{ab}$. In this normalization, our coupling constant is re-scaled as $g\rightarrow g/\sqrt{2}$ compared to the 'standard' normalization.}.

The Z-field action Eq.~\eqref{eq:Z_action} is most straightforwardly derived by canonically transforming the Yang-Mills action on the light cone \cite{Scherk1975}. The latter reads 
\begin{multline}
S_{\mathrm{YM}}\left[A^{\bullet},A^{\star}\right]=\int dx^{+}\int d^{3}\mathbf{x}\,\,\Bigg\{ 
-\mathrm{Tr}\,\hat{A}^{\bullet}\square\hat{A}^{\star}
-2ig\,\mathrm{Tr}\,\partial_{-}^{-1}\partial_{\bullet} \hat{A}^{\bullet}\left[\partial_{-}\hat{A}^{\star},\hat{A}^{\bullet}\right] \\
-2ig\,\mathrm{Tr}\,\partial_{-}^{-1}\partial_{\star}\hat{A}^{\star}\left[\partial_{-}\hat{A}^{\bullet},\hat{A}^{\star}\right]
-2g^{2}\,\mathrm{Tr}\,\left[\partial_{-}\hat{A}^{\bullet},\hat{A}^{\star}\right]\partial_{-}^{-2}\left[\partial_{-}\hat{A}^{\star},\hat{A}^{\bullet}\right]
\Bigg\}
\,.\label{eq:YM_LC_action}
\end{multline}
To derive the light cone Yang-Mills action Eq.~\eqref{eq:YM_LC_action}, one starts with the fully covariant form of the Yang-Mills action and re-express it in terms of the double-null coordinates using
\begin{gather} 
 {\hat A}^{+}=\frac{1}{\sqrt{2}} \left({\hat A}^0+{\hat A}^3\right) \,, \, {\hat A}^{-}=\frac{1}{\sqrt{2}} \left({\hat A}^0-{\hat A}^3\right) \,, \nonumber\\
 {\hat A}^{\bullet}=-\frac{1}{\sqrt{2}} \left({\hat A}^1+i{\hat A}^2\right) \,,\,{\hat A}^{\star}=-\frac{1}{\sqrt{2}} \left({\hat A}^1-i{\hat A}^2\right) \,.
\end{gather}
Imposing the light-cone gauge ${\hat A}\cdot\eta={\hat A}^{+}=0$ reduces the unphysical degrees of freedom by one. The resulting action turns out to be quadratic in ${\hat A}^{-}$, which can be integrated out of the partition function. By doing this we get the action in Eq.~\eqref{eq:YM_LC_action} which depends only on two field components ${\hat A}^{\bullet}$, ${\hat A}^{\star}$. These correspond to the two physical polarization states of a gluon in the on-shell limit. In our convention, $\bullet$ represents a plus helicity and $\star$ represents a minus helicity gluon. As a result, we can schematically express the action as
\begin{equation} S_{\mathrm{YM}}\left[A^{\bullet},A^{\star}\right]=\int\! dx^{+} \Big( \mathcal{L}_{+ -} + \mathcal{L}_{+ + -} + \mathcal{L}_{- - +} + \mathcal{L}_{+ + - -}\Big) \, ,
    \label{eq:SYM_com}
\end{equation}
where $\mathcal{L}_{+ -}$ is the kinetic term
\begin{equation}
    \mathcal{L}_{+ -}\left[A^{\bullet},A^{\star}\right]= \int d^{3}\mathbf{x}\, \left\{-\mathrm{Tr}\,\hat{A}^{\bullet}\square\hat{A}^{\star}\right\}\,,
\end{equation}
and $\mathcal{L}_{+ + -},  \mathcal{L}_{- - +} $ are the two triple gluon interaction vertices and $\mathcal{L}_{+ + - -}$ is the four point vertex.

\begin{figure}
    \centering
    \includegraphics[width=14cm]{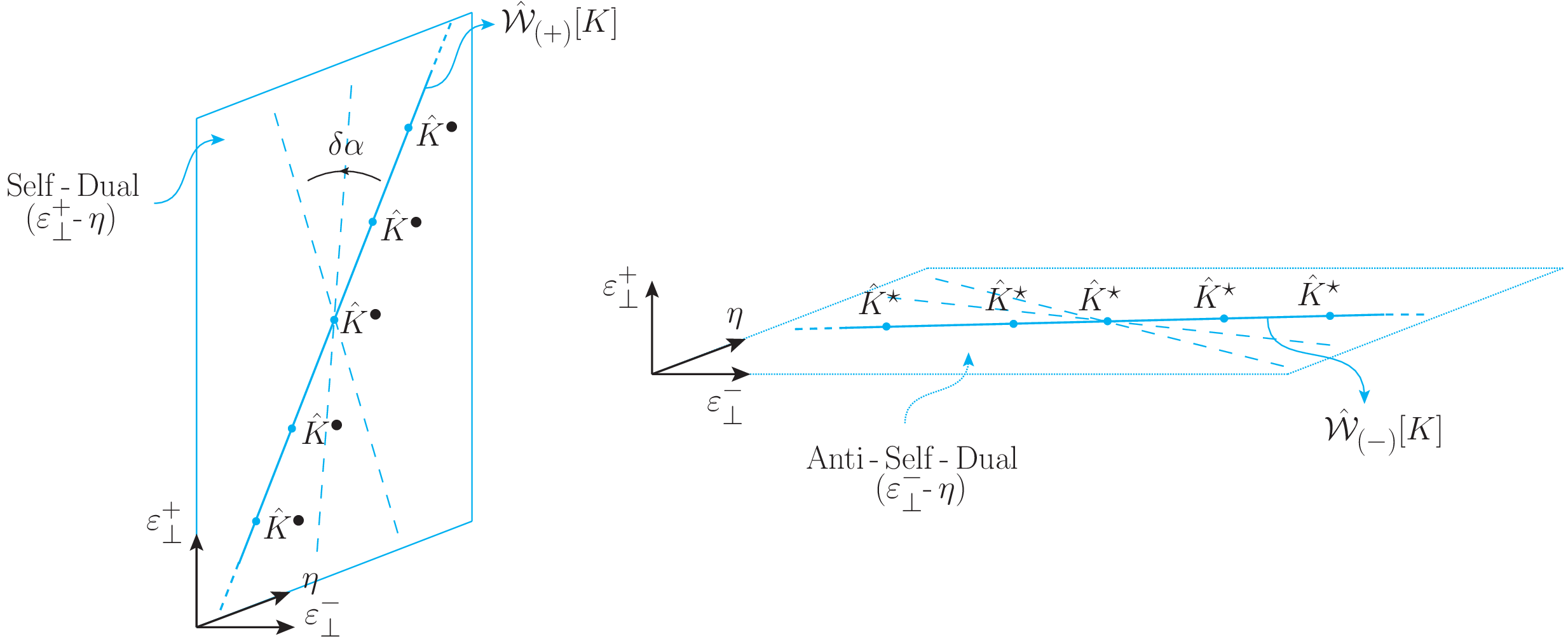}
    \caption{
    \small  
    On the left, we have the graphical representation of the straight-infinite Wilson lines contributing to the functional  $\mathcal{W}_{(+)}^a[K](x)$ of some $\hat{K}^{\bullet}$ fields. They are along the four vector $\varepsilon_{\alpha}^{+\, \mu} = \varepsilon_{\perp}^{+\, \mu }- \alpha \eta^{\mu}$ where $\alpha$ is the slope parameter. Varying the slope by $\delta\alpha$ as shown in the figure rotates the Wilson line from solid to dashed. This parameter is integrated over in the definition of $\mathcal{W}_{(+)}^a[K](x)$. The Wilson line 
    lives on the plane spanned by $\varepsilon_{\perp}^{+\, \mu } $ , $\eta^{\mu}$ which we call the Self-Dual plane. Similarly, on the right, we graphically represent the Wilson line $\mathcal{W}_{(+)}^a[K](x)$ of $\hat{K}^{\star}$ fields. It, however, lives on the 
    plane spanned by $\varepsilon_{\perp}^{-\, \mu } $ , $\eta^{\mu}$ which we call the Anti-Self-Dual plane.}
    \label{fig:WL_BSD}
\end{figure}

The Z-field action Eq.~\eqref{eq:Z_action}  is obtained by canonically transforming the Yang-Mills fields to a new pair of fields
\begin{equation}
    \left\{\hat{A}^{\bullet},\hat{A}^{\star}\right\} \rightarrow \Big\{\hat{Z}^{\bullet}\big[{A}^{\bullet},{A}^{\star}\big],\hat{Z}^{\star}\big[{A}^{\bullet},{A}^{\star}\big]\Big\} \, ,
    \label{eq:general_transf}
\end{equation}
via the following generating functional for canonical transformations
\begin{equation}
   \mathcal{G}[A^\bullet,Z^\star](x^+) =
    -\int\! d^3\mathbf{x}\,\,\,\Tr\,
     \hat{\mathcal{W}}^{\,-1}_{(-)}[Z](x)\,\,
     \partial_- \hat{\mathcal{W}}_{(+)}[A](x) \,,
    \label{eq:generatingfunc3}
\end{equation}
where $\mathcal{W}_{(\pm)}^a[K](x)$  represent functionals of some  ${\hat K}$ field (at fixed Minkowski point $x$) obtained from a straight infinite Wilson line  along the vector $\varepsilon_{\alpha}^{\pm}$, with the following explicit definition
\begin{equation}
     \mathcal{W}^{a}_{(\pm)}[K](x)=\int_{-\infty}^{\infty}d\alpha\,\mathrm{Tr}\left\{ \frac{1}{2\pi g}t^{a}\partial_{-}\, \mathbb{P}\exp\left[ig\int_{-\infty}^{\infty}\! ds\, \varepsilon_{\alpha}^{\pm}\cdot \hat{K}\left(x+s\varepsilon_{\alpha}^{\pm}\right)\right]\right\} \, .
\label{eq:WL_gen}
\end{equation}
The path-ordered exponential is along the four vector $\varepsilon_{\alpha}^{\pm\, \mu}$ defined as
\begin{equation}
    \varepsilon_{\alpha}^{\pm\, \mu} = \varepsilon_{\perp}^{\pm\, \mu }- \alpha \eta^{\mu} \, ,
    \label{eq:epsilon_alpha}
\end{equation}
where $\varepsilon_{\perp}^{\pm\, \mu }$ and $\eta^{\mu}$ were given in Eq.~\eqref{eq:DNC_basis}. The Wilson lines $\mathcal{W}_{(\pm)}^a[K](x)$ live on the planes spanned by $\varepsilon_{\perp}^{\pm\,\mu} $, $ \eta^{\mu}$, respectively, along a generic vector of "slope" $\alpha$ which gets integrated over (see Figure \ref{fig:WL_BSD}). 

In what follows, we shall simply call the functionals $\mathcal{W}_{(\pm)}$ "Wilson lines", although, strictly speaking, they rather resemble surface integrals.

The $\varepsilon_{\perp}^{+\,\mu} - \eta^{\mu}$ plane will be often referred to as a Self-Dual plane whereas the $\varepsilon_{\perp}^{-\,\mu} - \eta^{\mu}$ plane as an Anti-Self-Dual (we collect the definitions of Self-Dual and Anti-Self-Dual planes in Appendix \ref{sec:App01}). $\mathcal{W}^{\,-1}$ is the inverse of the functional $\mathcal{W}$ defined as $\mathcal{W}[\mathcal{W}^{-1}[K]]=K$. 

Notice, the generating functional Eq.~\eqref{eq:generatingfunc3} is defined over the constant light-cone time $x^+$ hyper-surface due to which the pair of new fields ${\hat Z}^{\bullet}(x^+;\mathbf{y}), {\hat Z}^{\star}(x^+;\mathbf{y})$ and the old ${\hat A}^{\bullet}(x^+;\mathbf{x}), {\hat A}^{\star}(x^+;\mathbf{x})$ have the same $x^+$.  Using Eq.~\eqref{eq:generatingfunc3}, we can write the following explicit relations for the conjugate momenta 
\begin{equation}
\partial_{-}A^{\star}_a(x^+,\mathbf{y}) =  \frac{\delta \, \mathcal{G}[A^{\bullet},Z^{\star} ](x^+)}{\delta A_a^{\bullet}\left(x^+,\mathbf{y}\right)} \,, \qquad \qquad \partial_{-}Z^{\bullet}_a(x^+,\mathbf{y}) = - \frac{\delta \, \mathcal{G}[A^{\bullet},Z^{\star} ](x^+)}{\delta Z_a^{\star}\left(x^+,\mathbf{y}\right)} \,. \label{eq:AtoZ_ct}
\end{equation}
It turns out that this transformation essentially maps the kinetic term $(+ -)$ and both the triple gluon vertices $(+ + -)$ and $(- - +)$ in the Yang-Mills action Eq.~\eqref{eq:YM_LC_action} to the sole kinetic term $(+ -)$ in the new action $S\left[Z^{\bullet}, Z^{\star}\right]$ Eq.~\eqref{eq:Z_action}. The transformation is canonical to preserve the integral measure in the partition function up to a field independent factor. 

The solution of the transformation Eq.~\eqref{eq:AtoZ_ct} has the following form
\begin{equation}
    A_a^{\bullet}(x^+;\mathbf{x})=\sum_{n=1}^{\infty}
    \int\! d^3\mathbf{y}_1\dots d^3\mathbf{y}_n \sum_{i=1}^{n}\, \Xi_{i,n-i}^{ab_1\dots b_n}(\mathbf{x};\mathbf{y}_1,\dots,\mathbf{y}_n) \prod_{k=1}^{i}Z_{b_k}^{\bullet}(x^+;\mathbf{y}_k)
    \prod_{l=i+1}^{n}Z_{b_l}^{\star}(x^+;\mathbf{y}_l) \,,
    \label{eq:Abullet_to_Z}
\end{equation}
\begin{equation}
    A_a^{\star}(x^+;\mathbf{x})=\sum_{n=1}^{\infty}
    \int\! d^3\mathbf{y}_1\dots d^3\mathbf{y}_n \sum_{i=1}^{n}\, \Lambda_{i,n-i}^{ab_1\dots b_n}(\mathbf{x};\mathbf{y}_1,\dots,\mathbf{y}_n) \prod_{k=1}^{i}Z_{b_k}^{\star}(x^+;\mathbf{y}_k)
    \prod_{l=i+1}^{n}Z_{b_l}^{\bullet}(x^+;\mathbf{y}_l) \,,
    \label{eq:Astar_to_Z}
\end{equation}
where the kernels, in momentum space, read (for derivation see Appendix \ref{sec:App0})
\begin{multline}
   \widetilde\Xi_{i,n-i}^{a\{b_1\dots b_i\} \{ b_{i+1} \dots b_n\}}(\mathbf{P};\{\mathbf{p}_1,\dots,\mathbf{p}_i\},\{\mathbf{p}_{i+1} \dots \mathbf{p}_n\}) = - (-g)^{n-1} \, 
    \delta^{3} (\mathbf{p}_{1} + \cdots +\mathbf{p}_{n} - \mathbf{P})\,\,  \mathrm{Tr} (t^{a} t^{b_{1}} \cdots t^{b_{n}})\\ 
    \frac{{\widetilde v}^{\star}_{(1 \cdots n)1}}{{\widetilde v}^{\star}_{1(1 \cdots n)}} \,\frac{1}{{\widetilde v}^{\star}_{21}{\widetilde v}^{\star}_{32} \cdots {\widetilde v}^{\star}_{(i-1)(i-2)}{\widetilde v}^{\star}_{(i\dots n)(i-1)}}    \left(\frac{p_i^+}{p_{i \dots n}^+}\right)  \, 
    \frac{1}{{\widetilde v}_{(i+1)i}{\widetilde v}_{(i+2)(i+1)} \cdots {\widetilde v}_{n(n-1)}}  \, ,
    \label{eq:xi_kernel_mom}
\end{multline}
\begin{multline}
\widetilde\Lambda_{i,n-i}^{a\{b_1\dots b_i\} \{ b_{i+1} \dots b_n\}}(\mathbf{P};\{\mathbf{p}_1,\dots,\mathbf{p}_i\},\{\mathbf{p}_{i+1} \dots \mathbf{p}_n\})
=    (-g)^{n-1} \, 
    \delta^{3} (\mathbf{p}_{1} + \cdots +\mathbf{p}_{n} - \mathbf{P})\,\,  \mathrm{Tr} (t^{a} t^{b_{1}} \cdots t^{b_{n}})  
      \, \\
 \sum_{k=1}^{i}   \left(\frac{p_{1\cdots k}^+}{p_{1\cdots n}^+}\right)  \,\,\frac{1}{{\widetilde v}^{\star}_{(k+1 \dots i+1)(1\cdots k)}{\widetilde v}^{\star}_{(i+2)(k+1 \cdots i+1)} {\widetilde v}^{\star}_{(i+3)(i+2)}\cdots {\widetilde v}^{\star}_{n(n-1)}}  \\ 
    \frac{{\widetilde v}_{(1 \cdots k)1}}{{\widetilde v}_{1(1 \cdots k)}} \, 
    \frac{1}{{\widetilde v}_{21}{\widetilde v}_{32} \cdots {\widetilde v}_{k(k-1)}}  
      \,     \left(\frac{p_{i+1}^+}{p_{k+1\cdots i+1}^+}\right)^2   \,\,\frac{{\widetilde v}_{(k+1\cdots i+1)(k+1)}}{{\widetilde v}_{(k+1)(k+1\cdots i+1)}}   
     \, 
    \frac{1}{{\widetilde v}_{(k+2)(k+1)}\cdots {\widetilde v}_{(i+1)i}}   \, .
    \label{eq:lambda_kernel_mom}
\end{multline}
Above, we use a shorthand notation for the sum of momenta $\mathbf{p}_1+\dots +\mathbf{p}_n\equiv \mathbf{p}_{1\dots n}$.  The curly braces on the color and momentum indices imply symmetry with respect to the interchange of these separately for the minus and the plus helicity fields. The quantities $\widetilde{v}_{ij}$, $\widetilde{v}^{\star}_{ij}$, first introduced in \cite{Motyka2009} (also see \cite{Cruz-Santiago2015} for several useful properties of these  symbols), are analogous to conventionally used spinor products $\left<ij\right>$, $\left[ij\right]$, with the following explicit definitions :
\begin{equation}
    \widetilde{v}_{ij}=
    p_i^+\left(\frac{p_{j}^{\star}}{p_{j}^{+}}-\frac{p_{i}^{\star}}{p_{i}^{+}}\right), \qquad 
\widetilde{v}^*_{ij}=
    p_i^+\left(\frac{p_{j}^{\bullet}}{p_{j}^{+}}-\frac{p_{i}^{\bullet}}{p_{i}^{+}}\right)\, .
\label{eq:vtilde}
\end{equation}
Notice, however, that these symbols can be used also for off-shell momenta, thus they provide a particular off-shell continuation of the spinor products.
They appear quite naturally in the Wilson line approach, because 
\begin{equation}
    \widetilde{v}^*_{ij}=-(\varepsilon_i^+\cdot p_j)\,, \quad 
    \widetilde{v}_{ij}=-(\varepsilon_i^-\cdot p_j )\, ,
    \label{eq:vtilde1}
\end{equation}
where $\varepsilon^{\pm}_i$ is the polarization vector for a momentum $\mathbf{p}_i\equiv\left(p_i^{+},p_i^{\bullet},p_i^{\star}\right)$ obtained from Eq.~\eqref{eq:epsilon_alpha}. The latter determines the direction of the Wilson line. 

\begin{figure}
    \centering
    \includegraphics[width=16cm]{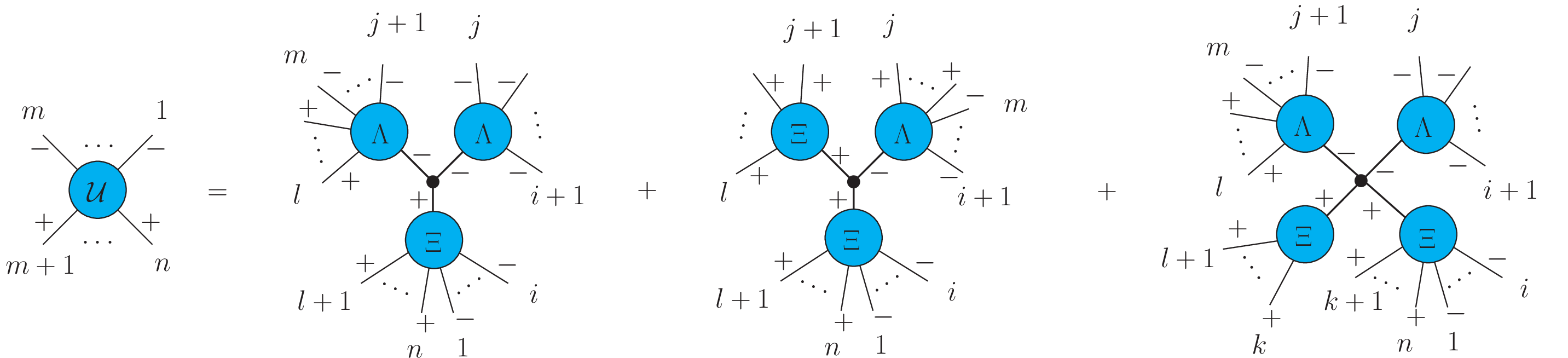}
    \caption{
    \small  
    L.H.S.: Color ordered $n$-point interaction vertex in the Z-field action with $m$ minus helicity fields. R.H.S.: Contributions to the vertex originating from the substitution of the color ordered solutions $\hat{A}^{\bullet}[{Z}^{\bullet},{Z}^{\star}],\hat{A}^{\star}[{Z}^{\bullet},{Z}^{\star}]$ Eqs.~\eqref{eq:Abullet_to_Z}-\eqref{eq:Astar_to_Z} of the canonical transformation to the color ordered Yang-Mills interaction vertices represented via $\bullet$ with three and four legs. Summing over all the contributions gives Eq.~\eqref{eq:Zver_YM}.  $\Xi$ and $\Lambda$ are the kernels of the solutions $\hat{A}^{\bullet}[{Z}^{\bullet},{Z}^{\star}],\hat{A}^{\star}[{Z}^{\bullet},{Z}^{\star}]$ respectively.}
    \label{fig:Z_ver_n}
\end{figure}

Substituting Eqs.~\eqref{eq:Abullet_to_Z}-\eqref{eq:Astar_to_Z} to the Yang-Mills action Eq.~\eqref{eq:YM_LC_action} derives the Z-field action Eq.~\eqref{eq:Z_action} (for the cancellation of both the triple gluon vertices see \cite{Kakkad:2021uhv}). Therefore, using this substitution, one can write the general expression for the vertices in the Z-field action Eq.~\eqref{eq:Z_action}. To do this, consider an $n$-point vertex with $m$ minus helicity fields shown below
\begin{equation}
    \mathcal{L}_{\underbrace{-\,\cdots\,-}_{m}\underbrace{+ \,\cdots\, +}_{n-m}}= 
   \int\!d^{3}\mathbf{p}_{1}\dots d^{3}\mathbf{p}_{n} \,\, \mathcal{U}^{b_1 \dots b_{n}}_{-\dots-+\dots+}\left(\mathbf{p}_{1},\cdots \mathbf{p}_{n}\right) 
   \prod_{i=1}^{m}Z^{\star}_{b_i} (x^+;\mathbf{p}_{i})
   \prod_{j=1}^{n-m}Z^{\bullet}_{b_j} (x^+;\mathbf{p}_{j}) \, .
   \label{eq:Z_vertex_lagr_mom}
\end{equation}
For the sake of simplicity, we color-decompose the vertex as follows
\begin{equation}
    \mathcal{U}_{-\dots-+\dots+}^{b_{1}\dots b_{n}}\left(\mathbf{p}_{1},\dots,\mathbf{p}_{n}\right)= \!\!\sum_{\underset{\text{\scriptsize permutations}}{\text{noncyclic}}}
 \mathrm{Tr}\left(t^{b_1}\dots t^{b_n}\right)
 \mathcal{U}\left(1^-,\dots,m^-,(m+1)^+,\dots,n^+\right)
\,,
\label{eq:Zvertex_color_decomp}
\end{equation}
where the numbers $i$ in the color-ordered vertex represent the momentum $\mathbf{p}_{i}$ and to these numbers we assign the helicities of the associated legs. Following the substitution, the generic expression for the color-ordered vertex reads (shown diagrammatically in Figure \ref{fig:Z_ver_n})
\begin{multline}
    \mathcal{U}\left(1^-,\dots,m^-,(m+1)^+,\dots,n^+\right) =  \sum_{i=0}^{m-2}\sum_{j=i+1}^{m-1}\sum_{l=m}^{n-1}
    \mathcal{V}_3\left(\,[i\!+\!1,\dots,j]^-,[j\!+\!1,\dots,l]^-,[l\!+\!1,\dots,i]^+\right) \\
    \widetilde{ \Lambda}_{j-i,0}\left((i\!+\!1)^-,\dots,j^-\right) \,\,
    \widetilde{ \Lambda}_{m-j,l-m}\left((j\!+\!1)^-,\dots,m^-,(m\!+\!1)^+,\dots,l^+\right) \,\, 
    \widetilde{ \Xi}_{n-l,i}\left((l\!+\!1)^+,\dots,n^+,1^-,\dots,i^-\right) \\
   + \sum_{i=0}^{m-1}\sum_{j=m}^{n-2}\sum_{l=j+1}^{n-1}
    \overline{\mathcal{V}}_3\left(\,[i\!+\!1,\dots,j]^-,[j\!+\!1,\dots,l]^+,[l\!+\!1,\dots,i]^+\right) \\
    \widetilde{ \Lambda}_{m-i,j-m}\left((i\!+\!1)^-,\dots,m^-, (m\!+\!1)^+,\dots,j^+\right) \,\,
    \widetilde{ \Xi}_{l-j,0}\left((j\!+\!1)^+,\dots,l^+\right) \,\, 
    \widetilde{ \Xi}_{n-l,i}\left((l\!+\!1)^+,\dots,n^+,1^-,\dots,i^-\right) \\
   + \sum_{i=0}^{m-2}\sum_{j=i+1}^{m-1}\sum_{l=m}^{n-2}\sum_{k=l+1}^{n-1}
    \mathcal{V}_4\left(\,[i\!+\!1,\dots,j]^-,[j\!+\!1,\dots,l]^-,[l\!+\!1,\dots,k]^+,[k\!+\!1,\dots,i]^+\right) \\
    \widetilde{ \Lambda}_{j-i,0}\left((i\!+\!1)^-,\dots,j^-\right) \,\,
    \widetilde{ \Lambda}_{m-j,l-m}\left((j\!+\!1)^-,\dots,m^-,(m\!+\!1)^+,\dots,l^+\right) \,\, \widetilde{ \Xi}_{k-l,0}\left((l\!+\!1)^+,\dots,k^+\right) \\
    \widetilde{ \Xi}_{n-k,i}\left((k\!+\!1)^+,\dots,n^+,1^-,\dots,i^-\right)\,.
    \label{eq:Zver_YM}
\end{multline}
Above, $\mathcal{V}_3(i^-,j^-,k^+)$, $\overline{\mathcal{V}}_3(i^-,j^+,k^+)$, and $\mathcal{V}_4(i^-,j^-,k^+, l^+)$ represent the color-ordered Yang-Mills vertices $(- - +)$, $(- + +)$ and $(- - + +)$ respectively. $[i,\dots,j] \equiv \mathbf{p}_{i} + \dots +\mathbf{p}_{j}$. We also used the color-ordered versions of the kernels Eqs.~\eqref{eq:xi_kernel_mom}-\eqref{eq:lambda_kernel_mom}.

\begin{figure}
    \centering
    \includegraphics[width=10cm]{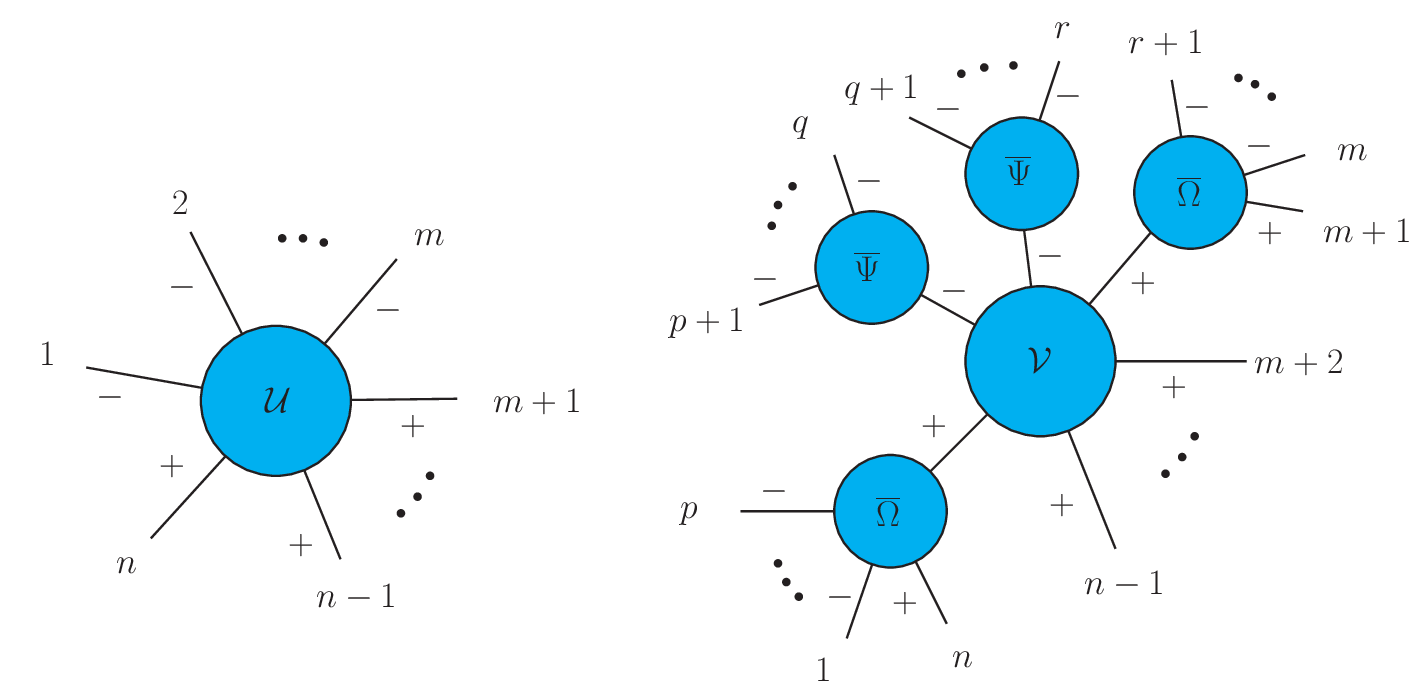}
    \caption{
    \small  
    On the left, we have the color ordered $n$-point interaction vertex in the Z-field action with $m$ minus helicity fields. On the right, we have the contributions to the vertex originating from the substitution of the color ordered solutions $\hat{B}^{\bullet}[{Z}^{\bullet},{Z}^{\star}],\hat{B}^{\star}[{Z}^{\star}]$ Eqs.~\eqref{eq:BbulletZ_exp}-\eqref{eq:BstarZ_exp} of the canonical transformation Eq.~\eqref{eq:BtoZtransform} to the color ordered interaction vertices of the MHV action Eq.~\eqref{eq:MHV_action}. Summing over all the contributions gives Eq.~\eqref{eq:Zver_MHV}.  $\overline{\Omega}$ and $\overline{\Psi}$ are the kernels of the solutions $\hat{B}^{\bullet}[{Z}^{\bullet},{Z}^{\star}],\hat{B}^{\star}[{Z}^{\star}]$ respectively. This image was modified from our paper \cite{Kakkad:2021uhv}.}
    \label{fig:Z_ver_o}
\end{figure}

The above expression can be simplified to the vertices derived in \cite{Kakkad:2021uhv} via canonical transformation of the MHV action (see Figure \ref{fig:Z_ver_o})
\begin{multline}
    \mathcal{U}\left(1^-,2^-,\dots,m^-,(m\!+\!1)^+,\dots,n^+\right) = 
    \sum_{p=0}^{m-2}\sum_{q=p+1}^{m-1}\sum_{r=q+1}^{m}\\
    \mathcal{V}_{\mathrm{MHV}}\left(\,[p\!+\!1,\dots,q]^-,[q\!+\!1,\dots,r]^-,[r\!+\!1,\dots,m\!+\!1]^+,(m\!+\!2)^+,\dots,(n\!-\!1)^+,[n,1,\dots,p]^+\right) \\
    \overline{ \Omega}\left(n^+,1^-,\dots,p^-\right) \,\,
    \overline{ \Psi}\left((p\!+\!1)^-,\dots,q^-\right) \,\, 
    \overline{ \Psi}\left((q\!+\!1)^-,\dots,r^-\right) \,\, 
    \overline{ \Omega}\left((r\!+\!1)^-,\dots,m^-,(m\!+\!1)^+\right) \,\,,
    \label{eq:Zver_MHV}
\end{multline}
where $\mathcal{V}_{\mathrm{MHV}}$ is the color-ordered MHV vertex. In our convention, it reads
\begin{equation}
\mathcal{V}_{\mathrm{MHV}}\left(1^-,2^-,3^+,\dots,n^+\right)= 
\frac{(-g)^{n-2}}{(n-2)!}  \left(\frac{p_{1}^{+}}{p_{2}^{+}}\right)^{2}
\frac{\widetilde{v}_{21}^{\star 4}}{\widetilde{v}_{1n}^{\star}\widetilde{v}_{n\left(n-1\right)}^{\star}\widetilde{v}_{\left(n-1\right)\left(n-2\right)}^{\star}\dots\widetilde{v}_{21}^{\star}}
\,,
\label{eq:MHV_vertex_colororder}
\end{equation}
and  $ \overline{ \Psi}$, $ \overline{ \Omega}$ are the kernels Eqs.~\eqref{eq:psiBar_kernel}-\eqref{eq:omegaBar_kernel} of the solution of the transformation Eq.~\eqref{eq:Bfield_transform} that derives the Z-field action from the MHV action (see Appendix \ref{sec:App0}). 

In the on-shell limit, the formula Eq.~\eqref{eq:Zver_MHV} is more appropriate to reveal the underlying structure of the interaction vertices in our action in the twistor space. To see this, consider the 6-point NMHV $(- - - + + +)$ vertex. It reads
\begin{multline}
    \mathcal{U}\left(1^-,2^-,3^-,4^+,5^+,6^+\right) = 
     \mathcal{V}_{\mathrm{MHV}}\left(\,[1,2]^-,3^-,4^+,5^+,6^+\right) 
    \overline{ \Psi}_2\left(1^-,2^-\right)    \\
   + \mathcal{V}_{\mathrm{MHV}}\left(1^-,\,[2,3]^-,4^+,5^+,6^+\right) 
    \overline{ \Psi}_2\left(2^-,3^-\right)
    + \mathcal{V}_{\mathrm{MHV}}\left(\,1^-,2^-,[3,4]^+,5^+,6^+\right) 
    \overline{ \Omega}_2\left(4^+,3^-\right) \\
    + \mathcal{V}_{\mathrm{MHV}}\left(\,2^-,3^-,4^+,5^+,[6,1]^+\right) 
    \overline{ \Omega}_2\left(6^+,1^-\right)\,\,.
    \label{eq:Zver_6NMHV}
\end{multline}

The first two terms in the expression above are a convolution of 5-point MHV with $\overline{ \Psi}_2$ kernel. The latter, in terms of the energy denominator, reads
\begin{equation}
    \overline{\widetilde \Psi}_{2}\left(i^{\pm}, j^{\pm} \right) =-g \,\, 
    \frac{p_{ij}^+}{p_{i}^+} \, 
    \frac{1}{{\widetilde v}_{ji}  }
      \,  = g \,\, 
    \frac{p_{ij}^+}{p_{i}^+} \, 
    \frac{\widetilde{v}^{\star}_{ij}}{p_{ij}^+(\hat{p}_i + \hat{p}_j -\hat{p}_{ij})  }
      \, =   \frac{\mathcal{V}_{\mathrm{MHV}}\left(\,i^-,j^-,[i,j]^+\right) }{p_{ij}^+(\hat{p}_i + \hat{p}_j -\hat{p}_{ij})  } \,.
      \label{eq:psi_off}
\end{equation}
Above, the first expression follows from Eq.~\eqref{eq:psiBar_kernel}. In going from the first expression to the second we used the identity  $p_{ij}^+(\hat{p}_i + \hat{p}_j -\hat{p}_{ij}) = - \widetilde{v}_{ji}\widetilde{v}^{\star}_{ij} $
where
\begin{equation}
    \hat{q} = q^{\bullet} q^{\star}/ q^{+}\,.
\end{equation}
And, in going from the second expression to the third we used the momentum space expression for the $(- - +)$ triple point MHV vertex in the Yang-Mills action Eq.~\eqref{eq:YM_LC_action}. So far, the above expressions are off-shell. Imposing the on-shellness of the external particles we have
\begin{equation}
  p_{i}^2\rightarrow 0 \quad \quad \Rightarrow \quad \quad p_i^- = p_i^{\bullet} p_i^{\star}/ p_i^{+}\equiv  \hat{p}_i  \, .
  \label{eq:hat}
\end{equation}
Using it in Eq.~\eqref{eq:psi_off} we get
\begin{equation}
    \left.\overline{\widetilde \Psi}_{2}\left(i^{\pm}, j^{\pm} \right)\right|_{\mathrm{on-shell}}  =   \frac{\mathcal{V}_{\mathrm{MHV}}\left(\,i^-,j^-,[i,j]^+\right) }{p_{ij}^+(p_i^- + p_j^- -\hat{p}_{ij})  } = 2\frac{\mathcal{V}_{\mathrm{MHV}}\left(\,i^-,j^-,[i,j]^+\right) }{p_{ij}^2  }\,,
     \label{eq:psi_on}
\end{equation}
where we used the momentum conserving delta for the minus component of the momentum $\delta(p_i^- + p_j^- - p_{ij}^-)$ to go from the first expression to the second.

Repeating the above procedure for the other terms in Eq.~\eqref{eq:Zver_6NMHV}, using Eq.~\eqref{eq:omegaBar_kernel} for the last two terms, we get
\begin{multline}
   \left. \mathcal{U}\left(1^-,2^-,3^-,4^+,5^+,6^+\right)\right|_{\mathrm{on-shell}}  = 
     2\, \mathcal{V}_{\mathrm{MHV}}\left(\,[1,2]^-,3^-,4^+,5^+,6^+\right) 
    \frac{\mathcal{V}_{\mathrm{MHV}}\left(\,1^-,2^-,[1,2]^+\right) }{p_{12}^2  }   \\
   +2\, \mathcal{V}_{\mathrm{MHV}}\left(1^-,\,[2,3]^-,4^+,5^+,6^+\right) 
    \frac{\mathcal{V}_{\mathrm{MHV}}\left(\,2^-,3^-,[2,3]^+\right) }{p_{23}^2  } \\
    + 2\,\mathcal{V}_{\mathrm{MHV}}\left(\,1^-,2^-,[3,4]^+,5^+,6^+\right) 
    \frac{\mathcal{V}_{\mathrm{MHV}}\left(\,3^-,[3,4]^-,4^+\right) }{p_{34}^2  } \\
    + 2\, \mathcal{V}_{\mathrm{MHV}}\left(\,2^-,3^-,4^+,5^+,[6,1]^+\right)
    \frac{\mathcal{V}_{\mathrm{MHV}}\left(\,[1,6]^-,1^-,6^+\right) }{p_{16}^2  } \,\,.
    \label{eq:Zver_6NMHV2}
\end{multline}
The above expression demonstrates that in the on-shell limit, the 6-point NMHV interaction vertex in our action can be factorized into all possible ways of connecting a 5-point MHV to a 3-point MHV via a scalar propagator. This picture generalizes to the other interaction vertices in our action as follows. In the on-shell limit the kernels $\overline{ \Psi}_n\left(1^-,\dots,n^-\right)$ and $\overline{ \Omega}_n\left(1^+,2^-,\dots,n^-\right)$ in Eq.~\eqref{eq:Zver_MHV} can be unfolded into trees consisting of a bunch of 3-point MHV vertices connected via scalar propagators. Based on these observations we can make the following conclusions
\begin{itemize}
    \item In the on-shell limit, the interaction vertices in our action 
    can be decomposed to only MHV vertices connected via propagators. In the twistor space, it corresponds to a set of terms each of which represents a bunch of $\mathbb{CP}^1$'s (degree-one curves),  one for each MHV vertex, connected via propagators.
    \item In the on-shell limit, the interaction vertices in our action, except for the MHV and the $\overline{\mathrm{MHV}}$ vertices, cannot localize on a single higher degree curve and must localize on a subset of degenerations of a higher degree curve into a bunch of $\mathbb{CP}^1$'s.  Consider for the sake of simplicity the 6-point NMHV tree level amplitude. There are three contributions to this amplitude in our action \cite{Kakkad:2021uhv} (see Figure~\ref{fig:NMHV6}). According to the completely connected approach recalled in Section~\ref{sec:Intro}, the amplitude localizes on a degree-two curve. The equivalence between the completely connected and completely disconnected approach (the CSW rules) states that the amplitude can be represented as a set of all possible degenerations of the degree-two curve into a pair of $\mathbb{CP}^1$'s. In terms of diagrams, it means that the amplitude can be computed using the CSW rules by considering all possible contributions using MHV vertices. The first two diagrams in Figure~\ref{fig:NMHV6} correspond to the contributions involving a pair of 4-point MHVs and these therefore account for a subset of degenerations involving $\mathbb{CP}^1$'s. The third contribution is from the 6-point NMHV interaction vertex which, as we demonstrated above, corresponds to the remaining diagram in the CSW approach involving a 3-point MHV connected to a 5-point MHV and therefore the interaction vertex in our action accounts for the remaining subset of degenerations. This is true for the other vertices as well. In general, the interaction vertices in our action only consist of a subset of degenerations of a higher degree curve into a bunch of $\mathbb{CP}^1$'s, or equivalently, these consist of a subset of Feynman diagrams required to compute an amplitude in the CSW approach and can therefore not localize on a single higher degree curve in twistor space. The MHV and the $\overline{\mathrm{MHV}}$ vertices are exceptions. The former is trivial to realize and for the latter, an exercise similar to the one we did above for the 6-point NMHV interaction vertex will demonstrate that these consist of the entire subset of degenerations or Feynman diagrams in the CSW approach. As a result, these alone give the corresponding amplitudes in the on-shell limit and thus would localize on a single higher degree curve in the twistor space. 
    \begin{figure}[h]
    \centering
 \includegraphics[width=13cm]{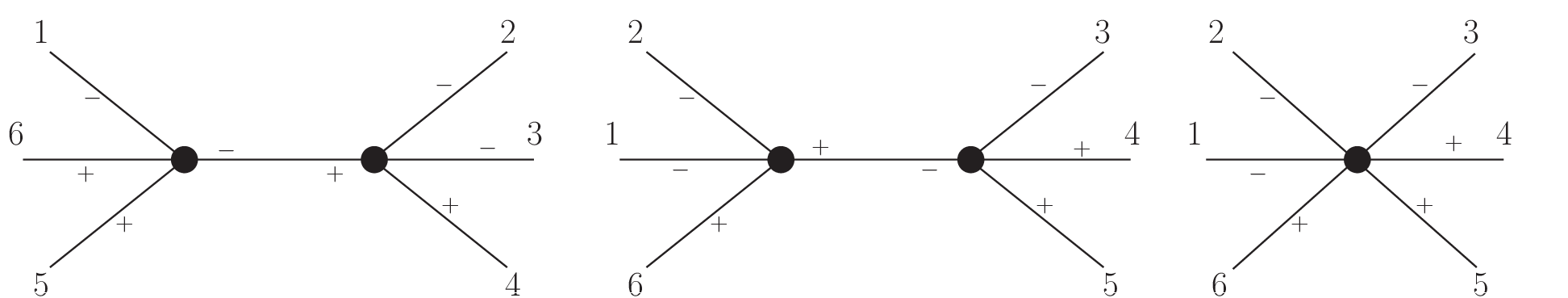}
    \caption{\small 
    Contributions obtained when computing the split-helicity 6-point NMHV $(- - - + + +)$ amplitude using the Z-field action Eq.~\eqref{eq:Z_action}. This image was taken from our paper \cite{Kakkad:2021uhv}.}
    \label{fig:NMHV6}
\end{figure}

\end{itemize}
Owing to the above observation, we believe that in the on-shell limit, the Z-field action should provide in the Minkowski space an explicit realization of the possibility of computing tree-level amplitudes using interaction vertices which in the twistor space corresponds to using curves of different degrees as was conjectured in \cite{gukov2004equivalence}.

\begin{figure}
    \centering
    \includegraphics[width=11cm]{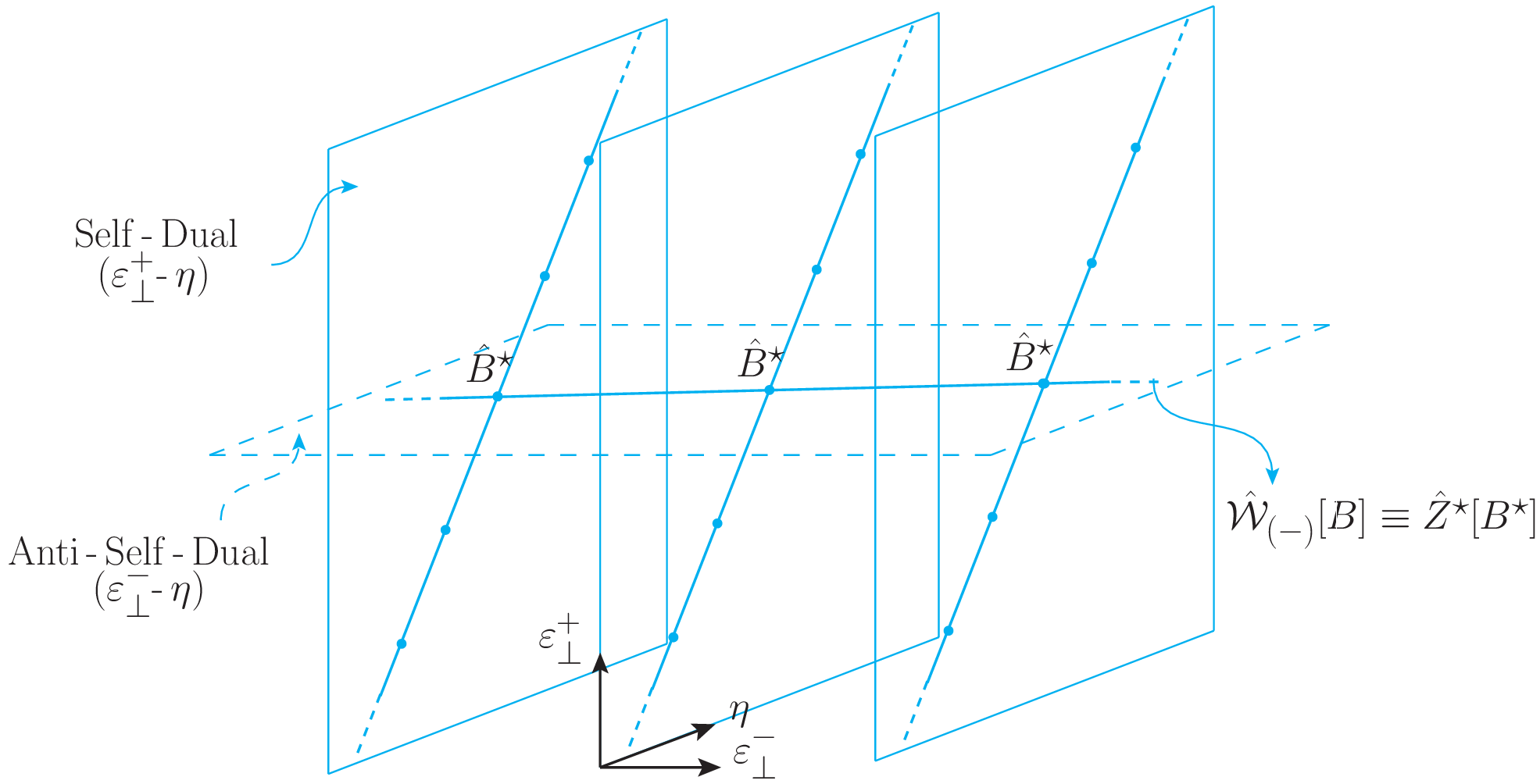}
    \caption{
    \small  $Z^{\star}$ field as a straight infinite Wilson line of the negative helicity field in the MHV action $B^{\star}$ Eq.~\eqref{eq:MHV_action} on the Anti-Self-Dual plane.  The latter itself is a straight infinite Wilson line of both the Yang-Mills fields $A^{\bullet}$ and $A^{\star}$ on the Self-Dual plane.}
    \label{fig:Zstar_WL}
\end{figure}

\begin{figure}
    \centering
    \includegraphics[width=11cm]{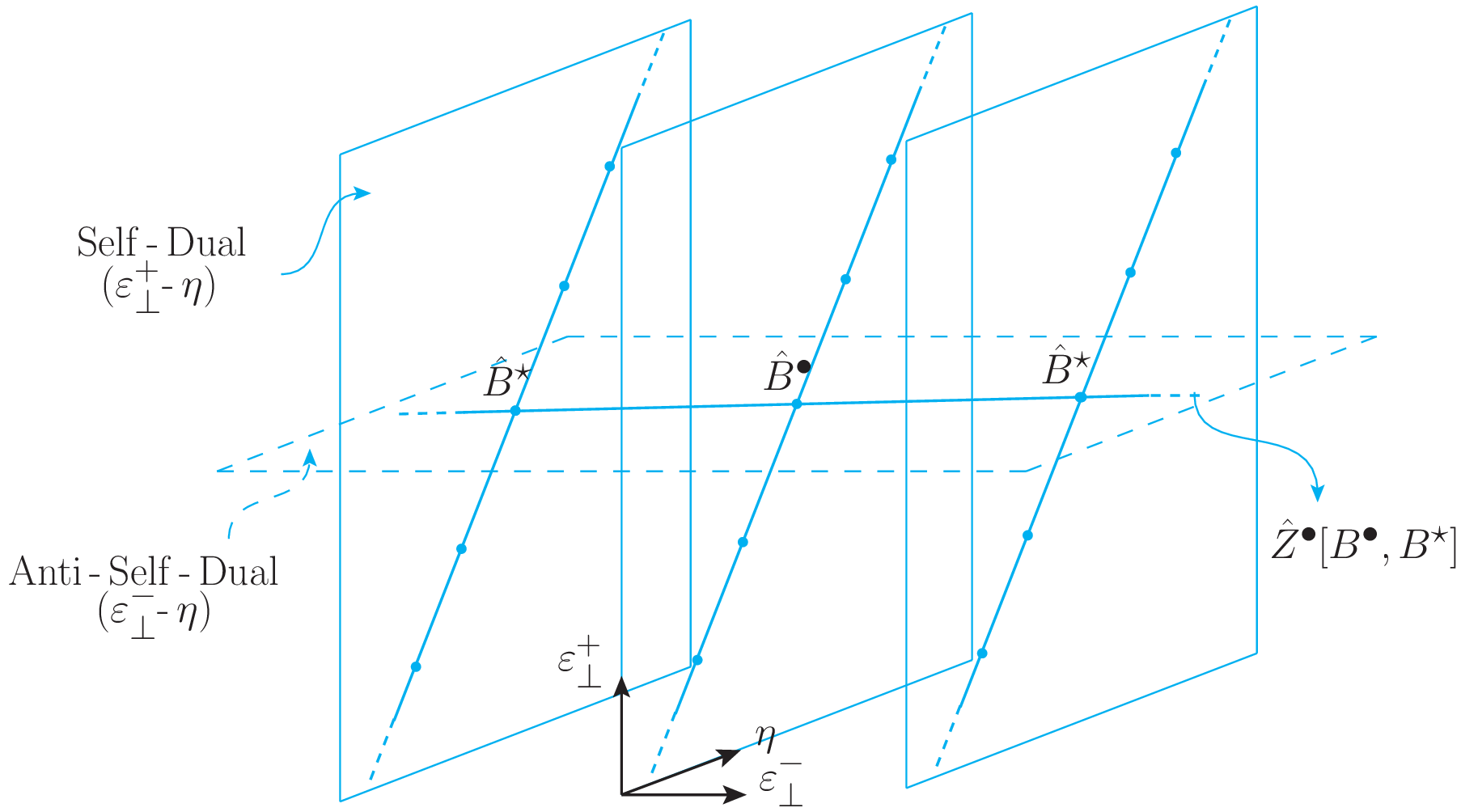}
    \caption{
    \small $Z^{\bullet}$ field as a straight infinite Wilson line of both the negative and positive helicity fields in the MHV action $B^{\star}$ and $B^{\bullet}$ Eq.~\eqref{eq:MHV_action} on the Anti-Self-Dual plane.  The $B^{\bullet}$ field is itself a straight infinite Wilson line of the positive helicity fields in the Yang-Mills action $A^{\bullet}$ on the Self-Dual plane and the $B^{\star}$ field is a similar straight infinite Wilson line where one of the $A^{\bullet}$ fields has been replaced by an $A^{\star}$ field. }
    \label{fig:Zbul_WL}
\end{figure}

Another interesting aspect of the Z-field action Eq.~\eqref{eq:Z_action} is that the fields $\{Z^{\bullet}, Z^{\star} \}$ in the action represent geometric objects expressed in terms of straight infinite Wilson line-based functionals spanning the Self-Dual and the Anti-Self-Dual planes.
This observation follows from the following set of equations (see Appendix \ref{sec:App0})
\begin{equation}
    Z^{\star}_a[B^{\star}](x)=\mathcal{W}_{(-)}^a[B](x)\,,\qquad
    Z_a^{\bullet}[B^{\bullet},B^{\star}](x) = 
    \int\! d^3\mathbf{y} \,
     \left[ \frac{\partial^2_-(y)}{\partial^2_-(x)} \,
     \frac{\delta \mathcal{W}^a_{(-)}[B](x^+;\mathbf{x})}{\delta {B}_c^{\star}(x^+;\mathbf{y})} \right] 
     {B}_c^{\bullet}(x^+;\mathbf{y})
      \, ,
      \label{eq:Zfield_Bpos}
\end{equation}
\begin{equation}
    B^{\bullet}_a[A^{\bullet}](x)=\mathcal{W}_{(+)}^a[A](x)\,,\qquad
    B_a^{\star}[A^{\bullet},A^{\star}](x) = 
    \int\! d^3\mathbf{y} \,
     \left[ \frac{\partial^2_-(y)}{\partial^2_-(x)} \,
     \frac{\delta \mathcal{W}^a_{(+)}[A](x^+;\mathbf{x})}{\delta {A}_c^{\bullet}(x^+;\mathbf{y})} \right] 
     {A}_c^{\star}(x^+;\mathbf{y})
      \, .
      \label{eq:Bfield_Apos}
\end{equation}
The first expression on the left in Eq.~\eqref{eq:Zfield_Bpos} expresses the $Z^{\star}$ field as a straight infinite Wilson line based functional of $B^{\star}$ field on the Anti-Self-Dual plane and the expression on the right expresses $Z^{\bullet}$ field as a similar functional where one of the $B^{\star}$ fields has been replaced by a $B^{\bullet}$ field (the fields $\{B^{\bullet}, B^{\star} \}$ appear in the MHV action Eq.~\eqref{eq:MHV_action}). In a similar fashion, Eq.~\eqref{eq:Bfield_Apos} expresses the $B^{\bullet}$ field as straight infinite Wilson line based functional of $A^{\bullet}$ field on the Self-Dual plane and the $B^{\star}$ field as a similar functional where one of the $A^{\bullet}$ fields has been replaced by an $A^{\star}$ field. Thus in terms of the gauge fields $\{A^{\bullet}, A^{\star} \}$, the Z-fields have a geometric structure shown in Figures \ref{fig:Zstar_WL},\ref{fig:Zbul_WL}.

\section{Delannoy numbers in tree amplitudes}
\label{sec:DEL}

In \cite{Kakkad:2021uhv} we computed several tree level \textit{split helicity} $(+ + \dots + - - \dots -)$ amplitudes using the Z-field action Eq.~\eqref{eq:Z_action}. In this section, we highlight the main results of those computations and present a newly discovered feature associated with the computation of split helicity tree amplitudes using our action. 

First, notice that amplitudes where all the gluons have have same helicity i.e. $(+ + \dots +)$, $( - - \dots -)$ as well as the ones where one of the gluons has a different helicity i.e. $(- + \dots +)$, $( + - \dots -)$ are all zero (to all loop orders) in our action. This is in agreement with the on-shell tree level results. It is well-known, however, that these amplitudes are non-zero at the loop level -- we shall discuss the loop amplitudes in detail in the next section. 

Let us now focus on the non-vanishing tree amplitudes. The amplitudes where all the gluons except two have plus helicities, i.e. MHV amplitudes $(- - + + \dots +)$ are obtained in our action via imposing the on-shell condition for all the legs of the corresponding single vertex in the first row of our action Eq.~\eqref{eq:Z_action}. The same is true for their conjugates  $(- - \dots - + +)$, the $\overline{\mathrm{MHV}}$ amplitudes. That is, the vertices in the first column of the Z-field action give the tree level $\overline{\mathrm{MHV}}$ amplitudes in the on-shell limit 
\begin{equation}
\left.\mathcal{U}\left(1^-,2^-,3^+,\dots,n^+\right)\right|_{\mathrm{on-shell}} = (-g)^{n-2}  \left(\frac{p_{1}^{+}}{p_{2}^{+}}\right)^{2}
\frac{\widetilde{v}_{21}^{\star 4}}{\widetilde{v}_{1n}^{\star}\widetilde{v}_{n\left(n-1\right)}^{\star}\widetilde{v}_{\left(n-1\right)\left(n-2\right)}^{\star}\dots\widetilde{v}_{21}^{\star}}
\,,
\end{equation}
\begin{equation}
    \left.\mathcal{U}\left(1^-,2^-,\dots,(n-2)^-,(n-1)^+,n^+\right)\right|_{\mathrm{on-shell}} =
 g^{n-2} \left(\frac{p_{n-1} ^{+}}{p_{n}^{+}}\right)^{2}
\frac{\widetilde{v}_{n(n-1)}^{4}}{\widetilde{v}_{1n}\widetilde{v}_{n(n-1)}\widetilde{v}_{\left(n-1\right)\left(n-2\right)}\dots \widetilde{v}_{21} } \, .
 \label{eq:MHVbar_onshell}
\end{equation}

\begin{figure}[h]
    \centering
 \includegraphics[width=13cm]{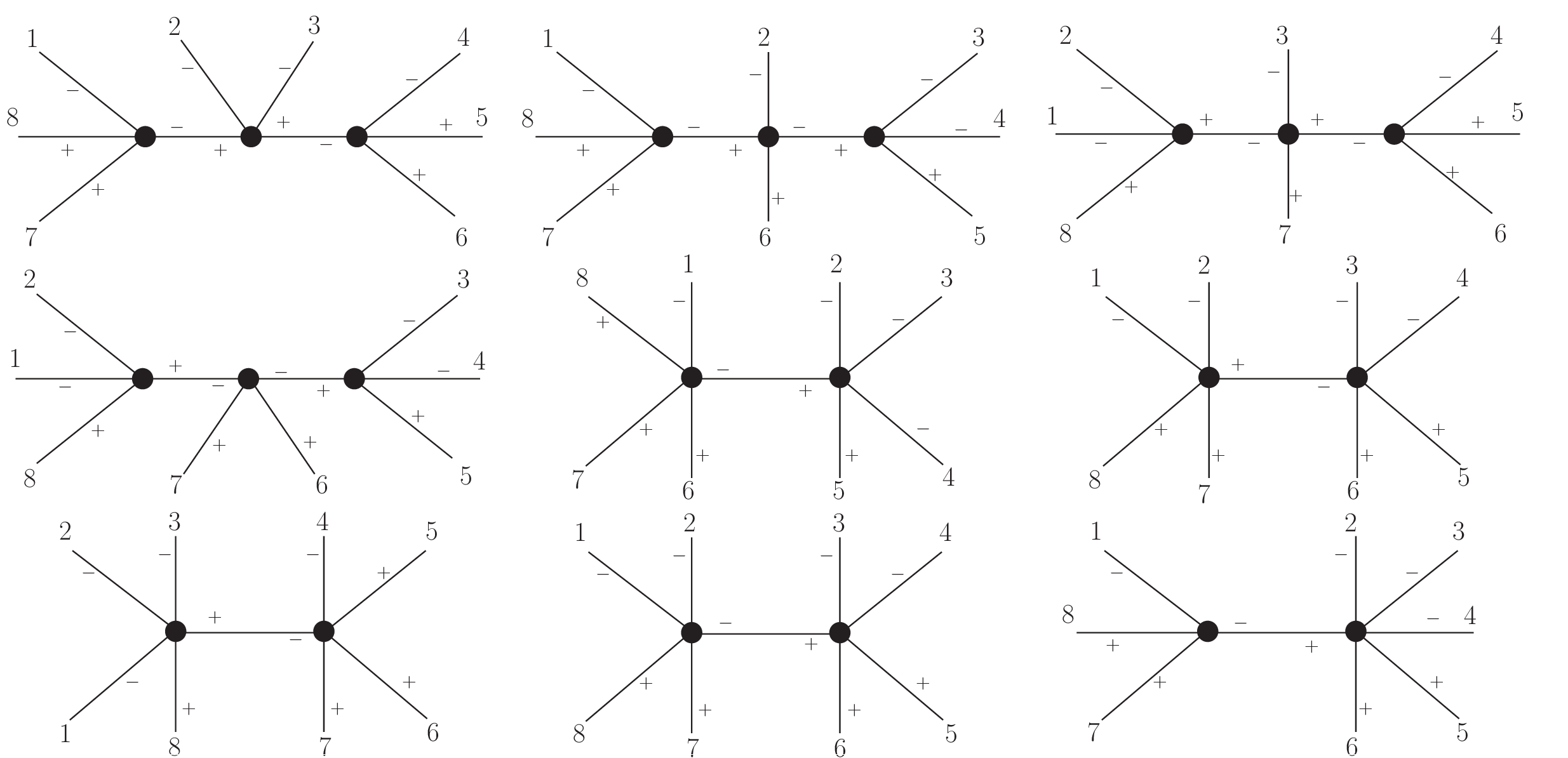}\\
 \includegraphics[width=13cm]{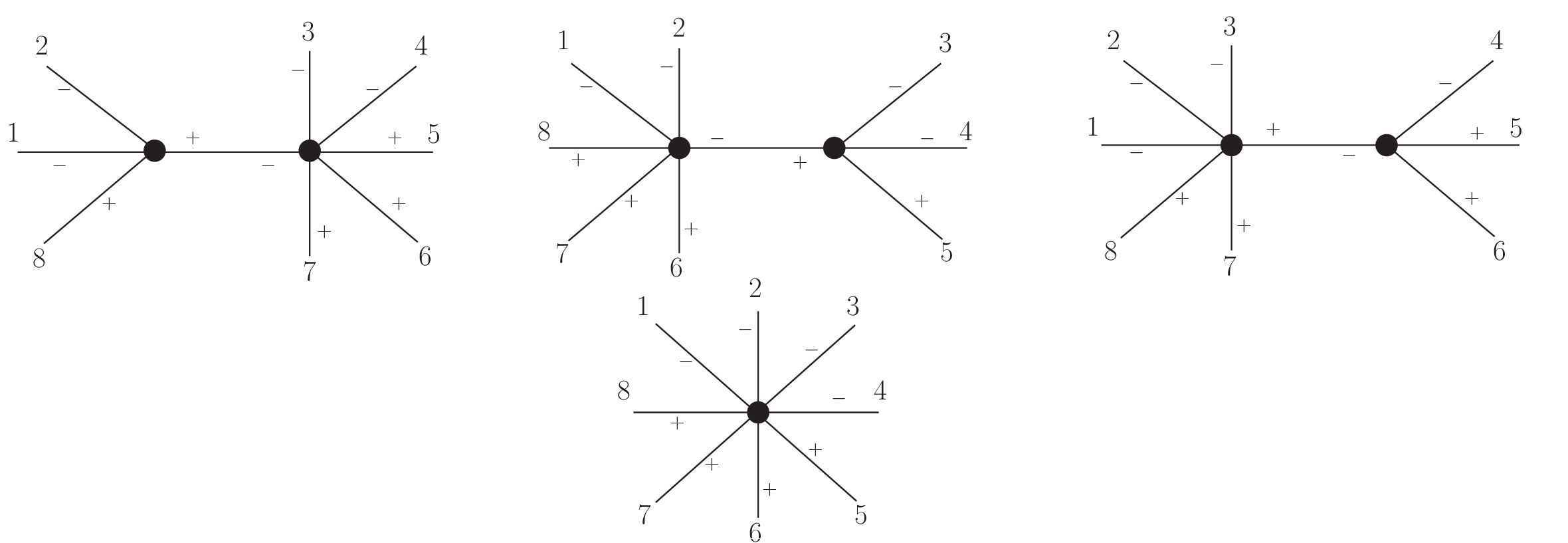}
    \caption{\small 
     Contributions obtained when computing the split-helicity 8-point NNMHV $(- - - - + + + +)$ amplitude using the Z-field action Eq.~\eqref{eq:Z_action}. This image was taken from our paper \cite{Kakkad:2021uhv}.}
    \label{fig:NNMHV8}
\end{figure}

Beyond these simple cases, in \cite{Kakkad:2021uhv} we computed amplitudes up to 8 points with different helicity configurations and found agreement with the standard results \cite{Dixon:2010ik}. The total number of diagrams we got in each case is tabulated in Table \ref{tab:NOD_AT}. The 3 diagrams required for the 6-point NMHV are shown in Figure \ref{fig:NMHV6}. The maximum number of diagrams we had was 13 for the 8-point Next-to-Next-to-MHV (NNMHV). These are shown in Figure \ref{fig:NNMHV8} (for details of computation see \cite{Kakkad:2021uhv}).

\begin{table}
\parbox{.46\linewidth}{
\centering
\begin{tabular}{c|ccccl}
$A^{\mathrm{tree}}_{n,m}$ & 2 & 3 & 4  & 5 & $\dots$ \\ \hline
2     & 1 & 1 & 1  & 1  & $\mathrm{MHV} $\\
3     & 1 & 3 & 5  & 7  & $\mathrm{NMHV} $\\
4     & 1 & 5 & 13 &  & $\mathrm{NNMHV} $\\
5     & 1 & 7 &  &  & $\mathrm{NNNMHV} $
\end{tabular}
\caption{\small No of diagrams contributing to the tree level split helicity amplitude with $n$ plus helicity and $m$ minus helicity legs. Above $\mathrm{N}^k\mathrm{MHV} \equiv (\mathrm{Next}\,\text{-}\,\mathrm{to})^k\mathrm{MHV}$ consists of $k+2$ negative helicities and the rest positive.}
 \label{tab:NOD_AT}
 }
\hfill
\parbox{.46\linewidth}{
\centering
\begin{tabular}{c|cccc}
D(n,m) & 0 & 1 & 2  & 3  \\ \hline
0     & 1 & 1 & 1  & 1  \\
1     & 1 & 3 & 5  & 7  \\
2     & 1 & 5 & 13 & 25 \\
3     & 1 & 7 & 25 & 63
\end{tabular}
\caption{\small No of Delannoy paths to go from origin of a 2D lattice to the point $(n,m)$ using just three moves: east $\rightarrow$, north $\uparrow$, and north-east $\nearrow \,$.}
\label{tab:NODp}
}
\end{table}

\begin{figure}
    \centering
    \includegraphics[width=4.3cm]{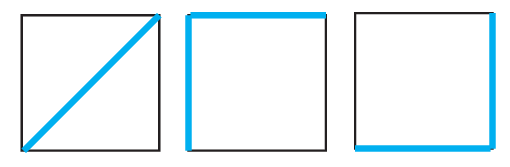}
    \caption{
    \small  The set of 3 Delannoy paths to go from origin of a 2D lattice to the point $(1,1)$ using just three moves: east $\rightarrow$, north $\uparrow$, and north-east $\nearrow \,$.}
    \label{fig:11DN}
\end{figure}

\begin{figure}
    \centering
    \includegraphics[width=9cm]{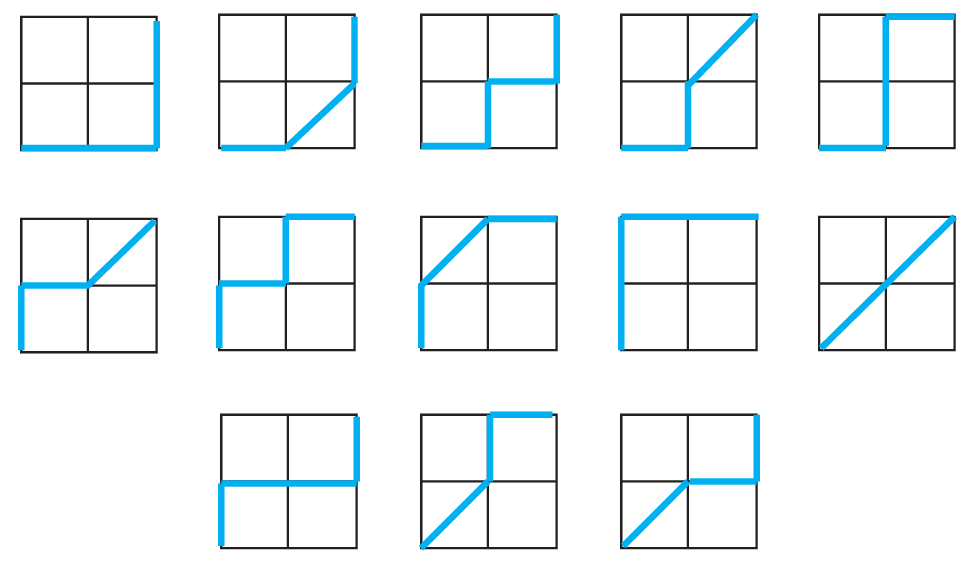}
    \caption{
    \small 
  The set of 13 Delannoy paths to go from origin of a 2D lattice to the point $(2,2)$ using just three moves: east $\rightarrow$, north $\uparrow$, and north-east $\nearrow \,$.}
    \label{fig:22DN}
\end{figure}

It turns out that the number of diagrams required to compute tree-level split helicity pure gluonic amplitudes using the Z-field action Eq.~\eqref{eq:Z_action} follows the \textit{Delannoy numbers} $D(n,m)$. These numbers represent the total number of paths that one can develop to go from the origin $(0,0)$ of a 2D lattice with unit spacing to the point $(n,m)$ using just three moves: east $\rightarrow$, north $\uparrow$, and north-east $\nearrow \,$. The generic formula for $D(n,m)$ reads
\begin{equation}
    D(n,m)= \sum_{i=0}^{\mathrm{min}(n,m)} \binom{m}{i} \binom{n+m-i}{m} = \sum_{i=0}^{\mathrm{min}(n,m)} 2^i \binom{m}{i} \binom{n}{i}\,.
    \label{eq:delannoy}
\end{equation}
Using this we computed the $D(n,m)$ for the first few values of $n$ and $m$. These are tabulated in Table~\ref{tab:NODp}. Consider for example $D(1,1)$ and $D(2,2)$. These represent the 3 and 13 paths to go from $(0,0)$ to $(1,1)$ and $(2,2)$, respectively, using the above mentioned three moves. We represent these paths diagrammatically in Figure \ref{fig:11DN} and \ref{fig:22DN} respectively. 

Thus, we observe the following correspondence
\begin{equation}
     \# \,\,\mathrm{contribution}\,\, {A}_{\underbrace{+ \,\cdots\, +}_{n+2},\underbrace{-\,\cdots\,-}_{m+2}}^{ \mathrm{tree}} = D(n,m) \,,
    \label{eq:delannoy_Z}
\end{equation}
between the entries of Table~\ref{tab:NOD_AT} and Table~\ref{tab:NODp}. This correspondence between the number of Delannoy paths for a given pair of $(n,m)$ and the number of diagrams required to compute the tree level amplitude with $n+2$ plus and $m+2$ plus minus helicity legs is not a coincidence. In fact, both the requirements of a 2D lattice as well the restricted set of 3 Delannoy moves necessary to generate the Delannoy numbers can be mapped respectively to the type of interaction vertices and the Feynman rules followed by the Z-field action when computing split helicity tree amplitudes. This is achieved as follows:
\begin{itemize}
    \item \textit{The} 2D \textit{lattice}: The interaction vertices of the Z-field action form a 2D lattice with the 4-point MHV as the origin and  increasing the plus helicity by one along one direction, say the horizontal, and the minus helicity along the perpendicular direction, say the vertical. Thus, all the MHV vertices lie along the horizontal axis whereas the $\mathrm{\overline{MHV}}$ lies along the vertical. The remaining vertices form the bulk between the axes. Notice that the origin of this lattice is $A^{\mathrm{tree}}_{2,2}$. Hence the shift in the correspondence between  $D(n,m)$ and $A^{\mathrm{tree}}_{n+2,m+2}$.
    \item \textit{The Delannoy moves}:  It is the restricted set of three moves on a 2D lattice that gives rise to the Delannoy number series\footnote{ Choosing a different set of moves/restrictions would give rise to a different number series, consider for instance the Schr{\"o}der numbers.}.
    The correspondence between the set of Delannoy paths and the set of diagrams required to compute tree level split helicity amplitude then implies that the Feynman rules for the latter must be equally restricted. Precisely, there must be exactly three moves whose iteration should results in all the Feynman diagrams. Given the form of our action Eq.~\ref{eq:Z_action}, this is indeed the case. For any given tree amplitude, there is always one contribution that comes from an interaction vertex with the same helicity configuration as the amplitude. This accounts for one move. The remaining contributions involve combining interaction vertices with lower number of legs. But given that the scalar propagator in our action contracts opposite helicity fields, there are two ways to connect a pair of interaction vertices: plus helicity leg of the first vertex connected with the negative helicity of the other and vice versa. These account for the remaining two moves. These three moves are most easily realized in the case of 6-point NMHV (See Figure~\ref{fig:NMHV6}). Iterating these moves we developed the 13 diagrams for the 8-point NNMHV shown in Figure~\ref{fig:NNMHV8}. The correspondence, therefore, implies that each Delannoy path for $D(n,m)$ must get uniquely mapped to a Feyanman diagram contributing to $A^{\mathrm{tree}}_{n+2,m+2}$ tree amplitude.
\end{itemize}

\begin{table}[]
\centering
\begin{tabular}{cccl}
(n,m)                      & $\# \,\,\mathrm{contribution}\,\, {A}_{n+2,m+2}^{ \mathrm{tree}}$ & $D(n,m)$                  &                           \\ \cline{1-3}
\multicolumn{1}{c|}{(5,0)} & \multicolumn{1}{c|}{1}                                            & \multicolumn{1}{c|}{1}  & MHV                       \\
\multicolumn{1}{c|}{(4,1)} & \multicolumn{1}{c|}{9}                                            & \multicolumn{1}{c|}{9}  & NMHV                      \\
\multicolumn{1}{c|}{(3,2)} & \multicolumn{1}{c|}{25}                                           & \multicolumn{1}{c|}{25} & NNMHV                     \\
\multicolumn{1}{c|}{(2,3)} & \multicolumn{1}{c|}{25}                                           & \multicolumn{1}{c|}{25} & $\mathrm{N}^3$MHV         \\
\multicolumn{1}{c|}{(1,4)} & \multicolumn{1}{c|}{9}                                            & \multicolumn{1}{c|}{9}  & $\mathrm{N}^4$MHV         \\
\multicolumn{1}{c|}{(0,5)} & \multicolumn{1}{c|}{1}                                            & \multicolumn{1}{c|}{1}  & $\mathrm{\overline{MHV}}$
\end{tabular}
\caption{\small
The table compares the number of Feynman diagrams required to compute all the split-helicity 9-point tree amplitudes using the Z-field action with the Delannoy numbers.}
\label{tab:Tab3}
\end{table}

Most recently, this correspondence has been verified for the 9-point tree amplitudes in \cite{Kulig2024}. Precisely, the author computed all the non-vanishing 9-point split helicity tree amplitudes using the Z-field action. The number of contributing diagrams for each case ${A}_{n+2,m+2}^{ \mathrm{tree}}$ is shown in Table \ref{tab:Tab3}. These, for a given pair of $(n,m)$, exactly match with the number of Delannoy paths $D(n,m)$ computed using Eq.~\eqref{eq:delannoy}. 

At this point, it is important to stress that this correspondence between the number of Feynman diagrams required to compute tree level pure gluonic amplitudes using the Z-field action Eq.~\eqref{eq:Z_action} and the Delannoy numbers $D(n,m)$ holds only for the split-helicity case.

\section{Z-field theory at one loop}
\label{sec:GAOL}

The Z-field action Eq.~\eqref{eq:Z_action}, as illustrated above, significantly improves the efficiency of computing tree amplitudes compared to the MHV action, which implements solely the MHV vertices \cite{Mansfield2006}. However, just like the MHV action, our classical Z-field action is unable to provide all the necessary contributions for computing one-loop amplitudes. This issue has been extensively discussed in the literature concerning the MHV action \cite{Brandhuber2006,Brandhuber2007,Ettle2007,Ettle2008,Kakkad_2022}. The underlying reason is that both actions are derived by eliminating triple gluon vertices from the Yang-Mills action Eq.~\eqref{eq:YM_LC_action}. While the MHV action eliminates only the self-dual sector, in deriving the Z-field action, we eliminate both the self-dual and anti-self-dual sectors of the Yang-Mills action. Although eliminating them seems beneficial for tree amplitudes, these sectors are necessary for providing the rational contributions -- cut-non-constructible parts -- of the one-loop amplitudes. Consequently, one-loop helicity amplitudes $(+ + \dots +)$, $(- + \dots +)$, $( - - \dots -)$, and $(- - \dots - +)$ all evaluate to zero in the Z-field action, while others have missing rational terms.

\begin{figure}
    \centering
    \parbox[c]{3.3cm}{ \includegraphics[width=3.3cm]{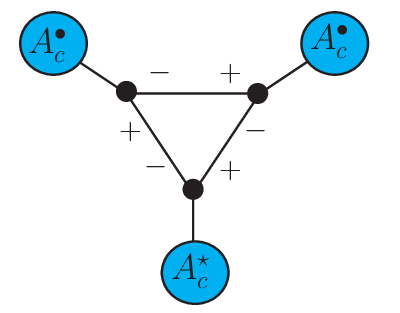}}\qquad $\xrightarrow{\hat{A}_c^i(x) \longrightarrow \hat{A}_c^i[Z_c](x)}$ \qquad\qquad
    \parbox[c]{7cm}{
    \includegraphics[width=7cm]{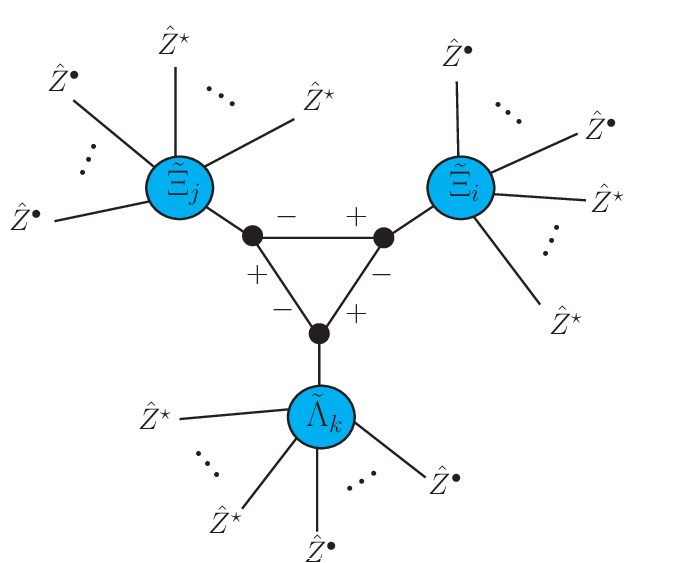}}
    \caption{\small
    On the left, we have the $(+ + -)$ triangle originating from the log term in the one-loop effective Yang-Mills action Eq.~\eqref{eq:OLEA_YM1}. Upon transformation, the loop remains unaltered however the fields outside the loop undergo the substitution $\hat{A}(x) \longrightarrow \hat{A}[Z](x)$.
    }
    \label{fig:zloop_st}
\end{figure}

In \cite{Kakkad_2022} we proposed to deal with this problem in the MHV theory by first constructing the one-loop effective action via integration of the quadratic field fluctuations at the Yang-Mills action level and then performing Mansfield's transformation  \cite{Mansfield2006} to obtain the classical MHV action plus the loop contributions. That way all the one-loop contributions are taken into account. 
It is straightforward to extend this approach to systematically develop quantum corrections to the Z-field theory. The only change in the derivation would be to replace Manfield's transformation with the canonical transformation Eq.~\eqref{eq:AtoZ_ct} that derives the Z-field action. We preform this derivation in Appendix~\ref{sec:AppB}. The resulting one-loop corrected Z-field action reads
\begin{multline}
   \Gamma[Z] = S[Z] 
    + \frac{i}{2} \Tr\ln \left[ \frac{\delta^2 S_{\mathrm{YM}}[A]}
    {\delta A^{\star I}\delta A^{\bullet K}} \, \frac{\delta^2 S_{\mathrm{YM}}[A]}
    {\delta A^{\star K}\delta A^{\bullet J}} \right. \\
    \left.- \frac{\delta^2 S_{\mathrm{YM}}[A]}
    {\delta A^{\star I}\delta A^{\bullet K}} \, \frac{\delta^2 S_{\mathrm{YM}}[A]}
    {\delta A^{\star K}\delta A^{\star L}} \left( \frac{\delta^2 S_{\mathrm{YM}}[A]}
    {\delta A^{\bullet L}\delta A^{\star M}} \right)^{-1} \frac{\delta^2 S_{\mathrm{YM}}[A]}
    {\delta A^{\bullet M}\delta A^{\bullet J}}\right]_{A=A[Z]}\,.
    \label{eq:OLEA_Z}
\end{multline}
Above $S[Z]$ is the Z-field action Eq.~\eqref{eq:Z_action} and  $S_{\mathrm{YM}}[A]$ is the Yang-Mills action Eq.~\eqref{eq:YM_LC_action}. $\hat{A}[Z] = \{ \hat{A}^{\bullet}[{Z}^{\bullet},{Z}^{\star}],\hat{A}^{\star}[{Z}^{\bullet},{Z}^{\star}] \}$ represents the solution Eqs.~\eqref{eq:Abullet_to_Z}-\eqref{eq:Astar_to_Z} of the canonical transformation. 
We introduced the collective indices $I,J,K \dots$ which run over the color and position. Repeated indices are summed over. The log term accounts for the loops (for details see  \cite{Kakkad_2022}). As evident, the double differentiated legs of the Yang-Mills action form the loop whereas the undifferentiated leg outside the loop undergoes field substitution  $\hat{A}(x) \longrightarrow \hat{A}[Z](x)$. We represent this diagrammatically for the $(+ + -)$ triangular one-loop diagram in Figure~\ref{fig:zloop_st}. The substitution  $\hat{A}[Z](x)$ in the log term does not affect the loop topologies due to which the loops involve only the Yang-Mills vertices as before.  It however accounts for all the tree level contributions outside the loop involving both $(+ + -)$, $(+ - -)$ triple gluon vertices that were eliminated via the canonical transformation \cite{Kotko2017, Kakkad2020, Kakkad:2021uhv}. Although this is advantageous when computing higher multiplicity loop amplitudes, it does not make the Z-field vertices explicit in the loop. Except for this, the result Eq.~\eqref{eq:OLEA_Z} is one-loop complete i.e. it has no missing loop contribution. We demonstrate this by computing $(- - - -)$
 one-loop amplitude in Appendix \ref{sec:AppB}, which could not be computed using the Z-filed action Eq.~\eqref{eq:Z_action}. Beyond this simple case, we also computed all the other 4-point one-loop amplitudes: $(+ + ++)$,  $(+ + +-)$, $(---+)$, and $(--++)$ and found agreement with the known results (see Chapter 5 of  \cite{kakkad2023scattering}).

Our major focus, however, in this work is to develop quantum corrections such that the Z-field interaction vertices are explicit in the loop. Before we discuss the details let us summarize both approaches.
\begin{itemize}
    \item  \textit{Approach} 1: There are only Yang-Mills vertices in the loop diagrams, while the vertices of the Z-field theory provide the tree-level connections to build actual amplitudes. It can be shown that parts of loop diagrams could be combined to build loop diagrams with Z-field theory vertices inside the loop, but this is rather cumbersome.
    \item \textit{Approach} 2: There are Z-field theory vertices directly in the loop, but in order to account for the rational contributions extra vertices are required in the loop. The Z-field theory vertices still provide the tree level connections to build amplitudes.
\end{itemize}

Each of the above methods has its pros and cons. The first one is quite direct and operational, but not fully satisfactory as the Z-field theory vertices are not apparent inside the loop. The second method seems more coherent but produces quite a few tadpole diagrams.

The construction of the first method results in Eq.~\eqref{eq:OLEA_Z}. As mentioned previously, it essentially parallels the one for the MHV theory in \cite{Kakkad_2022} (see Appendix \ref{sec:AppB} for details of the derivation).
Below, in Subsection~\ref{sec:OLEA} we develop in detail the second method.

\subsection{One-Loop Effective Action}
\label{sec:OLEA}

For a comprehensive review of the one-loop effective action approach for the Self-Dual and full Yang-Mills theory see Ref.~\cite{Kakkad_2022}, where we also construct the effective action for the MHV theory. Here we restrict the discussion to the Z-field theory only.

Our starting point is the Yang-Mills generating functional
\begin{equation}
    \mathcal{Z}[J]=\int[dA]\, e^{i\left(S_{\mathrm{YM}}[A] + \int\!d^4x\, \Tr \hat{J}_j(x) \hat{A}^j(x)\right) } \,,
    \label{eq:gen_YM}
\end{equation}
where $J$ represents the external current and the index $j = \bullet, \star$ runs over the transverse field components. The action $S_{\mathrm{YM}}$ is the Yang-Mills action on the light cone Eq.~\eqref{eq:YM_LC_action}.
We perform now the field transformation
\begin{equation}
    \mathcal{Z}[J]=\int[dA]\, e^{i\left(S_{\mathrm{YM}}[A] + \int\!d^4x\, \Tr \hat{J}_j(x) \hat{A}^j(x)\right) } \longrightarrow \int[dZ]\, e^{i\left(S[Z] + \int\!d^4x\, \Tr \hat{J}_j(x) \hat{A}^j[Z](x)\right) } \,,
   \label{eq:SZ_genr}
\end{equation}
where $S[Z]$ is the Z-field action.
Let us turn attention, that we transform also the linear (source) term. It is essential in our approach. A naive (but standard) approach is to take the Z-field action and add a new source term $\int\!d^4x\hat{J}'_i(x)\hat{Z}_i(x)$. Doing that, however, would lead to missing all-plus and all-minus helicity amplitudes, as well as the cut-non-constructible parts of other amplitudes. The reason for that is virtually the same as in the MHV theory, see the discussion in \cite{Kakkad_2022}.

Following the standard procedure, we compute the generating functional using the saddle point approximation. Expanding up to the second order in fields around a "minimum" we obtain
\begin{multline}
    S[Z] + \int\!d^4x\, \Tr \hat{J}_i(x) \hat{A}^i[Z](x)  
    = S[Z_c] + \int\!d^4x\, \Tr \hat{J}_i(x)\hat{A}^i[Z_c](x) \\ + \int\!d^4x\,\Tr\left(\hat{Z}^i(x)-\hat{Z}_c^i(x)\right)
    \left(\frac{\delta S[Z_c]}{\delta \hat{Z}^i(x)}+\int\!d^4y\,{\hat J}_k(y)\frac{\delta \hat{A}^k[Z_c](y)}{\delta \hat{Z}^i(x)}\right) \\
    +\frac{1}{2}\int\!d^4xd^4y\,\Tr\left(\hat{Z}^i(x)-\hat{Z}_c^i(x)\right)\left(\frac{\delta^2 S[Z_c]}{\delta\hat{Z}^i(x)\delta\hat{Z}^j(y)}\right. \\
    \left.+\int\!d^4z\,{\hat J}_k(z)\frac{\delta^2 \hat{A}^k[Z_c](z)}{\delta \hat{Z}^i(x)\delta\hat{Z}^j(y)}\right)\left(\hat{Z}^j(y)-\hat{Z}_c^j(y)\right) \,.
    \label{eq:Der_olea_sz}
\end{multline}
The field configurations at the saddle point $Z_c$ is given by the classical EOMs
\begin{equation}
    \left.\frac{\delta S[Z_c^{\bullet}, Z_c^{\star}]}{\delta \hat{Z}^{\bullet}(x)}+\int\!d^4y\,\left[{\hat J}_{\bullet}(y)\frac{\delta \hat{A}^{\bullet}[Z_c^{\bullet}, Z_c^{\star}](y)}{\delta \hat{Z}^{\bullet}(x)}+{\hat J}_{\star}(y)\frac{\delta \hat{A}^{\star}[Z_c^{\bullet}, Z_c^{\star}](y)}{\delta \hat{Z}^{\bullet}(x)}\right]\right|_{{\hat Z}={\hat Z}_{c}} =0\,,
    \label{eq:Z_bul_EOM}
\end{equation}
\begin{equation}
   \left.\frac{\delta S[Z_c^{\bullet}, Z_c^{\star}]}{\delta \hat{Z}^{\star}(x)}+\int\!d^4y\,\left[{\hat J}_{\bullet}(y)\frac{\delta \hat{A}^{\bullet}[Z_c^{\bullet}, Z_c^{\star}](y)}{\delta \hat{Z}^{\star}(x)}+{\hat J}_{\star}(y)\frac{\delta \hat{A}^{\star}[Z_c^{\bullet}, Z_c^{\star}](y)}{\delta \hat{Z}^{\star}(x)}\right]\right|_{{\hat Z}={\hat Z}_{c}} =0\,.
    \label{eq:Z_star_EOM}
\end{equation}
The classical solution $Z_c$ has to be understood as a functional of external sources, $Z_c=Z_c[J]$.
Performing the Gaussian integration we get
\begin{equation}
   \mathcal{Z}[J] \approx  \Big[ \det\,\mathrm{M}\Big]^{-\frac{1}{2}}
   \exp\left\{i\left(S[Z_c] 
    + \int\!d^4x\, \Tr \hat{J}_i(x)\hat{A}^i[Z_c](x) \right) \right\} \,,
    \label{eq:det_SZ}
\end{equation}
where the matrix under the determinant is
\begin{equation}
  \mathrm{M}[J]=\mathrm{M}^{\text{Z-field}}[J]+\mathrm{M}^{\text{src}}[J] \,,  
  \label{eq:Mfull}
\end{equation}
with
 \begin{equation}
\mathrm{M}^{\text{Z-field}}_{IK}[J]  =  \left(\begin{matrix}
     \frac{\delta^2 S[Z_c]}{\delta Z^{\bullet I}\delta Z^{\star K}} 
     &
     \frac{\delta^2 S[Z_c]}{\delta Z^{\bullet I}\delta Z^{\bullet K}}\\ \\
\frac{\delta^2 S[Z_c]}{\delta Z^{\star I}\delta Z^{\star K}}
      &
      \frac{\delta^2 S[Z_c]}{\delta Z^{\star I}\delta Z^{\bullet K}} 
\end{matrix}\right) \,,
\label{eq:M_Zfield}
\end{equation}
and
\begin{equation}
\mathrm{M}^{\text{src}}_{IK}[J] = \left(\begin{matrix}
    J_{\star L}\frac{\delta^2 A^{\star L}[Z_c]}{\delta Z^{\bullet I}\delta Z^{\star K}} + J_{\bullet L}\frac{\delta^2 A^{\bullet L}[Z_c]}{\delta Z^{\bullet I}\delta Z^{\star K}} 
     & J_{\star L}\frac{\delta^2 A^{\star L}[Z_c]}{\delta Z^{\bullet I}\delta Z^{\bullet K}} +  J_{\bullet L}\frac{\delta^2 A^{\bullet L}[Z_c]}{\delta Z^{\bullet I}\delta Z^{\bullet K}}\\ \\
 J_{\star L}\frac{\delta^2 A^{\star L}[Z_c]}{\delta Z^{\star I}\delta Z^{\star K}} + J_{\bullet L}\frac{\delta^2 A^{\bullet L}[Z_c]}{\delta Z^{\star I}\delta Z^{\star K}}
      &J_{\star L}\frac{\delta^2 A^{\star L}[Z_c]}{\delta Z^{\star I}\delta Z^{\bullet K}} + J_{\bullet L}\frac{\delta^2 A^{\bullet L}[Z_c]}{\delta Z^{\star I}\delta Z^{\bullet K}}
\end{matrix}\right) \,.
\label{eq:M_Zsrc}
\end{equation}
Above we use the collective indices introduced earlier. 
 The first matrix $\mathrm{M}^{\text{Z-field}}$ accommodates the vertices of the Z-field theory alone. If using the naive approach for computing one loop effective action (i.e. without transforming the source term) only this matrix would contribute to the determinant. However, because we also transformed the current term, we also have the second, source-dependent matrix $\mathrm{M}^{\text{src}}$. 
 We shall explicitly obtain the terms appearing in both matrices further below, but for now we do not need its explicit form.
Rewriting the determinant as an exponential of the trace we have
\begin{equation}
   \mathcal{Z}[J] \approx  
   \exp\left\{iS\left[Z_c\left[J\right]\right] 
    + i\int\!d^4x\, \Tr \hat{J}_i(x)\hat{A}^i[Z_c[J]](x) -\frac{1}{2}\Tr\ln\mathrm{M}[J] \right\} \,.
    \label{eq:det_SZln}
\end{equation}
The loop diagrams are generated by the log term. If one is interested only in the loop topologies and not the Green functions, it is convenient to consider the one-loop effective action that is obtained by the Legendre transformation
\begin{equation}
   \Gamma[Z_c] = W[J] - \int\!d^4x\, \Tr \hat{J}_i(x) \hat{A}^{i}_c[Z_c](x) \,, \quad \mathrm{where} \quad W[J] = -i \ln \left[ \mathcal{Z}[J]\right]\,.
   \label{eq:gam_zn}
\end{equation}
The above expression for $\Gamma[Z_c]$ is a functional of classical fields only whereas, the generating function $\mathcal{Z}[J]$ is a functional of only sources; recall that in the latter, for the fields $Z_c$ one should replace the classical current-dependent solutions to Eqs.\eqref{eq:Z_bul_EOM}-\eqref{eq:Z_star_EOM}, which is rather challenging. This would be appropriate if computing the generating functional $\mathcal{Z}[J]$.
 Here, instead, we are interested in the one-loop effective action $\Gamma[Z_c]$, that gives the loop contributions.
 Therefore, we have to eliminate the explicit current dependence in favour of the classical fields. 
 The most straightforward way to do that is by rewriting the classical EOMs Eqs.\eqref{eq:Z_bul_EOM}-\eqref{eq:Z_star_EOM} in terms of the Yang-Mills EOM as shown below 
\begin{equation}
  \frac{\delta S_{\mathrm{YM}}[A[Z_c]]}{\delta A^{\star L}}= -J_{\star L}\,, \quad\quad
    \frac{\delta S_{\mathrm{YM}}[A[Z_c]]}{\delta A^{\bullet L}}= - J_{\bullet L} \,.
    \label{eq:J_currZ1}
\end{equation}
Above notation means that one takes the Yang-Mills action on the light cone \eqref{eq:YM_LC_action}, derives it functionally over a field, and substitutes the field transformation solutions $\hat{A}(x) \longrightarrow \hat{A}[Z](x)$ using Eqs.~\eqref{eq:Abullet_to_Z}-\eqref{eq:Astar_to_Z}.
Unfortunately, this explicitly introduces the Yang-Mills vertices outside the loops, but they are necessary to make the logarithm term quantum-complete.
As discussed in the introduction to this Section, alternatively, we could follow the procedure of \cite{Kakkad_2022} used for the quantum MHV action, i.e. start with the Yang-Mills action, derive the one-loop effective action with Yang-Mills vertices, apply the field transformation to obtain Eq.\eqref{eq:OLEA_Z} (see Appendix \ref{sec:AppB}). That way, however, the Z-field theory vertices would not be explicit in the loop.

Now, substituting $ \mathcal{Z}[J]$ to the expression Eq.~\eqref{eq:gam_zn} we get
\begin{equation}
   \Gamma[Z_c] = S\left[Z_c\right] 
    + i\frac{1}{2}\Tr\ln\mathrm{M}[J[Z_c]]\,,
    \label{eq:olea_zth}
\end{equation}
where $J[Z_c]$ is the solution of equation Eq.~\eqref{eq:J_currZ1}.

We now proceed to discussing the content of the one loop partition function (or effective action). 
The log term is fairly complicated. In order to study its content, let us note that it can be rewritten as (dropping the functional dependence on the fields $Z_c$ for compactness)
\begin{equation}
    \Tr\ln\mathrm{M}= \Tr\ln\mathrm{M}_{\bullet\star} +
    \Tr\ln\left(\mathrm{M}_{\star\bullet}
    - \mathrm{M}_{\star\star}\mathrm{M}^{-1}_{\bullet\star}\mathrm{M}_{\bullet\bullet}\right) \,,
    \label{eq:lnM_decomp}
\end{equation}
where $\mathrm{M}_{\bullet\star}=\mathrm{M}_{\star\bullet}$ correspond to the diagonal blocks of the matrix Eq.~\eqref{eq:Mfull}, $\mathrm{M}_{\star\star}$ to the bottom-left and $\mathrm{M}_{\bullet\bullet}$ to top-right. For example, for the first term we have, more explicitly,
\begin{equation}
   \Tr\ln\left(\mathrm{M}_{\star\bullet}\right)_{IK}= 
   \Tr\ln\left\{ 
   \frac{\delta^2 S[Z_c]}{\delta Z^{\bullet I}\delta Z^{\star K}}
   - 
   \frac{\delta S_{\mathrm{YM}}[A[Z_c]]}{\delta A^{\star L}}\frac{\delta^2 A^{\star L}[Z_c]}{\delta Z^{\bullet I}\delta Z^{\star K}} -  \frac{\delta S_{\mathrm{YM}}[A[Z_c]]}{\delta A^{\bullet L}}\frac{\delta^2 A^{\bullet L}[Z_c]}{\delta Z^{\bullet I}\delta Z^{\star K}} 
   \right\} \,.
\end{equation}

Using the collective index notation for
the Z-field action $S[Z_c]$ 
\begin{equation}
    S[Z_c]
    = -Z_c^{\star L}\square_{LJ}Z_c^{\bullet J} -\sum_{n=4}^{\infty} 
     \, \sum_{m=2}^{n-2}\,
    \,\, \mathcal{U}^{ \{J_1 \dots J_m\}\{ J_{m+1}\dots J_n\} }_{\underbrace{-\,\cdots\,-}_{m}\underbrace{+ \,\cdots\, +}_{n-m}} 
   \prod_{i=1}^{m}Z_c^{\star J_i}
   \prod_{k=m+1}^{n}Z_c^{\bullet J_k} \, .
   \label{eq:S[Z]CI}
\end{equation}

 we get (see Appendix \ref{sec:AppA})
 \begin{multline}
   \Tr\ln\left(\mathrm{M}_{\bullet \star}\right)_{IK}= 
  \Tr\ln\Bigg[-\square \Bigg\{\delta_{IK} 
    + \sum_{n=4}^{\infty} 
     \, \sum_{m=2}^{n-2}
    \,m(n-m)\,\, \mathcal{U}^{ \{J_1 \dots J_m\}\{ J_{m+1}\dots J_n\} }_{-\,\cdots\,- + \,\cdots\, +} \,\left(\frac{1}{\square} \right)_{IP}\delta_{J_1}^K \delta_{J_{m+1}}^P\,
   \prod_{i=2}^{m}Z_c^{\star J_i} \\
   \prod_{k=m+2}^{n}Z_c^{\bullet J_k}
   + 
   \frac{\delta S_{\mathrm{YM}}[A[Z_c]]}{\delta A^{\star L}}\frac{\delta^2 A^{\star L}[Z_c]}{\delta Z^{\bullet P}\delta Z^{\star K}}\left(\frac{1}{\square} \right)_{IP} + \frac{\delta S_{\mathrm{YM}}[A[Z_c]]}{\delta A^{\bullet L}}\frac{\delta^2 A^{\bullet L}[Z_c]}{\delta Z^{\bullet P}\delta Z^{\star K}}\left(\frac{1}{\square} \right)_{IP} \Bigg\}\Bigg] \, , \\
  = \Tr\ln \{-\square_{IK}\} + \Tr\ln\Bigg\{\delta_{IK} 
    + \sum_{n=4}^{\infty} 
     \, \sum_{m=2}^{n-2}
    \,m(n-m)\,\, \mathcal{U}^{ \{J_1 \dots J_m\}\{ J_{m+1}\dots J_n\} }_{-\,\cdots\,- + \,\cdots\, +} \,\left(\frac{1}{\square} \right)_{IP}\delta_{J_1}^K \delta_{J_{m+1}}^P\,
   \prod_{i=2}^{m}Z_c^{\star J_i} \\
   \prod_{k=m+2}^{n}Z_c^{\bullet J_k}
   + 
   \frac{\delta S_{\mathrm{YM}}[A[Z_c]]}{\delta A^{\star L}}\frac{\delta^2 A^{\star L}[Z_c]}{\delta Z^{\bullet P}\delta Z^{\star K}}\left(\frac{1}{\square} \right)_{IP} + \frac{\delta S_{\mathrm{YM}}[A[Z_c]]}{\delta A^{\bullet L}}\frac{\delta^2 A^{\bullet L}[Z_c]}{\delta Z^{\bullet P}\delta Z^{\star K}}\left(\frac{1}{\square} \right)_{IP} \Bigg\}\,.
   \label{eq:Mbs_log}
\end{multline}
The first logarithm on the R.H.S. of Eq.~\eqref{eq:Mbs_log} is field independent and can be discarded. 
Note, that the remaining terms are equipped with a propagator which following the Feynman rules is necessary to connect the vertices as well as form loops when traced over. Expanding the logarithm 
up to the second order in fields, the expansion reads (see details in Appendix \ref{sec:AppA})
\begin{multline}
   \Tr \Bigg[ \Bigg\{ -\frac{\Lambda_{1,1}^{L \{K\}\{ P\}} }{\square_{IP}} \square_{LJ}{Z}_c^{\bullet J}  -\square_{LJ}\frac{\Xi_{1,1}^{L \{P\}\{ K\} } }{\square_{IP}}Z_c^{\star J} +  4  \,\frac{\mathcal{U}^{ \{K  J_1\}\{P J_2\} }_{-\,- + \, +}}{\square_{IP}} \,
   Z_c^{\star J_1}
   Z_c^{\bullet J_2} \\
    - \Bigg( 2 \frac{\Lambda_{1,2}^{L \{K\}\{ P J_1\}} }{\square_{IP}}\square_{LJ_2} + \frac{\Lambda_{1,1}^{L \{K\}\{ P\}} }{\square_{IP}}\square_{LJ} \Xi_{2,0}^{J \{J_1 J_2\} } + \frac{\Lambda_{1,1}^{L \{K\}\{ P\}} }{\square_{IP}}\left(V_{-++}\right)_{LJ_1 J_2}\Bigg){Z}_c^{\bullet J_1}{Z}_c^{\bullet J_2}\\
    - \Bigg(\frac{\Lambda_{1,1}^{L \{K\}\{ P\}} }{\square_{IP}} \square_{LJ} \Xi_{1,1}^{J \{J_1 \}\{J_2 \} } + 2 \frac{\Lambda_{2,1}^{L \{K J_2\}\{ P \}} }{\square_{IP}}\square_{LJ_1} + \frac{\Lambda_{1,1}^{L \{K\}\{ P\}} }{\square_{IP}} 2\left(V_{--+}\right)_{LJ_2 J_1 }\Bigg) {Z}_c^{\bullet J_1}{Z}_c^{\star J_2} \\
    - \Bigg( \frac{\Xi_{1,1}^{L \{P\}\{ K\} } }{\square_{IP}} \square_{LJ} \Lambda_{1,1}^{J \{J_1\}\{  J_2\} } + \frac{\Xi_{1,1}^{L \{P\}\{ K\} } }{\square_{IP}} 2\left(V_{-++}\right)_{J_1 L J_2} + 2\frac{\Xi_{2,1}^{L \{P J_2\}\{ K  \} } }{\square_{IP}} \square_{LJ_1}\Bigg) {Z}_c^{\star J_1} {Z}_c^{\bullet J_2} \\
    -\Bigg( \frac{\Xi_{1,1}^{L \{P\}\{ K\} } }{\square_{IP}}\square_{LJ} \Lambda_{2,0}^{J \{J_1 J_2\} } +\frac{\Xi_{1,1}^{L \{P\}\{ K\} } }{\square_{IP}}\left(V_{--+}\right)_{J_1 J_2 L} + 2\frac{\Xi_{1,2}^{L \{P\}\{ K J_1 \} } }{\square_{IP}}\square_{LJ_2} \Bigg){Z}_c^{\star J_1} {Z}_c^{\star J_2} \Bigg\} \\
    -\frac{1}{2} \Bigg\{ \frac{\Lambda_{1,1}^{L_1 \{K_1\}\{ P\}} }{\square_{IP}} \square_{L_1 J_1} \frac{\Lambda_{1,1}^{L_2 \{K\}\{ P_1\}} }{\square_{K_1 P_1}} \square_{L_2 J_2}{Z}_c^{\bullet J_1}{Z}_c^{\bullet J_2} + \frac{\Lambda_{1,1}^{L_1 \{K_1\}\{ P\}} }{\square_{IP}} \square_{L_1 J_1 } \square_{L_2 J_2}\frac{\Xi_{1,1}^{L_2 \{P_1\}\{ K\} } }{\square_{K_1 P_1}}{Z}_c^{\bullet J_1} Z_c^{\star J_2} \\
    +\square_{L_1 J_1}\frac{\Xi_{1,1}^{L_1 \{P\}\{ K_1\} } }{\square_{IP}}\frac{\Lambda_{1,1}^{L_2 \{K\}\{ P_1\}} }{\square_{K_1P_1}} \square_{L_2 J_2 }Z_c^{\star J_1}{Z}_c^{\bullet J_2} + \square_{L_1 J_1}\frac{\Xi_{1,1}^{L_1 \{P\}\{ K_1\} } }{\square_{IP}}\square_{L_2 J_2}\frac{\Xi_{1,1}^{L_2 \{P_1\}\{ K\} } }{\square_{K_1P_1}}Z_c^{\star J_1}Z_c^{\star J_2} \Bigg\}\Bigg] \, ,
   \label{eq:S[Z]+-2nd}
\end{multline}
where $\mathcal{U}^{ \{K  J_1\}\{P J_2\} }_{-\,- + \, +}$ is the 4-point MHV interaction vertex in the Z-field action Eq.~\eqref{eq:S[Z]CI} and $\Xi_{i,n-i}^{L \{J_1 \dots J_i\}\{ J_{i+1}\dots J_n\} }$, $\Lambda_{i,n-i}^{L \{J_1 \dots J_i\}\{ J_{i+1}\dots J_n\} }$ are the kernels of the solution Eqs.~\eqref{eq:Abullet_to_Z}-\eqref{eq:Astar_to_Z} expressed in terms of the collective indices as shown below
\begin{equation}
A^{\bullet L}[Z_c] = 
\sum_{n=1}^{\infty} 
     \, \sum_{i=1}^{n}\, \Xi_{i,n-i}^{L \{J_1 \dots J_i\}\{ J_{i+1}\dots J_n\} }\,\prod_{k=1}^{i}{Z}_c^{\bullet J_k} \prod_{l=i+1}^{n}{Z}_c^{\star J_l}\,,
    \label{eq:A_bul_ZCI}
\end{equation}
\begin{equation}
A^{\star L}[Z_c] = 
\sum_{n=1}^{\infty} 
     \, \sum_{i=1}^{n}\,\Lambda_{i,n-i}^{L \{J_1 \dots J_i\}\{ J_{i+1}\dots J_n\} } \,\prod_{k=1}^{i}{Z}_c^{\star J_k} \prod_{l=i+1}^{n}{Z}_c^{\bullet J_l}\,.
    \label{eq:A_star_ZCI}
\end{equation}
 
\begin{figure}
    \centering
    \includegraphics[width=3.5cm]{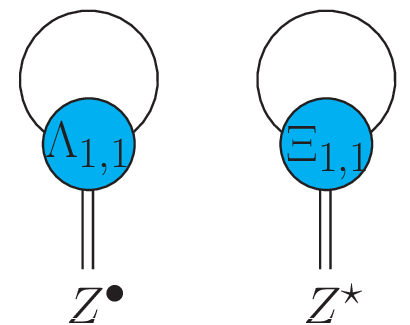}
    
    \vspace{0.2cm}
    \includegraphics[width=12.2cm]{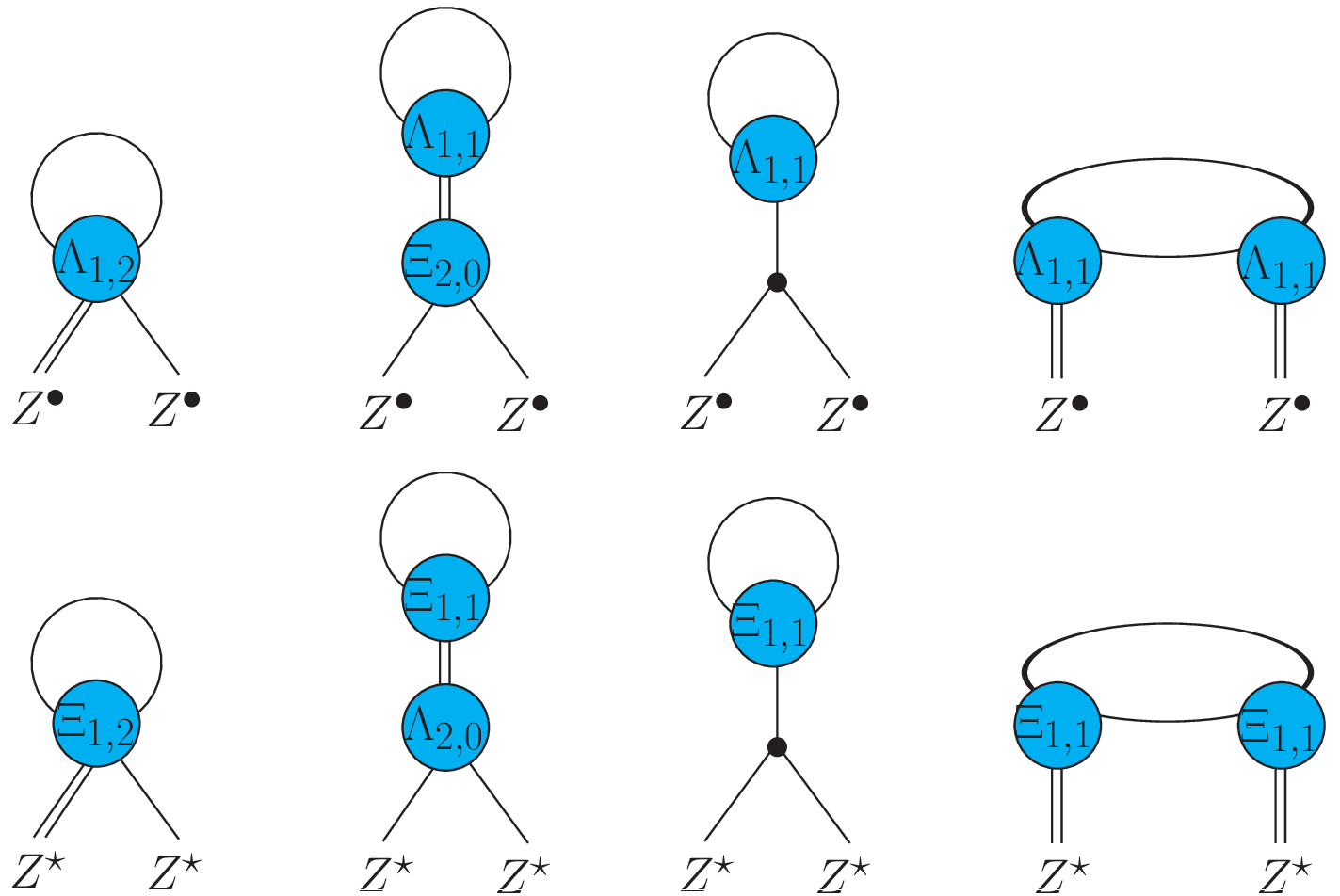}

    \vspace{0.2cm}
    \includegraphics[width=15.5cm]{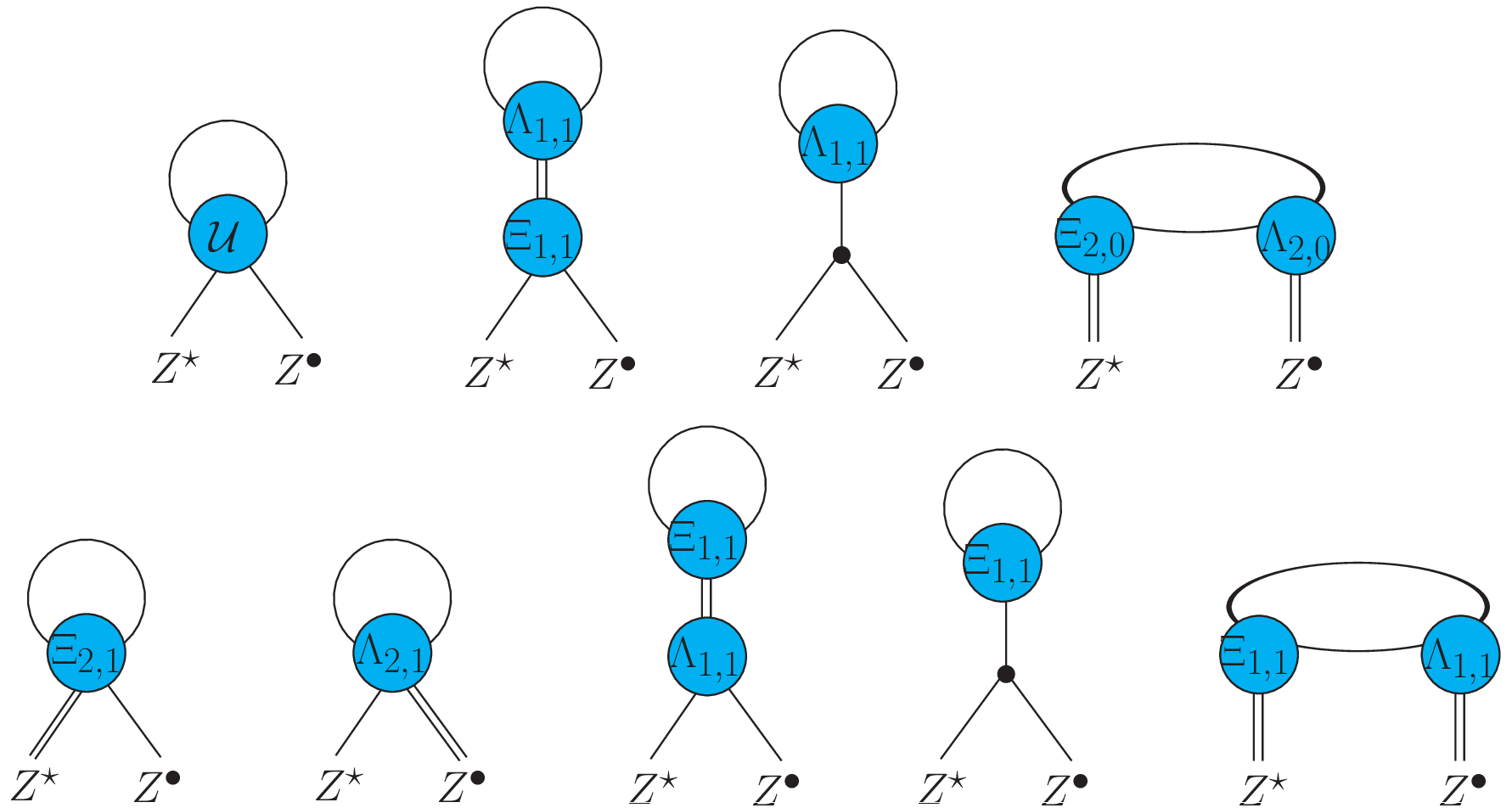}
    \caption{
    \small  
  All the one-loop contributions up to two points originating from the log term in Eq.~\eqref{eq:olea_zth}. The first two terms are the tadpoles. $\Xi_{i,j}$ and $\Lambda_{i,j}$ are the kernels in Eqs. \eqref{eq:xi_kernel_mom}-\eqref{eq:lambda_kernel_mom} and $Z^{\bullet}$, $Z^{\star}$ is the plus and minus helicity fields, respectively, in the action Eq.~\eqref{eq:Z_action}. The double line represents that the incoming leg of the kernel has been contracted with the inverse propagator: $\square \, \Lambda_{i,j}$ and $\square \, \Xi_{i,j}$. The remaining are the one-loop terms contributing to the three types of gluon self-enegies: $(+ +)$, $(- -)$, and $(- +)$ respectively}
    \label{fig:Zth_loop_2pt}
\end{figure}

The second logarithm in Eq.~\eqref{eq:lnM_decomp} is even more complicated. Up to second order in fields it reads (see Appendix~\ref{sec:AppA})
\begin{multline}
     \left. \Tr\ln\left(\mathrm{M}_{\bullet\star}
    - \mathrm{M}_{\bullet\bullet}\mathrm{M}^{-1}_{\star\bullet}\mathrm{M}_{\star\star}\right) \right|_{\mathrm{2nd}} = \Tr \Bigg[ \Bigg\{ -\frac{\Lambda_{1,1}^{L \{K\}\{ P\}} }{\square_{IP}} \square_{LJ}{Z}_c^{\bullet J}  -\square_{LJ}\frac{\Xi_{1,1}^{L \{P\}\{ K\} } }{\square_{IP}}Z_c^{\star J}  \\
    - \Bigg( 2 \frac{\Lambda_{1,2}^{L \{K\}\{ P J_1\}} }{\square_{IP}}\square_{LJ_2} + \frac{\Lambda_{1,1}^{L \{K\}\{ P\}} }{\square_{IP}}\square_{LJ} \Xi_{2,0}^{J \{J_1 J_2\} } + \frac{\Lambda_{1,1}^{L \{K\}\{ P\}} }{\square_{IP}}\left(V_{-++}\right)_{LJ_1 J_2}\Bigg){Z}_c^{\bullet J_1}{Z}_c^{\bullet J_2}\\
    - \Bigg(\frac{\Lambda_{1,1}^{L \{K\}\{ P\}} }{\square_{IP}} \square_{LJ} \Xi_{1,1}^{J \{J_1 \}\{J_2 \} } + 2 \frac{\Lambda_{2,1}^{L \{K J_2\}\{ P \}} }{\square_{IP}}\square_{LJ_1} + \frac{\Lambda_{1,1}^{L \{K\}\{ P\}} }{\square_{IP}} 2\left(V_{--+}\right)_{LJ_2 J_1 }\Bigg) {Z}_c^{\bullet J_1}{Z}_c^{\star J_2} \\
    - \Bigg( \frac{\Xi_{1,1}^{L \{P\}\{ K\} } }{\square_{IP}} \square_{LJ} \Lambda_{1,1}^{J \{J_1\}\{  J_2\} } + \frac{\Xi_{1,1}^{L \{P\}\{ K\} } }{\square_{IP}} 2\left(V_{-++}\right)_{J_1 L J_2} + 2\frac{\Xi_{2,1}^{L \{P J_2\}\{ K  \} } }{\square_{IP}} \square_{LJ_1}\Bigg) {Z}_c^{\star J_1} {Z}_c^{\bullet J_2} \\
    -\Bigg( \frac{\Xi_{1,1}^{L \{P\}\{ K\} } }{\square_{IP}}\square_{LJ} \Lambda_{2,0}^{J \{J_1 J_2\} } +\frac{\Xi_{1,1}^{L \{P\}\{ K\} } }{\square_{IP}}\left(V_{--+}\right)_{J_1 J_2 L} + 2\frac{\Xi_{1,2}^{L \{P\}\{ K J_1 \} } }{\square_{IP}}\square_{LJ_2} \Bigg){Z}_c^{\star J_1} {Z}_c^{\star J_2} \\
    +  4  \,\frac{\mathcal{U}^{ \{K  J_1\}\{P J_2\} }_{-\,- + \, +}}{\square_{IP}} \,
   Z_c^{\star J_1}
   Z_c^{\bullet J_2} -4\,\frac{\Xi_{2,0}^{L_1 \{PK_1\}  }}{\square_{IP}}\square_{L_1 J_1} \, \frac{\Lambda_{2,0}^{L_2 \{P_1K\} }}{\square_{K_1P_1}}\square_{L_2 J_2}Z_c^{\star J_1}Z_c^{\bullet J_2} \Bigg\}\\
    -\frac{1}{2} \Bigg\{ \frac{\Lambda_{1,1}^{L_1 \{K_1\}\{ P\}} }{\square_{IP}} \square_{L_1 J_1} \frac{\Lambda_{1,1}^{L_2 \{K\}\{ P_1\}} }{\square_{K_1 P_1}} \square_{L_2 J_2}{Z}_c^{\bullet J_1}{Z}_c^{\bullet J_2} + \frac{\Lambda_{1,1}^{L_1 \{K_1\}\{ P\}} }{\square_{IP}} \square_{L_1 J_1 } \square_{L_2 J_2}\frac{\Xi_{1,1}^{L_2 \{P_1\}\{ K\} } }{\square_{K_1 P_1}}{Z}_c^{\bullet J_1} Z_c^{\star J_2} \\
    +\square_{L_1 J_1}\frac{\Xi_{1,1}^{L_1 \{P\}\{ K_1\} } }{\square_{IP}}\frac{\Lambda_{1,1}^{L_2 \{K\}\{ P_1\}} }{\square_{K_1P_1}} \square_{L_2 J_2 }Z_c^{\star J_1}{Z}_c^{\bullet J_2} + \square_{L_1 J_1}\frac{\Xi_{1,1}^{L_1 \{P\}\{ K_1\} } }{\square_{IP}}\square_{L_2 J_2}\frac{\Xi_{1,1}^{L_2 \{P_1\}\{ K\} } }{\square_{K_1P_1}}Z_c^{\star J_1}Z_c^{\star J_2} \Bigg\}\Bigg] \, .
    \label{eq:Log2_2}
\end{multline}
Except for one term (the last term in the 6th line above), all the others are the same in Eq.~\eqref{eq:Log2_2} and Eq.~\eqref{eq:S[Z]+-2nd}. Thus adding them and tracing over the differentiated legs we get
\begin{multline}
      \Tr\ln\mathrm{M}\Bigg|_{2nd} = -2\frac{\Lambda_{1,1}^{L \{I\}\{ P\}} }{\square_{IP}} \square_{LJ}{Z}_c^{\bullet J}  -2\square_{LJ}\frac{\Xi_{1,1}^{L \{P\}\{ I\} } }{\square_{IP}}Z_c^{\star J}  \\
    - \Bigg( 4 \frac{\Lambda_{1,2}^{L \{I\}\{ P J_1\}} }{\square_{IP}}\square_{LJ_2} + 2\,\frac{\Lambda_{1,1}^{L \{I\}\{ P\}} }{\square_{IP}}\square_{LJ} \Xi_{2,0}^{J \{J_1 J_2\} } + 2\,\frac{\Lambda_{1,1}^{L \{I\}\{ P\}} }{\square_{IP}}\left(V_{-++}\right)_{LJ_1 J_2} \\ + \frac{\Lambda_{1,1}^{L_1 \{K_1\}\{ P\}} }{\square_{IP}} \square_{L_1 J_1} \frac{\Lambda_{1,1}^{L_2 \{I\}\{ P_1\}} }{\square_{K_1 P_1}} \square_{L_2 J_2}\Bigg){Z}_c^{\bullet J_1}{Z}_c^{\bullet J_2}\\
    -\Bigg( 2\,\frac{\Xi_{1,1}^{L \{P\}\{ I\} } }{\square_{IP}}\square_{LJ} \Lambda_{2,0}^{J \{J_1 J_2\} } +2\,\frac{\Xi_{1,1}^{L \{P\}\{ I\} } }{\square_{IP}}\left(V_{--+}\right)_{J_1 J_2 L} + 4\,\frac{\Xi_{1,2}^{L \{P\}\{ I J_1 \} } }{\square_{IP}}\square_{LJ_2} \\
    +  \square_{L_1 J_1}\frac{\Xi_{1,1}^{L_1 \{P\}\{ K_1\} } }{\square_{IP}}\square_{L_2 J_2}\frac{\Xi_{1,1}^{L_2 \{P_1\}\{ I\} } }{\square_{K_1P_1}}\Bigg){Z}_c^{\star J_1} {Z}_c^{\star J_2} \\
    - \Bigg(  -8  \,\frac{\mathcal{U}^{ \{I  J_1\}\{P J_2\} }_{-\,- + \, +}}{\square_{IP}} \,
   +4\,\frac{\Xi_{2,0}^{L_1 \{PK_1\}  }}{\square_{IP}}\square_{L_1 J_1} \, \frac{\Lambda_{2,0}^{L_2 \{P_1I\} }}{\square_{K_1P_1}}\square_{L_2 J_2} + 2\,\frac{\Lambda_{1,1}^{L \{I\}\{ P\}} }{\square_{IP}} \square_{LJ} \Xi_{1,1}^{J \{J_2 \}\{J_1 \} } 
   + 4\, \frac{\Lambda_{2,1}^{L \{I J_1\}\{ P \}} }{\square_{IP}}\square_{LJ_2} \\
   + 2\,\frac{\Lambda_{1,1}^{L \{I\}\{ P\}} }{\square_{IP}} 2\left(V_{--+}\right)_{LJ_1 J_2 } + 2\,\frac{\Xi_{1,1}^{L \{P\}\{ I\} } }{\square_{IP}} \square_{LJ} \Lambda_{1,1}^{J \{J_1\}\{  J_2\} } + 2\, \frac{\Xi_{1,1}^{L \{P\}\{ I\} } }{\square_{IP}} 2\left(V_{-++}\right)_{J_1 L J_2} + 4\,\frac{\Xi_{2,1}^{L \{P J_2\}\{ I  \} } }{\square_{IP}} \square_{LJ_1} \\
     + \frac{\Lambda_{1,1}^{L_1 \{K_1\}\{ P\}} }{\square_{IP}} \square_{L_1 J_2 } \square_{L_2 J_1}\frac{\Xi_{1,1}^{L_2 \{P_1\}\{ I\} } }{\square_{K_1 P_1}} 
    +\square_{L_1 J_1}\frac{\Xi_{1,1}^{L_1 \{P\}\{ K_1\} } }{\square_{IP}}\frac{\Lambda_{1,1}^{L_2 \{I\}\{ P_1\}} }{\square_{K_1P_1}} \square_{L_2 J_2 } \Bigg) Z_c^{\star J_1}{Z}_c^{\bullet J_2}  \, .
    \label{eq:Log2_2point}
    \end{multline}
The above expression represents all the one-loop contributions up to two points originating from the log term in Eq.~\eqref{eq:olea_zth}. The first two terms are the tadpoles and the remaining are the terms contributing to the three types of gluon self-enegies: $(+ +)$, $(- -)$, and $(- +)$ respectively. We represent these diagrammatically in Figure \ref{fig:Zth_loop_2pt}. 

\begin{figure}
    \centering
    \includegraphics[width=14cm]{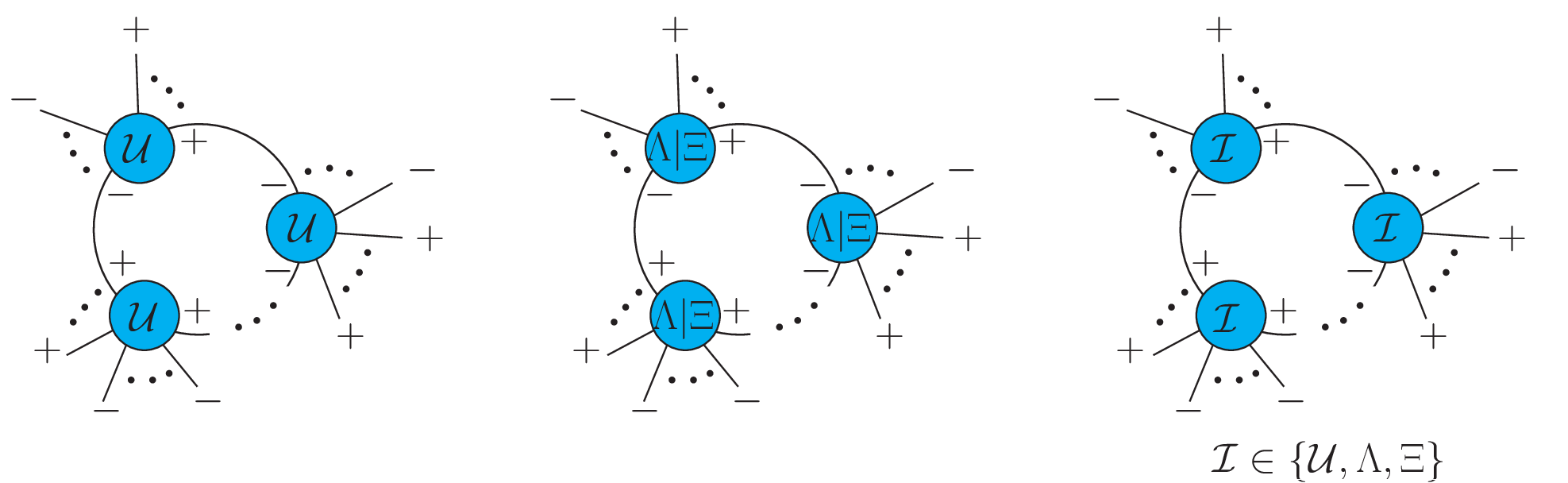}
    \caption{
    \small 
  The three types of loop diagrams originating from the log term in Eq.~\eqref{eq:olea_zth}. First involves only the Z-field interaction vertices $\mathcal{U}$ in the loop. The second involves only the kernels $\Lambda_{i,j}$ and/or $\Xi_{i,j}$ Eqs.~\eqref{eq:lambda_kernel_mom}-\eqref{eq:xi_kernel_mom} in the loop. The third mixes the Z-field interaction vertices with the kernels $\Lambda_{i,j}$ and/or $\Xi_{i,j}$ in the loop. 
  }
    \label{fig:Zth_loop}
\end{figure}

Notice, in Eq.~\eqref{eq:Log2_2point} there are essentially two types of contributions. One involves only the interaction vertex from the Z-field action. This is the first term in line 6 of Eq.~\eqref{eq:Log2_2point}. All the remaining terms involve the kernels $\Lambda_{i,j}$ and/or $\Xi_{i,j}$. All these terms would be missing had we developed loops using the Z-field action alone or equivalently the so-called "naive" approach discussed at the beginning of this section because in that case the only matrix entering the log term would be the one involving the second order derivatives of the Z-field action Eq.~\eqref{eq:M_Zfield}. In addition, there is one more class of terms that would be missing. These appear when we go to higher multiplicity one-loop contributuions. These terms mix the interaction vertices of the Z-field action with the kernels $\Lambda_{i,j}$ and/or $\Xi_{i,j}$. We represent these three sets of terms originating from the log term in Eq.~\eqref{eq:Mfull} in Figure \ref{fig:Zth_loop} for a generic higher multiplicity one-loop case. 

 We see, that, in addition to Z-field theory vertices (that include the MHV, $\overline{\text{MHV}}$, and $\text{N}^k\text{MHV}$ vertices), we have the  kernels $\Lambda_{i,j}$ and $\Xi_{i,j}$ that would give rise to the self-dual and anti-self dual vertices in the loop.
Let us note, that this is thanks to 
 the source-dependent matrix. 
The proof of that fact is provided in 
Appendix \ref{sec:AppB}. Below we provide a more intuitive discussion of the necessity to transform the current dependent term to account for the missing loop contributions.

\subsection{Equivalence of the two actions}
\label{sec:Equiv}

Given the structure of the log terms, we see, that the presence of the self-dual and anti-self dual vertices i.e. $V_{-++}$, $V_{--+}$ in the loop is not as explicit in Eq.~\eqref{eq:olea_zth} as it is in Eq.~\eqref{eq:OLEA_Z}.  In the latter, these vertices directly enter the loop formation whereas in the former, the Z-field interaction vertices and the kernels $\Lambda_{i,j}$ and $\Xi_{i,j}$ form the loops.
Note, however, as demonstrated in \cite{Kotko2017, Kakkad2020, Kakkad:2021uhv}, the kernels $\Lambda_{i,j}$ and $\Xi_{i,j}$ essentially resum all the tree level connections involving the self-dual and anti-self dual vertices $V_{-++}$, $V_{--+}$. Therefore, the presence of the second and the third type of the terms in Figure \ref{fig:Zth_loop} involving the kernels $\Lambda_{i,j}$ and $\Xi_{i,j}$ in the loop does provide a qualitative assurance that the interaction vertices of the self-dual and the anti-self-dual sectors of the Yang-Mills are indeed present in the log term of Eq.~\eqref{eq:olea_zth}.

Indeed, in Appendix \ref{sec:AppB}, we show that the two one-loop effective actions given in Eq.~\eqref{eq:olea_zth} and Eq.~\eqref{eq:OLEA_Z} are \textit{equivalent}. We start with the one-loop action Eq.~\eqref{eq:olea_zth} derived in the previous section, and then using the properties of the canonical transformation Eq.~\eqref{eq:AtoZ_ct}, we show that it can be re-written as the action in Eq.~\eqref{eq:OLEA_Z} modulo a field-independent volume divergent factor which does not contribute to amplitude computations. Given that the classical action $S[Z]$ is the same in both the actions, the equivalence implies that the log term in the two actions can be converted into one another. This means the self-dual and the anti-self-dual interaction vertices should indeed originate from the $\Lambda_{i,j}$ and $\Xi_{i,j}$ because the Z-field action does not have these. This, in turn, also reinforces the importance of the source-matrix Eq.~\eqref{eq:M_Zsrc} in the log term of Eq.~\eqref{eq:AtoZ_ct} which in our approach originates from transforming the source term in the generating functional for the Yang-Mills Greens function Eq.~\eqref{eq:gen_YM}.

An important outcome of the above result is that the quantum corrections of the self-dual and the anti-self-dual sectors of the Yang-Mills, i.e. one-loop amplitudes where all the gluons have the same helicities, are both contained in the term shown in the middle in Figure \ref{fig:Zth_loop}. In general, this term accounts for all the one-loop corrections involving only the self-dual and the anti-self-dual interaction vertices in the loop. The third term in Figure \ref{fig:Zth_loop} mixes these two sectors with the Z-field interaction vertices. 

We expect the the one-loop effective action Eq.~\eqref{eq:olea_zth} to be "quantum complete", i.e. there are no missing loop contributions. This follows from the equivalence of the Eq.~\eqref{eq:olea_zth} and Eq.~\eqref{eq:OLEA_Z}. It was shown in \cite{kakkad2023scattering} that the latter gives the correct 4-point amplitudes, including the rational ones.

Let us see directly how the full field-transformed partition function Eq.~\eqref{eq:SZ_genr} with the transformed current term brings in the contributions from both self-dual and anti-self dual sectors.

Consider an $n$-point Green's function generated by the functional Eq.~\eqref{eq:SZ_genr}
\begin{equation}
   \hat{\mathcal{G}}^{j_1\dots j_n}(x_1,\dots,x_n) = \langle0|\mathcal{T} \{{\hat{A}}^{j_1}[Z](x_1) {\hat{A}}^{j_2}[Z](x_2) \dots {\hat{A}}^{j_n}[Z](x_n) e^{i\,S[Z]}\} |0\rangle\,,
   \label{eq:Green1}
\end{equation}
where $j_i={\bullet,\star}$ are helicity projections of the fields. The ordinary gluon field $A$ are expanded in $Z$ fields using Eqs.~\eqref{eq:Abullet_to_Z}-\eqref{eq:Astar_to_Z}. To first order in $Z$ fields we have
\begin{equation}
    \hat{\mathcal{G}}^{j_1\dots j_n}_{(I)}(x_1,\dots,x_n) =
    \langle0|\mathcal{T} \{{\hat{Z}}^{j_1}(x_1) {\hat{Z}}^{j_2}(x_2) \dots {\hat{Z}}^{j_n}(x_n) e^{i\,S[Z]}\} |0\rangle\,.
\end{equation}
This contribution to the full Green's function accounts only for contributions (tree and loop) that are made of Z-field theory vertices. For example, the connected component of the following Green's function, that would generate all-same-helicity amplitude, is zero both at tree and loop level, 
\begin{equation}
    \hat{\mathcal{G}}^{\star\star \dots \star}_{(I)}(x_1,\dots,x_n) =
    \langle0|\mathcal{T} \{{\hat{Z}}^{\star}(x_1) {\hat{Z}}^{\star}(x_2) \dots {\hat{Z}}^{\star}(x_n) e^{i\,S[Z]}\} |0\rangle \doteq 0\,,
\end{equation}
as there is no way of connecting the set of same-helicity $Z$ fields by the vertices in the $Z$-field action.
But the full Green's function contains also other terms contributing to the same helicity configuration. For example, expanding the fields to second order we get a contribution
\begin{multline}
    \hat{\mathcal{G}}^{\star\star \dots \star}_{(IIa)}(x_1,\dots,x_n) = \\
    \int\!d^4y_{11}d^4y_{12}\dots d^4y_{n1}d^4y_{n2}\, \Tr\{t^{a_1}\dots t^{a_n}\}  \Lambda_{1,1}^{a_1 b_{11}b_{12}}(x_1;y_{11},y_{12})
    \dots
    \Lambda_{1,1}^{a_n b_{n1}b_{n2}}(x_n;y_{n1},y_{n2}) \\
    \langle0|\mathcal{T} \{ {Z}^{b_{11}\star}(y_{11})
    {Z}^{b_{12}\bullet}(y_{12})
    {Z}^{b_{21}\star}(y_{21}){Z}^{b_{22}\bullet}(y_{22})
     \dots
     {Z}^{b_{n1}\star}(y_{n1}){Z}^{b_{n2}\bullet}(y_{n2}) e^{i\,S[Z]}\} |0\rangle\,.
\end{multline}
Applying the Wick theorem to pairs of $Z$ fields of the opposite helicity we get the loop contribution to all-same-helicity amplitude. As a matter of fact these contributions can be mapped to the ordinary Yang-Mills diagrams contribution to same-helicity amplitudes. Indeed, application of the inverse propagator to external legs (LSZ reduction) will convert the $\Lambda_{1,1}$ kernels to triple gluon vertices (see for example discussion in \cite{Kakkad2020}).


\section{Discussion and Conclusions}
\label{sec:conclusion}

In this work, our primary objective is to develop quantum corrections to the Z-field action described in Eq.~\eqref{eq:Z_action},
which itself is not sufficient to compute loop amplitudes, as both the self-dual and anti-self dual sectors of the Yang-Mills theory are present only on the classical level.
 We introduce two \textit{equivalent} approaches to achieve this goal. The first approach extends our previous work \cite{Kakkad_2022}, where we apply the transformation given by Eq.~\eqref{eq:AtoZ_ct} to the one-loop effective Yang-Mills action Eq.~\eqref{eq:OLEA_YM1} to obtain the one-loop corrected Z-field action Eq.~\eqref{eq:OLEA_Z}, as detailed in Appendix \ref{sec:AppB}. Our result is "one-loop complete", with no missing loop contributions, as confirmed through the computation of all four-point one-loop amplitudes: $(++++)$, $(+++-)$, $(++--)$, $(+---)$, and $(----)$, in \cite{kakkad2023scattering}. As an example, we provide the computation of $(----)$ in Appendix \ref{sec:AppB}.

The above approach offers a significant advantage in computing one-loop amplitudes compared to ordinary Yang-Mills action.  This is because the transformation Eq.~\eqref{eq:AtoZ_ct} applied to the log term of the latter automatically encompasses all tree-level connections involving both the Yang-Mills triple gluon vertices outside the loops. Whereas in one-loop effective Yang-Mills action Eq.~\eqref{eq:OLEA_YM1}, these contributions must be individually developed by combining the triple gluon vertices in the classical Yang-Mills action with the log term. Additionally, the transformation substitutes the classical Yang-Mills action in Eq.~\eqref{eq:OLEA_YM1} with the classical Z-field action, which features more efficient higher multiplicity interaction vertices, thereby further reducing the number of contributions arising from the interplay between the classical action and the log term. However, the transformation does not modify the loop structure. As a result, only Yang-Mills vertices are involved in loop formation, while the Z-field interaction vertices are exclusively utilized for tree-level connections outside the loop. Therefore, the major focus of this work is on developing loop corrections such that the Z-field interaction vertices are explicit in the loop instead of the Yang-Mills vertices.

To achieve this, we developed a new approach which involves transforming the Yang-Mills partition function Eq.~\eqref{eq:gen_YM}, including the source term,  and then deriving the one-loop effective Z-field action Eq.~\eqref{eq:olea_zth} by integrating out the quadratic field fluctuations. This way, the log term of the latter explicitly includes the Z-field interaction vertices within the loops. A key aspect of this method is transforming the source term in the partition function Eq.~\eqref{eq:gen_YM}, leading to its non-linearity in Z-fields and the generation of the source matrix Eq.~\eqref{eq:M_Zsrc} in the log term. This matrix is essential for capturing the loop contributions that are missing in the Z-field action Eq.~\eqref{eq:Z_action}. We explicitly established this by verifying the equality between the two one-loop effective actions Eqs.~\eqref{eq:OLEA_Z} and \eqref{eq:olea_zth}. Since we confirm the completeness of the former through amplitude calculations, the equivalence implies the quantum completeness of the latter. The missing contributions stem from the source matrix described in Eq.~\eqref{eq:M_Zsrc}.

Although Eq.~\eqref{eq:olea_zth} is fully functional, its effectiveness in terms of computing one-loop amplitudes primarily relies on the representation of source matrix described Eq.~\eqref{eq:M_Zsrc} in the logarithmic term. In order to express the source matrix Eq.~\eqref{eq:M_Zsrc} solely in terms of fields, we replaced the sources using the Yang-Mills classical EOMs, with Yang-Mills classical fields replaced via the field transformation (see Eq.~\eqref{eq:J_currZ1}).  Alternatively, one could attempt to solve the complicated classical EOMs detailed in Eqs.~\eqref{eq:Z_bul_EOM}-\eqref{eq:Z_star_EOM} to express the sources in terms of Z-fields via the functional derivatives of the Z-field action. Another option would be to derive a new set of classical EOMs involving the MHV action and thereby expressing the sources in terms of the MHV action. In each scenario, the form of the source matrix Eq.~\eqref{eq:M_Zsrc} would differ, impacting not only the structure of loops but also the number of terms stemming from the log expansion and thus influencing the efficiency of computation by altering the number of loop diagrams. Nevertheless, the one-loop actions would remain equivalent.

\begin{figure}
    \centering
    \includegraphics[width=13cm]{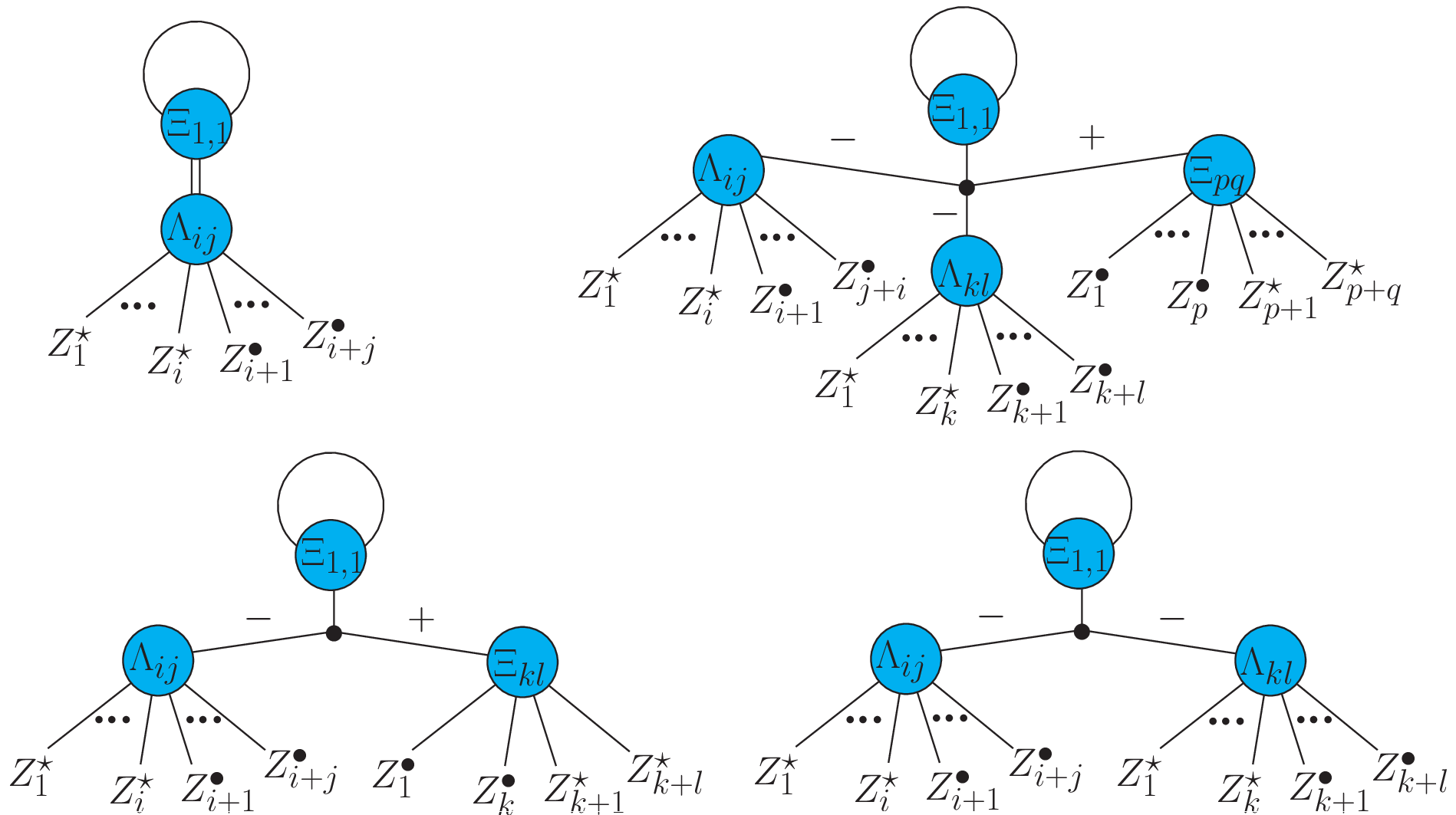}
    \caption{
    \small  In this figure, we represent the proliferation of the $\Xi_{1,1}$ tadpole shown in Eq.~\eqref{eq:POT}. The first diagram represents the $\Xi_{1,1}$ tadpole contracted with $\square_{LJ}A^{\star J}[Z_c]$ where the latter is further expanded in terms of the Z-fields using Eq.~\eqref{eq:Astar_to_Z}. The remaining terms originate from the contraction of the $\Xi_{1,1}$ tadpole with the Yang-Mills interaction vertices. Only the negative helicity leg of the latter undergoes the contraction. The remaining legs undergo the the expansion in terms of the Z-fileds following the Eqs.\eqref{eq:Abullet_to_Z}-\eqref{eq:Astar_to_Z}. The proliferation of the $\Lambda_{1,1}$ tadpole shown in Eq.~\eqref{eq:POT} proceeds in exactly the similar fashion.}
    \label{fig:Zth_POT}
\end{figure}

The representation we used for the source matrix Eq.~\eqref{eq:M_Zsrc} suffers from the proliferation of tadpoles. To see this consider the diagonal blocks of the source matrix Eq.~\eqref{eq:M_Zsrc}. These read:
\begin{equation}
  \left(\mathrm{M}^{\text{src}}_{IK}[J]\right)_{\bullet \star}= 
      J_{\star L}\frac{\delta^2 A^{\star L}[Z_c]}{\delta Z^{\bullet I}\delta Z^{\star K}} + J_{\bullet L}\frac{\delta^2 A^{\bullet L}[Z_c]}{\delta Z^{\bullet I}\delta Z^{\star K}} \,.
   \label{eq:Mn+-}
\end{equation}
After substituting for the sources using Eq.~\eqref{eq:J_currZ1} and factoring out the inverse propagator we get
\begin{equation}
   \left(\frac{1}{\square} \right)_{IP}\left(\mathrm{M}^{\text{src}}_{PK}[J]\right)_{\bullet \star} =  
   \left(\frac{1}{\square} \right)_{IP} \frac{\delta S_{\mathrm{YM}}[A[Z_c]]}{\delta A^{\star L}}\frac{\delta^2 A^{\star L}[Z_c]}{\delta Z^{\bullet P}\delta Z^{\star K}} + \left(\frac{1}{\square} \right)_{IP}\frac{\delta S_{\mathrm{YM}}[A[Z_c]]}{\delta A^{\bullet L}}\frac{\delta^2 A^{\bullet L}[Z_c]}{\delta Z^{\bullet P}\delta Z^{\star K}} \,.
\end{equation}
Consider the terms involving only the second order expansion of the A-field in terms of the Z-fields using Eqs.~\eqref{eq:Abullet_to_Z}-\eqref{eq:Astar_to_Z}. Tracing over the differentiated legs of these second order terms, as stated previously, gives rise to the tadpoles 
\begin{equation}
    \frac{\delta S_{\mathrm{YM}}[A[Z_c]]}{\delta A^{\star L}}\frac{\Lambda_{1,1}^{L \{K\}\{ P\}} }{\square_{IP}} + \frac{\delta S_{\mathrm{YM}}[A[Z_c]]}{\delta A^{\bullet L}}\frac{\Xi_{1,1}^{L \{P\}\{ K\} } }{\square_{IP}} \,.
\end{equation}
The two types of tadpoles are contracted with the functional derivatives of the Yang-Mills action. Substituting for these using Eqs~\eqref{eq:S+-Z}-\eqref{eq:S+-Z2} we get
\begin{multline}
   \Bigg\{-\square_{LJ}A^{\bullet J}[Z_c]-\left(V_{-++}\right)_{LJQ}A^{\bullet J}[Z_c] A^{\bullet Q}[Z_c] \\
     -2\left(V_{--+}\right)_{LQJ}A^{\star Q}[Z_c] A^{\bullet J}[Z_c] 
    -2\left(V_{--++}\right)_{LQJR}A^{\star I}[Z_c] A^{\bullet J}[Z_c] A^{\bullet R}[Z_c]\Bigg\}\frac{\Lambda_{1,1}^{L \{K\}\{ P\}} }{\square_{IP}} \\
+\Bigg\{ -\square_{LJ}A^{\star J}[Z_c]-2\left(V_{-++}\right)_{JLQ}A^{\star J}[Z_c] A^{\bullet Q}[Z_c] \\
    -\left(V_{--+}\right)_{QJL}A^{\star Q}[Z_c] A^{\star J}[Z_c] 
    -2\left(V_{--++}\right)_{QJLR}A^{\star Q}[Z_c] A^{\star J}[Z_c] A^{\bullet R}[Z_c] \Bigg\}\frac{\Xi_{1,1}^{L \{P\}\{ K\} } }{\square_{IP}} \,.
    \label{eq:POT}
\end{multline}
Each of the A-fields in the aforementioned expression can be further decomposed into Z-fields using Eqs~\eqref{eq:A_bul_ZCI}-\eqref{eq:A_star_ZCI}. Consequently, we observe that any one-loop amplitude calculation will entail a substantial number of contributions stemming from the expression above, predominantly comprising tadpoles (see Figure \ref{fig:Zth_POT}), some of which we consider as unphysical and thus should be disregarded. Unfortunately, we are not aware of any apriori constraint that would circumvent such terms.
Identifying an alternative representation for the source matrix that encompasses mostly physical contribution is one of the potential directions of a future research.

 Besides developing quantum corrections to the Z-field action, in this work, we report the newly observed correspondence between the Delannoy numbers $D(n,m)$ and the number of contributing diagrams required for computing split-helicity tree-level amplitudes with $n+2$ positive and $m+2$ negative helicities using the Z-field action. We attribute this correlation solely to the nature of interaction vertices present in the action.

An interesting future research direction would be to prove that our action provides, in the on-shell limit, a novel twistor space prescription utilizing curves of different degrees for computing pure gluonic amplitudes as was conjectured in \cite{gukov2004equivalence}. In this work, we provide a preliminary exposition of this idea.
\section{Acknowledgments}
\label{sec:ack}
We would like to thank Jacob Bourjaily, Lance Dixon, Radu Roiban, and Cristian Vergu for the discussions. We thank Mateusz Kulig for sharing his results on 9-point amplitudes.

H.K. acknowledges the hospitality of SLAC National Accelerator Laboratory, Stanford University, and the Department of Physics, Pennsylvania State University.

H.K. is supported by the National Science Centre, Poland grant no. 2021/41/N/ST2/02956. P.K. is supported by the National Science Centre, Poland grant no.  DEC-2020/39/O/ST2/03011.  A.M.S. is supported  by the U.S. Department of Energy Grant 
 DE-SC-0002145 and  in part by  National Science Centre in Poland, grant 2019/33/B/ST2/02588.

\bibliographystyle{JHEP}
\bibliography{library}

\newpage
\appendix
\section{Definition of Self-Dual and Anti-Self-Dual planes}
\label{sec:App01}
Self-Dual plane satisfies:
\begin{itemize}
    \item Any vector $P^\mu$ tangent to the plane is light-like  $P^2=0$, implying that the plane is a null plane. 
    \item A skew-symmetric bivector defined using a pair of tangent vectors say, $P^{\mu}, Q^{\nu}$  as follows
\begin{equation}
    G^{\mu\nu}=P^{\mu}Q^{\nu}-P^{\nu}Q^{\mu}\,,
\end{equation}
satisfies the Self-Duality condition 
\begin{equation}
    {G}^{\mu\nu} = \ast {G}^{\mu\nu}\, ,\quad \mathrm{where} \quad \ast {G}_{\mu\nu} = -i \epsilon_{\mu\nu\alpha\beta}{G}^{\alpha\beta}\,,
\end{equation}
where $\ast {G}_{\mu\nu}$ is the Hodge dual.
\end{itemize}
On the other hand, if ${G}^{\mu\nu}$, satisfies Anti-Self-Duality condition, then the plane is the Anti-Self-Dual. These planes are commonly referred to as $\alpha$, $\beta$ planes in the twistor space terminology \cite{Ward1977a}.
\section{Kernels of \texorpdfstring{$\widetilde{A}^{\bullet}[{Z}^{\bullet},{Z}^{\star}],\widetilde{A}^{\star}[{Z}^{\bullet},{Z}^{\star}]$}{Solu} in momentum space}
\label{sec:App0}

In order to obtain the explicit expressions for the kernels in the expansion of $\widetilde{A}^{\bullet}[{Z}^{\bullet} ,
 {Z}^{\star}] $ and $\widetilde{A}^{\star}[{Z}^{\bullet},{Z}^{\star}]$
\begin{equation}
    \widetilde{A}_a^{\bullet}(x^+;\mathbf{P})=\sum_{n=1}^{\infty}
    \int\! d^3\mathbf{p}_1\dots d^3\mathbf{p}_n \sum_{i=1}^{n}\, \widetilde\Xi_{i,n-i}^{ab_1\dots b_n}(\mathbf{P};\mathbf{p}_1,\dots,\mathbf{p}_n) \prod_{k=1}^{i}\widetilde{Z}_{b_k}^{\bullet}(x^+;\mathbf{p}_k)
    \prod_{l=i+1}^{n}\widetilde{Z}_{b_l}^{\star}(x^+;\mathbf{p}_l) \,,
    \label{eq:Abullet_to_ZP}
\end{equation}
\begin{equation}
    \widetilde{A}_a^{\star}(x^+;\mathbf{P})=\sum_{n=1}^{\infty}
    \int\! d^3\mathbf{p}_1\dots d^3\mathbf{p}_n \sum_{i=1}^{n}\, \widetilde\Lambda_{i,n-i}^{ab_1\dots b_n}(\mathbf{P};\mathbf{p}_1,\dots,\mathbf{p}_n) \prod_{k=1}^{i}\widetilde{Z}_{b_k}^{\star}(x^+;\mathbf{p}_k)
    \prod_{l=i+1}^{n}\widetilde{Z}_{b_l}^{\bullet}(x^+;\mathbf{p}_l) \,,
    \label{eq:Astar_to_ZP}
\end{equation}
one needs to, in principle, solve Eq.~\eqref{eq:AtoZ_ct}. There is, however, a simple method to do this. In \cite{Kakkad:2021uhv}, we showed that the canonical transformation Eq.~\eqref{eq:AtoZ_ct} from $A$-fields to the $Z$-fields is equal to the following pair of consecutive canonical transformations:
\begin{enumerate}[label={\it\roman*}$\,$)]
    \item \textit{Self-Dual part of the Yang-Mills action to a new kinetic term}:
    \begin{equation}
\mathcal{L}_{-+}[A^{\bullet},A^{\star}]+\mathcal{L}_{-++}[A^{\bullet},A^{\star}]
\,\, \longrightarrow \,\,
\mathcal{L}_{-+}[B^{\bullet},B^{\star}]
\,.
\label{eq:SD_MT}
\end{equation}
The L.H.S. represents the Self-Dual sector of the Yang-Mills action Eq.~\eqref{eq:YM_LC_action}. The above transformation is the so-called Mansfield's transformation \cite{Mansfield2006} and the equations governing it read
\begin{equation}
    B^{\bullet}_a[A^{\bullet}](x)=\mathcal{W}_{(+)}^a[A](x)\,,\qquad
    B_a^{\star}[A^{\bullet},A^{\star}](x) = 
    \int\! d^3\mathbf{y} \,
     \left[ \frac{\partial^2_-(y)}{\partial^2_-(x)} \,
     \frac{\delta \mathcal{W}^a_{(+)}[A](x^+;\mathbf{x})}{\delta {A}_c^{\bullet}(x^+;\mathbf{y})} \right] 
     {A}_c^{\star}(x^+;\mathbf{y})
      \, ,
      \label{eq:Bfield_transform}
\end{equation}
where $\mathcal{W}_{(+)}^a[A](x)$ is the Wilson line based functional introduced in Eq.~\eqref{eq:WL_gen}. The solutions of Eq.~\eqref{eq:Bfield_transform} in momentum space read
\begin{equation}
    \widetilde{A}^{\bullet}_a(x^+;\mathbf{P}) = \sum_{n=1}^{\infty} 
    \int d^3\mathbf{p}_1\dots d^3\mathbf{p}_n \, \widetilde{\Psi}_n^{a\{b_1\dots b_n\}}(\mathbf{P};\{\mathbf{p}_1,\dots ,\mathbf{p}_n\}) \prod_{i=1}^n\widetilde{B}^{\bullet}_{b_i}(x^+;\mathbf{p}_i)\,,
    \label{eq:A_bull_exp}
\end{equation}
\begin{equation}
    \widetilde{A}^{\star}_a(x^+;\mathbf{P}) = \sum_{n=1}^{\infty} 
    \int d^3\mathbf{p}_1\dots d^3\mathbf{p}_n \, {\widetilde \Omega}_{n}^{a b_1 \left \{b_2 \cdots b_n \right \} }(\mathbf{P}; \mathbf{p_1} ,\left \{ \mathbf{p_2} , \dots ,\mathbf{p_n} \right \}) \widetilde{B}^{\star}_{b_1}(x^+;\mathbf{p}_1)\prod_{i=2}^n\widetilde{B}^{\bullet}_{b_i}(x^+;\mathbf{p}_i)\, ,
    \label{eq:A_star_exp}
\end{equation}
where the kernels are
\begin{equation}
    {\widetilde \Psi}_{n}^{a \left \{b_1 \cdots b_n \right \} }(\mathbf{P}; \left \{\mathbf{p}_{1},  \dots ,\mathbf{p}_{n} \right \}) =- (-g)^{n-1} \,\,  
    \frac{{\widetilde v}^{\star}_{(1 \cdots n)1}}{{\widetilde v}^{\star}_{1(1 \cdots n)}} \, 
    \frac{\delta^{3} (\mathbf{p}_{1} + \cdots +\mathbf{p}_{n} - \mathbf{P})\,\,  \mathrm{Tr} (t^{a} t^{b_{1}} \cdots t^{b_{n}})}{{\widetilde v}^{\star}_{21}{\widetilde v}^{\star}_{32} \cdots {\widetilde v}^{\star}_{n(n-1)}}  
      \, ,
    \label{eq:psi_kernel}
\end{equation}
\begin{equation}
    {\widetilde \Omega}_{n}^{a b_1 \left \{b_2 \cdots b_n \right \} }(\mathbf{P}; \mathbf{p}_{1} , \left \{ \mathbf{p}_{2} , \dots ,\mathbf{p}_{n} \right \} ) = n \left(\frac{p_1^+}{p_{1\cdots n}^+}\right)^2 {\widetilde \Psi}_{n}^{a b_1 \cdots b_n }(\mathbf{P};  \mathbf{p}_{1},  \dots , \mathbf{p}_{n}) \, .
    \label{eq:omega_star_z}
\end{equation}
Substituting Eqs.~\eqref{eq:A_bull_exp}-\eqref{eq:A_star_exp} to the Yang-Mills action Eq.~\eqref{eq:YM_LC_action} derives the MHV action
\begin{equation}
S_{\mathrm{MHV}}\left[{B}^{\bullet}, {B}^{\star}\right]=\int dx^{+}\left(
-\int d^{3}\mathbf{x}\,\mathrm{Tr}\,\hat{B}^{\bullet}\square\hat{B}^{\star} 
+\mathcal{L}_{--+}+\dots+\mathcal{L}_{--+\dots+}+\dots\right)\,.\label{eq:MHV_action}
\end{equation}
    \item \textit{Anti-Self-Dual part of the MHV action to a new kinetic term}:
    \begin{equation}
\mathcal{L}_{-+}[B^{\bullet},B^{\star}]+\mathcal{L}_{--+}[B^{\bullet},B^{\star}]
\,\, \longrightarrow \,\,
\mathcal{L}_{-+}[Z^{\bullet},Z^{\star}]
\,.
\label{eq:BtoZtransform}
\end{equation}
The above transformation is exactly conjugate $\bullet \leftrightarrow \star$ to Eq.~\eqref{eq:SD_MT}. Consequently, the transformation
\begin{equation}
    Z^{\star}_a[B^{\star}](x)=\mathcal{W}_{(-)}^a[B](x)\,,\qquad
    Z_a^{\bullet}[B^{\bullet},B^{\star}](x) = 
    \int\! d^3\mathbf{y} \,
     \left[ \frac{\partial^2_-(y)}{\partial^2_-(x)} \,
     \frac{\delta \mathcal{W}^a_{(-)}[B](x^+;\mathbf{x})}{\delta {B}_c^{\star}(x^+;\mathbf{y})} \right] 
     {B}_c^{\bullet}(x^+;\mathbf{y})
      \, ,
      \label{eq:Zfield_transform}
\end{equation}
as well as the solutions
\begin{equation}
    \widetilde{B}^{\star}_a(x^+;\mathbf{P}) = \sum_{n=1}^{\infty} 
    \int d^3\mathbf{p}_1\dots d^3\mathbf{p}_n \, \overline{\widetilde{\Psi}}\,^{a\{b_1\dots b_n\}}_n(\mathbf{P};\{\mathbf{p}_1,\dots ,\mathbf{p}_n\}) \prod_{i=1}^n\widetilde{Z}^{\star}_{b_i}(x^+;\mathbf{p}_i)\,,
    \label{eq:BstarZ_exp}
\end{equation}
\begin{equation}
    \widetilde{B}^{\bullet}_a(x^+;\mathbf{P}) = \sum_{n=1}^{\infty} 
    \int d^3\mathbf{p}_1\dots d^3\mathbf{p}_n \, \overline{\widetilde \Omega}\,^{a b_1 \left \{b_2 \cdots b_n \right \}}_{n}(\mathbf{P}; \mathbf{p_1} ,\left \{ \mathbf{p_2} , \dots ,\mathbf{p_n} \right \}) \widetilde{Z}^{\bullet}_{b_1}(x^+;\mathbf{p}_1)\prod_{i=2}^n\widetilde{Z}^{\star}_{b_i}(x^+;\mathbf{p}_i)\,,
    \label{eq:BbulletZ_exp}
\end{equation}
with
\begin{equation}
    \overline{\widetilde \Psi}\,^{a \left \{b_1 \cdots b_n \right \}}_{n}(\mathbf{P}; \left \{\mathbf{p}_{1},  \dots ,\mathbf{p}_{n} \right \}) =- (-g)^{n-1} \,\,  
    \frac{{\widetilde v}_{(1 \cdots n)1}}{{\widetilde v}_{1(1 \cdots n)}} \, 
    \frac{\delta^{3} (\mathbf{p}_{1} + \cdots +\mathbf{p}_{n} - \mathbf{P})\,\,  \mathrm{Tr} (t^{a} t^{b_{1}} \cdots t^{b_{n}})}{{\widetilde v}_{21}{\widetilde v}_{32} \cdots {\widetilde v}_{n(n-1)}}  
      \, .
    \label{eq:psiBar_kernel}
\end{equation}
\begin{equation}
    \overline{\widetilde \Omega}\,^{a b_1 \left \{b_2 \cdots b_n \right \}}_{n}(\mathbf{P}; \mathbf{p}_{1} , \left \{ \mathbf{p}_{2} , \dots ,\mathbf{p}_{n} \right \} ) = n \left(\frac{p_1^+}{p_{1\cdots n}^+}\right)^2 \overline{\widetilde \Psi}\,^{a b_1 \cdots b_n }_{n}(\mathbf{P}; \mathbf{p}_{1},  \dots ,\mathbf{p}_{n}) \, .
    \label{eq:omegaBar_kernel}
\end{equation}
are also conjugate $\bullet \leftrightarrow \star$ (with obvious interchange of the fields 
\[
\{(\hat{B}^{\bullet},\hat{B}^{\star}), (\hat{A}^{\bullet},\hat{A}^{\star})\} 
 \rightarrow  \{(\hat{Z}^{\bullet},\hat{Z}^{\star}), (\hat{B}^{\bullet},\hat{B}^{\star})\}.
 \]
 Substituting Eqn.~\eqref{eq:BstarZ_exp}-\eqref{eq:BbulletZ_exp} to the MHV action Eq.~\eqref{eq:MHV_action} we derived the Z-field action Eq.~\eqref{eq:Z_action} in \cite{Kakkad:2021uhv}.
\end{enumerate}

Owing to the above, the simplest way to obtain the kernels in Eqs.~\eqref{eq:Abullet_to_ZP}-\eqref{eq:Astar_to_ZP} is to substitute Eqs.~\eqref{eq:BstarZ_exp}-\eqref{eq:BbulletZ_exp} to Eqs.~\eqref{eq:A_bull_exp}-\eqref{eq:A_star_exp}. To do this, we introduce the following notation for the color-ordered form of any generic kernel
\begin{equation}
    \widetilde{\kappa}\,^{a\{b_{1}\dots b_{m}\}}_m\left(\mathbf{P};\{\mathbf{p}_{1},\dots,\mathbf{p}_{m}\}\right)= \!\!\sum_{\underset{\text{\scriptsize permutations}}{\text{noncyclic}}}
 \mathrm{Tr}\left(t^{a}t^{b_1}\dots t^{b_m}\right)
 {\kappa}\left(1^{h_1},\dots,m^{h_m}\right)
\,,
\label{eq:K_color_decomp}
\end{equation}
where the curly braces on the color and momentum indices on L.H.S. represent the symmetry of the kernels with respect to the interchange of these indices. On R.H.S. we use numbers $i$ to represent the momentum $\mathbf{p}_{i}$ and to these numbers, we assign the helicities $h_i$ of the associated leg. And the sum is over non-cyclic permutations of $\{1,2,\dots m\}$.

Now, consider the kernel $\widetilde\Xi$ in the expansion of $\widetilde{A}_a^{\bullet}$
to $i$th order in $\widetilde{Z}^{\bullet}$ fields and $(n-i)$th order in $\widetilde{Z}^{\star}$ fields with the following color-ordering
\begin{equation}
    \widetilde\Xi_{i,n-i}^{ab_1\dots b_n}(\mathbf{P};\mathbf{p}_1,\dots,\mathbf{p}_n) = 
 \mathrm{Tr}\left(t^{a}t^{b_1}\dots t^{b_n}\right)
 {\Xi}\left(1^{+},\dots,i^{+}, (i+1)^{-}, \dots, n^- \right)
\,.
\label{eq:xi_com}
\end{equation}
When substituting  Eqs.~\eqref{eq:BbulletZ_exp} to Eqs.~\eqref{eq:A_bull_exp}, only one term contributes to the above color-ordered kernel. It is the term where $\widetilde{A}_a^{\bullet}$ is expanded to $i$th order in $\widetilde{B}^{\bullet}$ fields. Then each of $(i-1)$ $\widetilde{B}^{\bullet}$ fields are expanded to first order in $\widetilde{Z}^{\bullet}$ whereas the $i$th $\widetilde{B}^{\bullet}$ field is expanded to $(n-i+1)$th order consisting of one $\widetilde{Z}^{\bullet}$ field and $(n-i)$ $\widetilde{Z}^{\star}$ fields. The kernel therefore reads
\begin{equation}
    {\Xi}\left(1^{+},\dots,i^{+}, (i+1)^{-}, \dots, n^- \right) = {\Psi}\left(1^{+},\dots,(i-1)^+,[i, \dots, n]^{+} \right)\overline{\Omega}\left(i^{+}, (i+1)^{-}, \dots, n^- \right)
\,.
\label{eq:xi_com1}
\end{equation}
Or equivalently,
\begin{equation}
    \widetilde\Xi_{i,n-i}^{ab_1\dots b_n}(\mathbf{P};\mathbf{p}_1,\dots,\mathbf{p}_n) = 
 {\widetilde \Psi}_{i}^{a \left \{b_1 \cdots b_{i-1} b_k \right \} }(\mathbf{P}; \left \{\mathbf{p}_{1},  \dots,\mathbf{p}_{i-1} ,\mathbf{p}_{k} \right \}) \overline{\widetilde \Omega}\,^{b_k b_i \left \{b_{i+1} \cdots b_n \right \}}_{n-i+1}(\mathbf{p}_k; \mathbf{p}_{i} , \left \{ \mathbf{p}_{i+1} , \dots ,\mathbf{p}_{n} \right \} )
\,,
\label{eq:xi_com2}
\end{equation}
where $[i, \dots, n] \equiv \mathbf{p}_k \equiv \mathbf{p}_{i} +  \mathbf{p}_{i+1}  \dots + \mathbf{p}_{n}$. From above, we see that the kernel $\widetilde\Xi$ is symmetric under the interchange of the color and momentum indices associated with either the plus helicity fields (among themselves) or the minus helicity fields. Therefore, it is suitable to use
\begin{equation}
    \widetilde\Xi_{i,n-i}^{ab_1\dots b_n}(\mathbf{P};\mathbf{p}_1,\dots,\mathbf{p}_n) = \widetilde\Xi_{i,n-i}^{a\{b_1\dots b_i\} \{ b_{i+1} \dots b_n\}}(\mathbf{P};\{\mathbf{p}_1,\dots,\mathbf{p}_i\},\{\mathbf{p}_{i+1} \dots \mathbf{p}_n\})\,.
\end{equation}
Substituting Eqs.~\eqref{eq:psi_kernel},\eqref{eq:omegaBar_kernel} to Eq.~\eqref{eq:xi_com2} we get
\begin{multline}
   \widetilde\Xi_{i,n-i}^{a\{b_1\dots b_i\} \{ b_{i+1} \dots b_n\}}(\mathbf{P};\{\mathbf{p}_1,\dots,\mathbf{p}_i\},\{\mathbf{p}_{i+1} \dots \mathbf{p}_n\}) =  - (-g)^{n-1} \,\, 
    \delta^{3} (\mathbf{p}_{1} + \cdots +\mathbf{p}_{n} - \mathbf{P})\,\,  \mathrm{Tr} (t^{a} t^{b_{1}} \cdots t^{b_{n}})\\ 
    \frac{{\widetilde v}^{\star}_{(1 \cdots n)1}}{{\widetilde v}^{\star}_{1(1 \cdots n)}} \,\frac{1}{{\widetilde v}^{\star}_{21}{\widetilde v}^{\star}_{32} \cdots {\widetilde v}^{\star}_{(i-1)(i-2)}{\widetilde v}^{\star}_{(i\dots n)(i-1)}}    \left(\frac{p_i^+}{p_{i \dots n}^+}\right)  \, 
    \frac{1}{{\widetilde v}_{(i+1)i}{\widetilde v}_{(i+2)(i+1)} \cdots {\widetilde v}_{n(n-1)}}  \, .
    \label{eq:xi_kernel_P}
\end{multline}
The above expression represents the explicit form for the kernel in Eq.~\eqref{eq:Abullet_to_ZP}.

Now, for the kernel $ \widetilde\Lambda $ in the expansion of $\widetilde{A}_a^{\star}$ to $i$th order in $\widetilde{Z}^{\star}$ fields and $(n-i)$th order in $\widetilde{Z}^{\bullet}$ fields, when substituting  Eqs.~\eqref{eq:BstarZ_exp}-\eqref{eq:BbulletZ_exp} to Eqs.~\eqref{eq:A_star_exp}, we get a sum of terms that can contribute. First, $\widetilde{A}_a^{\star}$ is expanded to $(n-i+1)$th order consisting of one $\widetilde{B}^{\star}$ field and $(n-i)$ $\widetilde{B}^{\bullet}$ fields. Then the $\widetilde{B}^{\star}$ field can be expanded to $k$th order in $\widetilde{Z}^{\star}$ fields and the adjacent $\widetilde{B}^{\bullet}$ field can be expanded to $(i-k+1)$th order consisting of one $\widetilde{Z}^{\bullet}$ field and $(i-k)$ $\widetilde{Z}^{\star}$ fields. The remaining 
$\widetilde{B}^{\bullet}$ fields are all expanded to first order in $\widetilde{Z}^{\bullet}$. The kernel therefore reads
\begin{align}
    \widetilde\Lambda_{i,n-i}^{ab_1\dots b_n}(\mathbf{P};\mathbf{p}_1,\dots,\mathbf{p}_n) &= \widetilde\Lambda_{i,n-i}^{a\{b_1\dots b_i\} \{ b_{i+1} \dots b_n\}}(\mathbf{P};\{\mathbf{p}_1,\dots,\mathbf{p}_i\},\{\mathbf{p}_{i+1} \dots \mathbf{p}_n\})\,, \nonumber \\
     &=  {\widetilde \Omega}_{n-i+1}^{a c_1 \left \{c_2 b_{i+2} \cdots b_n \right \} }(\mathbf{P}; \mathbf{q}_{1} , \left \{ \mathbf{q}_{2}\mathbf{p}_{i+2} , \dots ,\mathbf{p}_{n} \right \} )\sum_{k=1}^{i} \Big[\overline{\widetilde \Psi}\,^{c_1 \left \{b_1 \cdots b_k \right \}}_{k}(\mathbf{q}_1; \left \{\mathbf{p}_{1},  \dots,\mathbf{p}_{k} \right \})\nonumber \\
    & \quad \times \overline{\widetilde \Omega}\,^{c_2 b_{i+1} \left \{b_{k+1} \cdots b_{i} \right \}}_{i-k+1}(\mathbf{q}_2; \mathbf{p}_{i+1} , \left \{ \mathbf{p}_{k+1} , \dots ,\mathbf{p}_{i} \right \} )\Big] \,.
\end{align}
The first expression above represents the symmetry with respect to the interchange of indices associated with the plus and minus helicity fields (among themselves) via the curly braces. Substituting  Eqs.~\eqref{eq:omega_star_z},~\eqref{eq:psiBar_kernel},~\eqref{eq:omegaBar_kernel}, we get
\begin{multline}
\widetilde\Lambda_{i,n-i}^{ab_1\dots b_n}(\mathbf{P};\mathbf{p}_1,\dots,\mathbf{p}_n) 
=    (-g)^{n-1} \, 
    \delta^{3} (\mathbf{p}_{1} + \cdots +\mathbf{p}_{n} - \mathbf{P})\,\,  \mathrm{Tr} (t^{a} t^{b_{1}} \cdots t^{b_{n}})  
      \, \\
 \sum_{k=1}^{i}   \left(\frac{p_{1\cdots k}^+}{p_{1\cdots n}^+}\right)  \,\,\frac{1}{{\widetilde v}^{\star}_{(k+1 \dots i+1)(1\cdots k)}{\widetilde v}^{\star}_{(i+2)(k+1 \cdots i+1)} {\widetilde v}^{\star}_{(i+3)(i+2)}\cdots {\widetilde v}^{\star}_{n(n-1)}}  \\ 
    \frac{{\widetilde v}_{(1 \cdots k)1}}{{\widetilde v}_{1(1 \cdots k)}} \, 
    \frac{1}{{\widetilde v}_{21}{\widetilde v}_{32} \cdots {\widetilde v}_{k(k-1)}}  
      \,     \left(\frac{p_{i+1}^+}{p_{k+1\cdots i+1}^+}\right)^2   \,\,\frac{{\widetilde v}_{(k+1\cdots i+1)(k+1)}}{{\widetilde v}_{(k+1)(k+1\cdots i+1)}}   
     \, 
    \frac{1}{{\widetilde v}_{(k+2)(k+1)}\cdots {\widetilde v}_{(i+1)i}}   \, . 
\end{multline}
The above expression represents the explicit form for the kernel in Eq.~\eqref{eq:Astar_to_ZP}.
\section{Expansion of the log term}
\label{sec:AppA}

In this appendix, we expand the following
\begin{equation}
    \Tr\ln\mathrm{M}= \Tr\ln\mathrm{M}_{\bullet\star} +
    \Tr\ln\left(\mathrm{M}_{\bullet \star}
    - \mathrm{M}_{\bullet\bullet}\mathrm{M}^{-1}_{\star\bullet}\mathrm{M}_{\star\star}\right) \,,
    \label{eq:lnM_dec}
\end{equation}
up to second order in fields and in the process highlight the approach following which one can compute higher point one-loop contributions. A priori, we expect this expansion to consist of tadpoles and bubble topologies.

Let us begin with the first term on the R.H.S. It reads
\begin{equation}
   \Tr\ln\left(\mathrm{M}_{\bullet\star}\right)_{IK}= 
   \Tr\ln\left\{ 
   \frac{\delta^2 S[Z_c]}{\delta Z^{\bullet I}\delta Z^{\star K}}
   + 
   J_{\star L}\frac{\delta^2 A^{\star L}[Z_c]}{\delta Z^{\bullet I}\delta Z^{\star K}} + J_{\bullet L}\frac{\delta^2 A^{\bullet L}[Z_c]}{\delta Z^{\bullet I}\delta Z^{\star K}} 
   \right\} \,.
   \label{eq:ln+-}
\end{equation}
$S[Z_c]$ is the Z-field action which in the collective index notation reads
\begin{equation}
    S[Z_c]
    = -Z_c^{\star L}\square_{LJ}Z_c^{\bullet J} -\sum_{n=4}^{\infty} 
     \, \sum_{m=2}^{n-2}\,
    \,\, \mathcal{U}^{ \{J_1 \dots J_m\}\{ J_{m+1}\dots J_n\} }_{\underbrace{-\,\cdots\,-}_{m}\underbrace{+ \,\cdots\, +}_{n-m}} 
   \prod_{i=1}^{m}Z_c^{\star J_i}
   \prod_{k=m+1}^{n}Z_c^{\bullet J_k} \, .
   \label{eq:S[Z]}
\end{equation}
Differentiating the above with respect to $Z^{\bullet I} $ and $  Z^{\star K}$ we get
\begin{equation}
    \frac{\delta^2 S[Z]}{\delta Z^{\bullet I}\delta Z^{\star K}}
    = -\square_{IK} -\sum_{n=4}^{\infty} 
     \, \sum_{m=2}^{n-2}
    \,m(n-m)\,\, \mathcal{U}^{ \{J_1 \dots J_m\}\{ J_{m+1}\dots J_n\} }_{-\,\cdots\,- + \,\cdots\, +} \,\delta_{J_1}^K \delta_{J_{m+1}}^I\,
   \prod_{i=2}^{m}Z_c^{\star J_i}
   \prod_{k=m+2}^{n}Z_c^{\bullet J_k} \, .
   \label{eq:S[Z]+-}
\end{equation}
For the auxiliary sources, we have the following relations
\begin{equation}
  \frac{\delta S_{\mathrm{YM}}[A[Z_c]]}{\delta A^{\star L}}=-J_{\star L}\,, \quad\quad
    \frac{\delta S_{\mathrm{YM}}[A[Z_c]]}{\delta A^{\bullet L}}=- J_{\bullet L} \,.
    \label{eq:J_currZ}
\end{equation}
Substituting the above along with Eq.~\eqref{eq:S[Z]+-} to Eq.~\eqref{eq:ln+-} and factoring out the inverse propagator we get
\begin{multline}
     \Tr\ln\Bigg[-\square \Bigg\{\delta_{IK} 
    + \sum_{n=4}^{\infty} 
     \, \sum_{m=2}^{n-2}
    \,m(n-m)\,\, \mathcal{U}^{ \{J_1 \dots J_m\}\{ J_{m+1}\dots J_n\} }_{-\,\cdots\,- + \,\cdots\, +} \,\left(\frac{1}{\square} \right)_{IP}\delta_{J_1}^K \delta_{J_{m+1}}^P\,
   \prod_{i=2}^{m}Z_c^{\star J_i}
   \prod_{k=m+2}^{n}Z_c^{\bullet J_k}\\
   + 
   \frac{\delta S_{\mathrm{YM}}[A[Z_c]]}{\delta A^{\star L}}\frac{\delta^2 A^{\star L}[Z_c]}{\delta Z^{\bullet P}\delta Z^{\star K}}\left(\frac{1}{\square} \right)_{IP} + \frac{\delta S_{\mathrm{YM}}[A[Z_c]]}{\delta A^{\bullet L}}\frac{\delta^2 A^{\bullet L}[Z_c]}{\delta Z^{\bullet P}\delta Z^{\star K}}\left(\frac{1}{\square} \right)_{IP} \Bigg\}\Bigg] \, .
   \label{eq:S[Z]+-f}
\end{multline}
For the first order derivative of the Yang-Mills action, we have the following expression
\begin{multline}
   \frac{\delta S_{\mathrm{YM}}[A[Z_c]]}
    {\delta A^{\star L}} 
    = -\square_{LJ}A^{\bullet J}[Z_c]-\left(V_{-++}\right)_{LJK}A^{\bullet J}[Z_c] A^{\bullet K}[Z_c] -2\left(V_{--+}\right)_{LKJ}A^{\star K}[Z_c] A^{\bullet J}[Z_c] \\
    -2\left(V_{--++}\right)_{LIJK}A^{\star I}[Z_c] A^{\bullet J}[Z_c] A^{\bullet K}[Z_c]
    \label{eq:S+-Z}
    \,,
\end{multline}
\begin{multline}
   \frac{\delta S_{\mathrm{YM}}[A[Z_c]]}
    {\delta A^{\bullet L}} 
    = -\square_{LJ}A^{\star J}[Z_c]-2\left(V_{-++}\right)_{JLK}A^{\star J}[Z_c] A^{\bullet K}[Z_c] -\left(V_{--+}\right)_{KJL}A^{\star K}[Z_c] A^{\star J}[Z_c] \\
    -2\left(V_{--++}\right)_{IJLK}A^{\star I}[Z_c] A^{\star J}[Z_c] A^{\bullet K}[Z_c]
    \label{eq:S+-Z2}
    \,,
\end{multline}
where
\begin{equation}
A^{\star L}[Z_c] = 
\sum_{n=1}^{\infty} 
     \, \sum_{i=1}^{n}\,\Lambda_{i,n-i}^{L \{J_1 \dots J_i\}\{ J_{i+1}\dots J_n\} } \,\prod_{k=1}^{i}{Z}_c^{\star J_k} \prod_{l=i+1}^{n}{Z}_c^{\bullet J_l}\,,
    \label{eq:A_star_Za}
\end{equation}
and
\begin{equation}
A^{\bullet L}[Z_c] = 
\sum_{n=1}^{\infty} 
     \, \sum_{i=1}^{n}\, \Xi_{i,n-i}^{L \{J_1 \dots J_i\}\{ J_{i+1}\dots J_n\} }\,\prod_{k=1}^{i}{Z}_c^{\bullet J_k} \prod_{l=i+1}^{n}{Z}_c^{\star J_l}\,.
    \label{eq:A_bul_Za}
\end{equation}
Differentiating Eqs.~\eqref{eq:A_star_Za}-\eqref{eq:A_bul_Za} with respect to $Z^{\bullet I} $ and $  Z^{\star K}$ we get
\begin{equation}
\frac{\delta^2 A^{\star L}[Z_c]}{\delta Z^{\bullet I}\delta Z^{\star K}} = 
\sum_{n=2}^{\infty} 
     \, \sum_{i=1}^{n-1}\,i (n-i) \Lambda_{i,n-i}^{L \{J_1 \dots J_i\}\{ J_{i+1}\dots J_n\} } \delta_{J_1}^K \delta_{J_{i+1}}^I\,\prod_{k=2}^{i}{Z}_c^{\star J_k} \prod_{l=i+2}^{n}{Z}_c^{\bullet J_l}\,,
    \label{eq:A_star_bs}
\end{equation}
and
\begin{equation}
\frac{\delta^2 A^{\bullet L}[Z_c]}{\delta Z^{\bullet I}\delta Z^{\star K}} = 
\sum_{n=2}^{\infty} 
     \, \sum_{i=1}^{n-1}\,i (n-i) \Xi_{i,n-i}^{L \{J_1 \dots J_i\}\{ J_{i+1}\dots J_n\} } \delta_{J_1}^I \delta_{J_{i+1}}^K\,\prod_{k=2}^{i}{Z}_c^{\bullet J_k} \prod_{l=i+2}^{n}{Z}_c^{\star J_l}\,.
    \label{eq:A_bul_bs}
\end{equation}

Now, let us consider the last two terms in Eq.~\eqref{eq:S[Z]+-f} one by one. For the first term, combining Eq.~\eqref{eq:S+-Z} and Eq.~\eqref{eq:A_star_bs} reads
\begin{multline}
   \frac{\delta S_{\mathrm{YM}}[A[Z_c]]}{\delta A^{\star L}}\frac{\delta^2 A^{\star L}[Z_c]}{\delta Z^{\bullet P}\delta Z^{\star K}}\left(\frac{1}{\square} \right)_{IP} =   
     \Bigg\{-\square_{LJ}A^{\bullet J}[Z_c]-\left(V_{-++}\right)_{LJQ}A^{\bullet J}[Z_c] A^{\bullet Q}[Z_c] \\
     -2\left(V_{--+}\right)_{LQJ}A^{\star Q}[Z_c] A^{\bullet J}[Z_c] 
    -2\left(V_{--++}\right)_{LQJR}A^{\star I}[Z_c] A^{\bullet J}[Z_c] A^{\bullet R}[Z_c]\Bigg\} \\
     \Bigg\{\sum_{n=2}^{\infty} 
     \,\sum_{i=1}^{n-1}\,i (n-i) \Lambda_{i,n-i}^{L \{J_1 \dots J_i\}\{ J_{i+1}\dots J_n\} } \delta_{J_1}^K \frac{\delta_{J_{i+1}}^P}{\square_{IP}}\,\prod_{k=2}^{i}{Z}_c^{\star J_k} \prod_{l=i+2}^{n}{Z}_c^{\bullet J_l} \Bigg\}\,.
\end{multline}
Substituting Eqs.~\eqref{eq:A_star_Za}-\eqref{eq:A_bul_Za} on the R.H.S we get
\begin{multline}
     \Bigg\{-\square_{LJ} {Z}_c^{\bullet J}  
    -\square_{LJ} \Xi_{1,1}^{J \{J_1 \}\{J_2 \} }{Z}_c^{\bullet J_1}{Z}_c^{\star J_2} -\square_{LJ} \Xi_{2,0}^{J \{J_1 J_2\} }{Z}_c^{\bullet J_1} {Z}_c^{\bullet J_2} \\
    -\left(V_{-++}\right)_{LJQ}{Z}_c^{\bullet J} {Z}_c^{\bullet Q} 
    -2\left(V_{--+}\right)_{LQJ}{Z}_c^{\star Q} {Z}_c^{\bullet J} -2\left(V_{--++}\right)_{LQJR}{Z}_c^{\star Q}{Z}_c^{\bullet J} {Z}_c^{\bullet R} + \dots \Bigg\}\\
    \Bigg\{\frac{\Lambda_{1,1}^{L \{K\}\{ P\}} }{\square_{IP}} + 2 \frac{\Lambda_{1,2}^{L \{K\}\{ P L_3\}} }{\square_{IP}}{Z}_c^{\bullet L_3} + 2 \frac{\Lambda_{2,1}^{L \{K L_2\}\{ P \}} }{\square_{IP}}{Z}_c^{\star L_2} + \dots \Bigg\}
\end{multline}
where we used the identity $\Xi_{1,0}^{J \{K \} } = \Lambda_{1,0}^{J \{K \} } = \delta^{JK}$. The dots represent higher order terms in fields.

Up to the second order in fields, the above expansion reads
\begin{multline}
  \left. \frac{\delta S_{\mathrm{YM}}[A[Z_c]]}{\delta A^{\star L}}\frac{\delta^2 A^{\star L}[Z_c]}{\delta Z^{\bullet P}\delta Z^{\star K}}\left(\frac{1}{\square} \right)_{IP}\,\,\right|_{\mathrm{2nd}} = \Bigg\{-\frac{\Lambda_{1,1}^{L \{K\}\{ P\}} }{\square_{IP}} \square_{LJ}{Z}_c^{\bullet J} \\
    - \Bigg[ 2 \frac{\Lambda_{1,2}^{L \{K\}\{ P J_1\}} }{\square_{IP}}\square_{LJ_2} + \frac{\Lambda_{1,1}^{L \{K\}\{ P\}} }{\square_{IP}}\square_{LJ} \Xi_{2,0}^{J \{J_1 J_2\} } + \frac{\Lambda_{1,1}^{L \{K\}\{ P\}} }{\square_{IP}}\left(V_{-++}\right)_{LJ_1 J_2}\Bigg]{Z}_c^{\bullet J_1}{Z}_c^{\bullet J_2}\\
    - \Bigg[\frac{\Lambda_{1,1}^{L \{K\}\{ P\}} }{\square_{IP}} \square_{LJ} \Xi_{1,1}^{J \{J_1 \}\{J_2 \} } + 2 \frac{\Lambda_{2,1}^{L \{K J_2\}\{ P \}} }{\square_{IP}}\square_{LJ_1} + \frac{\Lambda_{1,1}^{L \{K\}\{ P\}} }{\square_{IP}} 2\left(V_{--+}\right)_{LJ_2 J_1 }\Bigg] {Z}_c^{\bullet J_1}{Z}_c^{\star J_2} \Bigg\}\,.
    \label{eq:Js_Z2}
\end{multline}

Repeating the above steps for the other term, we get
\begin{multline}
    \frac{\delta S_{\mathrm{YM}}[A[Z_c]]}{\delta A^{\bullet L}}\frac{\delta^2 A^{\bullet L}[Z_c]}{\delta Z^{\bullet P}\delta Z^{\star K}}\left(\frac{1}{\square} \right)_{IP} = \Bigg\{ -\square_{LJ}A^{\star J}[Z_c]-2\left(V_{-++}\right)_{JLQ}A^{\star J}[Z_c] A^{\bullet Q}[Z_c] \\
    -\left(V_{--+}\right)_{QJL}A^{\star Q}[Z_c] A^{\star J}[Z_c] 
    -2\left(V_{--++}\right)_{QJLR}A^{\star Q}[Z_c] A^{\star J}[Z_c] A^{\bullet R}[Z_c] \Bigg\} \\
    \Bigg\{ \sum_{n=2}^{\infty} 
     \, \sum_{i=1}^{n-1}\,i (n-i) \Xi_{i,n-i}^{L \{J_1 \dots J_i\}\{ J_{i+1}\dots J_n\} }  \frac{\delta_{J_{1}}^P}{\square_{IP}} \delta_{J_{i+1}}^K\,\prod_{k=2}^{i}{Z}_c^{\bullet J_k} \prod_{l=i+2}^{n}{Z}_c^{\star J_l}\Bigg\}\,.
\end{multline}
After that we substitute Eqs.~\eqref{eq:A_star_Za}-\eqref{eq:A_bul_Za} on the R.H.S
\begin{multline}
    \Bigg\{-\square_{LJ}Z_c^{\star J} -\square_{LJ} \Lambda_{1,1}^{J \{J_1\}\{  J_2\} } {Z}_c^{\star J_1} {Z}_c^{\bullet J_2}-\square_{LJ} \Lambda_{2,0}^{J \{J_1 J_2\} } {Z}_c^{\star J_1} {Z}_c^{\star J_2} -2\left(V_{-++}\right)_{J_1 L J_2}Z_c^{\star J_1} Z_c^{\bullet J_2} \\
    -\left(V_{--+}\right)_{J_1 J_2 L} Z_c^{\star J_1} Z_c^{\star J_2}+ \dots \Bigg\} \Bigg\{  \frac{\Xi_{1,1}^{L \{P\}\{ K\} } }{\square_{IP}}+ 2\frac{\Xi_{1,2}^{L \{P\}\{ K L_1 \} } }{\square_{IP}}{Z}_c^{\star L_1} + 2\frac{\Xi_{2,1}^{L \{P L_1\}\{ K  \} } }{\square_{IP}}{Z}_c^{\bullet L_1} + \dots \Bigg\}\,.
\end{multline}
And finally, collecting terms up to second order in fields, we get
\begin{multline}
    \left. \frac{\delta S_{\mathrm{YM}}[A[Z_c]]}{\delta A^{\bullet L}}\frac{\delta^2 A^{\bullet L}[Z_c]}{\delta Z^{\bullet P}\delta Z^{\star K}}\left(\frac{1}{\square} \right)_{IP}\,\, \right|_{\mathrm{2nd}} = \Bigg\{ -\square_{LJ}\frac{\Xi_{1,1}^{L \{P\}\{ K\} } }{\square_{IP}}Z_c^{\star J} \\
    - \Bigg[ \frac{\Xi_{1,1}^{L \{P\}\{ K\} } }{\square_{IP}} \square_{LJ} \Lambda_{1,1}^{J \{J_1\}\{  J_2\} } + \frac{\Xi_{1,1}^{L \{P\}\{ K\} } }{\square_{IP}} 2\left(V_{-++}\right)_{J_1 L J_2} + 2\frac{\Xi_{2,1}^{L \{P J_2\}\{ K  \} } }{\square_{IP}} \square_{LJ_1}\Bigg] {Z}_c^{\star J_1} {Z}_c^{\bullet J_2} \\
    -\Bigg[ \frac{\Xi_{1,1}^{L \{P\}\{ K\} } }{\square_{IP}}\square_{LJ} \Lambda_{2,0}^{J \{J_1 J_2\} } +\frac{\Xi_{1,1}^{L \{P\}\{ K\} } }{\square_{IP}}\left(V_{--+}\right)_{J_1 J_2 L} + 2\frac{\Xi_{1,2}^{L \{P\}\{ K J_1 \} } }{\square_{IP}}\square_{LJ_2} \Bigg]{Z}_c^{\star J_1} {Z}_c^{\star J_2} \Bigg\} \,.
    \label{eq:Jb_Z2}
\end{multline}

Substituting Eqs.~\eqref{eq:Js_Z2}-\eqref{eq:Jb_Z2} to Eq.~\eqref{eq:S[Z]+-f}, up to second order in fields we have
\begin{multline}
     \Tr\ln\Bigg[ \delta_{IK} 
    + \Bigg\{ -\frac{\Lambda_{1,1}^{L \{K\}\{ P\}} }{\square_{IP}} \square_{LJ}{Z}_c^{\bullet J}  -\square_{LJ}\frac{\Xi_{1,1}^{L \{P\}\{ K\} } }{\square_{IP}}Z_c^{\star J} +  4  \,\frac{\mathcal{U}^{ \{K  J_1\}\{P J_2\} }_{-\,- + \, +}}{\square_{IP}} \,
   Z_c^{\star J_1}
   Z_c^{\bullet J_2} \\
    - \Bigg( 2 \frac{\Lambda_{1,2}^{L \{K\}\{ P J_1\}} }{\square_{IP}}\square_{LJ_2} + \frac{\Lambda_{1,1}^{L \{K\}\{ P\}} }{\square_{IP}}\square_{LJ} \Xi_{2,0}^{J \{J_1 J_2\} } + \frac{\Lambda_{1,1}^{L \{K\}\{ P\}} }{\square_{IP}}\left(V_{-++}\right)_{LJ_1 J_2}\Bigg){Z}_c^{\bullet J_1}{Z}_c^{\bullet J_2}\\
    - \Bigg(\frac{\Lambda_{1,1}^{L \{K\}\{ P\}} }{\square_{IP}} \square_{LJ} \Xi_{1,1}^{J \{J_1 \}\{J_2 \} } + 2 \frac{\Lambda_{2,1}^{L \{K J_2\}\{ P \}} }{\square_{IP}}\square_{LJ_1} + \frac{\Lambda_{1,1}^{L \{K\}\{ P\}} }{\square_{IP}} 2\left(V_{--+}\right)_{LJ_2 J_1 }\Bigg) {Z}_c^{\bullet J_1}{Z}_c^{\star J_2} \\
    - \Bigg( \frac{\Xi_{1,1}^{L \{P\}\{ K\} } }{\square_{IP}} \square_{LJ} \Lambda_{1,1}^{J \{J_1\}\{  J_2\} } + \frac{\Xi_{1,1}^{L \{P\}\{ K\} } }{\square_{IP}} 2\left(V_{-++}\right)_{J_1 L J_2} + 2\frac{\Xi_{2,1}^{L \{P J_2\}\{ K  \} } }{\square_{IP}} \square_{LJ_1}\Bigg) {Z}_c^{\star J_1} {Z}_c^{\bullet J_2} \\
    -\Bigg( \frac{\Xi_{1,1}^{L \{P\}\{ K\} } }{\square_{IP}}\square_{LJ} \Lambda_{2,0}^{J \{J_1 J_2\} } +\frac{\Xi_{1,1}^{L \{P\}\{ K\} } }{\square_{IP}}\left(V_{--+}\right)_{J_1 J_2 L} + 2\frac{\Xi_{1,2}^{L \{P\}\{ K J_1 \} } }{\square_{IP}}\square_{LJ_2} \Bigg){Z}_c^{\star J_1} {Z}_c^{\star J_2} +\dots\Bigg\}\Bigg] \, .
   \label{eq:S[Z]+-f11}
\end{multline}
Above, we factored out $\Tr\ln(-\square)$ as an infinite constant and we ignored it. The dots represent higher order terms in fields starting with 3-point. Notice, the last term in the first line above represents the term originating from second order differentiation of the Z-field action. The log can be expanded into a series using 
\begin{equation}
    \ln \left(\mathbb{1} +x\right)= \sum_{k=1}^{\infty} \frac{(-1)^{k+1}}{k} x^k\,.
    \label{eq:LOG_exp}
\end{equation}
Up to second order in fields, the expansion reads
\begin{multline}
   \Tr \Bigg[ \Bigg\{ -\frac{\Lambda_{1,1}^{L \{K\}\{ P\}} }{\square_{IP}} \square_{LJ}{Z}_c^{\bullet J}  -\square_{LJ}\frac{\Xi_{1,1}^{L \{P\}\{ K\} } }{\square_{IP}}Z_c^{\star J} +  4  \,\frac{\mathcal{U}^{ \{K  J_1\}\{P J_2\} }_{-\,- + \, +}}{\square_{IP}} \,
   Z_c^{\star J_1}
   Z_c^{\bullet J_2} \\
    - \Bigg( 2 \frac{\Lambda_{1,2}^{L \{K\}\{ P J_1\}} }{\square_{IP}}\square_{LJ_2} + \frac{\Lambda_{1,1}^{L \{K\}\{ P\}} }{\square_{IP}}\square_{LJ} \Xi_{2,0}^{J \{J_1 J_2\} } + \frac{\Lambda_{1,1}^{L \{K\}\{ P\}} }{\square_{IP}}\left(V_{-++}\right)_{LJ_1 J_2}\Bigg){Z}_c^{\bullet J_1}{Z}_c^{\bullet J_2}\\
    - \Bigg(\frac{\Lambda_{1,1}^{L \{K\}\{ P\}} }{\square_{IP}} \square_{LJ} \Xi_{1,1}^{J \{J_1 \}\{J_2 \} } + 2 \frac{\Lambda_{2,1}^{L \{K J_2\}\{ P \}} }{\square_{IP}}\square_{LJ_1} + \frac{\Lambda_{1,1}^{L \{K\}\{ P\}} }{\square_{IP}} 2\left(V_{--+}\right)_{LJ_2 J_1 }\Bigg) {Z}_c^{\bullet J_1}{Z}_c^{\star J_2} \\
    - \Bigg( \frac{\Xi_{1,1}^{L \{P\}\{ K\} } }{\square_{IP}} \square_{LJ} \Lambda_{1,1}^{J \{J_1\}\{  J_2\} } + \frac{\Xi_{1,1}^{L \{P\}\{ K\} } }{\square_{IP}} 2\left(V_{-++}\right)_{J_1 L J_2} + 2\frac{\Xi_{2,1}^{L \{P J_2\}\{ K  \} } }{\square_{IP}} \square_{LJ_1}\Bigg) {Z}_c^{\star J_1} {Z}_c^{\bullet J_2} \\
    -\Bigg( \frac{\Xi_{1,1}^{L \{P\}\{ K\} } }{\square_{IP}}\square_{LJ} \Lambda_{2,0}^{J \{J_1 J_2\} } +\frac{\Xi_{1,1}^{L \{P\}\{ K\} } }{\square_{IP}}\left(V_{--+}\right)_{J_1 J_2 L} + 2\frac{\Xi_{1,2}^{L \{P\}\{ K J_1 \} } }{\square_{IP}}\square_{LJ_2} \Bigg){Z}_c^{\star J_1} {Z}_c^{\star J_2} \Bigg\} \\
    -\frac{1}{2} \Bigg\{ \frac{\Lambda_{1,1}^{L_1 \{K_1\}\{ P\}} }{\square_{IP}} \square_{L_1 J_1} \frac{\Lambda_{1,1}^{L_2 \{K\}\{ P_1\}} }{\square_{K_1 P_1}} \square_{L_2 J_2}{Z}_c^{\bullet J_1}{Z}_c^{\bullet J_2} + \frac{\Lambda_{1,1}^{L_1 \{K_1\}\{ P\}} }{\square_{IP}} \square_{L_1 J_1 } \square_{L_2 J_2}\frac{\Xi_{1,1}^{L_2 \{P_1\}\{ K\} } }{\square_{K_1 P_1}}{Z}_c^{\bullet J_1} Z_c^{\star J_2} \\
    +\square_{L_1 J_1}\frac{\Xi_{1,1}^{L_1 \{P\}\{ K_1\} } }{\square_{IP}}\frac{\Lambda_{1,1}^{L_2 \{K\}\{ P_1\}} }{\square_{K_1P_1}} \square_{L_2 J_2 }Z_c^{\star J_1}{Z}_c^{\bullet J_2} + \square_{L_1 J_1}\frac{\Xi_{1,1}^{L_1 \{P\}\{ K_1\} } }{\square_{IP}}\square_{L_2 J_2}\frac{\Xi_{1,1}^{L_2 \{P_1\}\{ K\} } }{\square_{K_1P_1}}Z_c^{\star J_1}Z_c^{\star J_2} \Bigg\}+\dots\Bigg] \, .
   \label{eq:S[Z]+-f1}
\end{multline}

We now consider the second term in Eq.~\eqref{eq:lnM_dec}
\begin{equation}
     \Tr\ln\left(\mathrm{M}_{\bullet\star}
    - \mathrm{M}_{\bullet\bullet}\mathrm{M}^{-1}_{\star\bullet}\mathrm{M}_{\star\star}\right)\,.
\end{equation}
We already derived the expression for the expansion of the first matrix $\mathrm{M}_{\bullet\star}$ in the log above. Up to the second order in fields, the expansion is given in the square brackets in Eq.~\eqref{eq:S[Z]+-f1}. Let us therefore focus on the remaining three matrices. Factoring our the inverse propagators  from $\mathrm{M}_{\bullet\star}$ (\emph{cf.} Eq.~\eqref{eq:S[Z]+-f}) "equips" each matrix in the second term with a propagator
\begin{equation}
     \Tr\ln \Bigg[-\square\left\{\left(\frac{-1}{\square} \right)\mathrm{M}_{\bullet\star}
    - \left(\frac{-1}{\square} \right)\mathrm{M}_{\bullet\bullet}\left(\frac{-1}{\square} \mathrm{M}_{\star\bullet}\right)^{-1} \left(\frac{-1}{\square} \right)\mathrm{M}_{\star\star} \right\}\Bigg] \,.
    \label{eq:matr_in}
\end{equation}

Let us begin with the first matrix in the second term above
\begin{equation}
   \left(\frac{-1}{\square} \right)_{IP} \left(\mathrm{M}_{\bullet\bullet}\right)_{PK}= \left(\frac{-1}{\square} \right)_{IP}
   \left\{ 
   \frac{\delta^2 S[Z_c]}{\delta Z^{\bullet P}\delta Z^{\bullet K}}
   + 
   J_{\star L}\frac{\delta^2 A^{\star L}[Z_c]}{\delta Z^{\bullet P}\delta Z^{\bullet K}} + J_{\bullet L}\frac{\delta^2 A^{\bullet L}[Z_c]}{\delta Z^{\bullet P}\delta Z^{\bullet K}} 
   \right\} \,.
   \label{eq:MBB}
\end{equation}
Differentiating $S[Z_c]$ with respect to $Z^{\bullet P} $ and $  Z^{\bullet K}$ we get
\begin{equation}
    \frac{-1}{\square}_{IP} \frac{\delta^2 S[Z]}{\delta Z^{\bullet P}\delta Z^{\bullet K}}
    =  \sum_{n=4}^{\infty} 
     \, \sum_{m=2}^{n-2}
    \,(n-m)(n-m-1)\,\, \mathcal{U}^{ \{J_1 \dots J_m\}\{ J_{m+1}\dots J_n\} }_{\underbrace{-\,\cdots\,-}_{m}\underbrace{+ \,\cdots\, +}_{n-m}} \,\frac{\delta_{J_{m+1}}^P}{\square_{IP}} \delta_{J_{m+2}}^K\,
   \prod_{i=1}^{m}Z_c^{\star J_i}
   \prod_{k=m+3}^{n}Z_c^{\bullet J_k} \, .
   \label{eq:S[Z]++}
\end{equation}
Considering terms up to second order in fields we have
\begin{equation}
   \left. \frac{-1}{\square}_{IP} \frac{\delta^2 S[Z]}{\delta Z^{\bullet P}\delta Z^{\bullet K}} \right|_{\mathrm{2nd}}
    =  2\,  \,\frac{\mathcal{U}^{ \{J_1 J_2\}\{ PK\} }_{-\,- + \, +}}{\square_{IP}} \,
   Z_c^{\star J_1} Z_c^{\star J_2} \, .
   \label{eq:S[Z]++2n}
\end{equation}
For the second term in Eq.~\eqref{eq:MBB}, we require 
\begin{equation}
\frac{\delta^2 A^{\star L}[Z_c]}{\delta Z^{\bullet I}\delta Z^{\bullet K}} = 
\sum_{n=3}^{\infty} 
     \, \sum_{i=1}^{n-2}\, (n-i) (n-i-1) \Lambda_{i,n-i}^{L \{J_1 \dots J_i\}\{ J_{i+1}\dots J_n\} } \delta_{J_{i+1}}^I \delta_{J_{i+2}}^K \prod_{k=1}^{i}{Z}_c^{\star J_k} \prod_{l=i+3}^{n}{Z}_c^{\bullet J_l}\,.
    \label{eq:A_star_bb}
\end{equation}
Combining the above with the source $J_{\star L}$ using Eq.\eqref{eq:J_currZ} and Eq.~\eqref{eq:S+-Z}, up to second order in fields we get
\begin{equation}
    \left. \frac{\delta S_{\mathrm{YM}}[A[Z_c]]}{\delta A^{\star L}}\frac{\delta^2 A^{\star L}[Z_c]}{\delta Z^{\bullet P}\delta Z^{\bullet K}}\left(\frac{1}{\square} \right)_{IP}\,\,\right|_{\mathrm{2nd}} = \Bigg\{ -2\,\square_{LJ_1}   \frac{\Lambda_{1,2}^{L \{J_2 \}\{ PK\} }}{\square_{IP}} Z_c^{\bullet J_1}{Z}_c^{\star J_2}  \Bigg\}\,.
\end{equation}
For the last term in Eq.~\eqref{eq:MBB}, we require
\begin{equation}
\frac{\delta^2 A^{\bullet L}[Z_c]}{\delta Z^{\bullet I}\delta Z^{\bullet K}} = 
\sum_{n=2}^{\infty} 
     \, \sum_{i=2}^{n}\,i (i-1)\, \Xi_{i,n-i}^{L \{J_1 \dots J_i\}\{ J_{i+1}\dots J_n\} } \delta_{J_1}^I \delta_{J_2}^K\,\prod_{k=3}^{i}{Z}_c^{\bullet J_k} \prod_{l=i+1}^{n}{Z}_c^{\star J_l}\,.
    \label{eq:A_bul_bb}
\end{equation}
Repeating the same steps, up to the second order in fields we get

\begin{multline}
    \left. \frac{\delta S_{\mathrm{YM}}[A[Z_c]]}{\delta A^{\bullet L}}\frac{\delta^2 A^{\bullet L}[Z_c]}{\delta Z^{\bullet P}\delta Z^{\bullet K}}\left(\frac{1}{\square} \right)_{IP}\,\,\right|_{\mathrm{2nd}} = \Bigg\{ -2\,\frac{\Xi_{2,0}^{L \{PK\}  }}{\square_{IP}}\square_{LJ}Z_c^{\star J} \\ 
    -\Bigg(2\,\frac{\Xi_{2,0}^{L \{PK\}  }}{\square_{IP}}\square_{LJ}\Lambda_{1,1}^{J \{J_1 \}\{ J_{2}\} }  +4\,\frac{\Xi_{2,0}^{L \{PK\}  }}{\square_{IP}}\left(V_{-++}\right)_{J_1LJ_2} +6\,\square_{LJ_1} \frac{\Xi_{3,0}^{L \{PK J_2 \}}}{\square_{IP}}\Bigg) Z_c^{\star J_1}{Z}_c^{\bullet J_2}
    \\ -\Bigg(2\,\frac{\Xi_{2,0}^{L \{PK\}  }}{\square_{IP}}\square_{LJ}\Lambda_{2,0}^{J \{J_1 J_{2}\} }   
     +2\,\frac{\Xi_{2,0}^{L \{PK\}  }}{\square_{IP}}\left(V_{--+}\right)_{J_1 J_2 L} +2\,\square_{LJ_1}\frac{\Xi_{2,1}^{L \{PK\} \{J_2 \}}}{\square_{IP}}\Bigg) Z_c^{\star J_1}{Z}_c^{\star J_2} \Bigg\}\,.
\end{multline}

Now for the inverse of the matrix, we employ the following strategy. Already from Eq.~\eqref{eq:S[Z]+-f1}, we know that
\begin{multline}
  \left. \left(\frac{-1}{\square} \mathrm{M}_{\star\bullet}\right)_{IK}\right|_{\mathrm{2nd}} = \delta_{IK} 
    + \Bigg\{ -\frac{\Lambda_{1,1}^{L \{K\}\{ P\}} }{\square_{IP}} \square_{LJ}{Z}_c^{\bullet J}  -\square_{LJ}\frac{\Xi_{1,1}^{L \{P\}\{ K\} } }{\square_{IP}}Z_c^{\star J} +  4  \,\frac{\mathcal{U}^{ \{K  J_1\}\{P J_2\} }_{-\,- + \, +}}{\square_{IP}} \,
   Z_c^{\star J_1}
   Z_c^{\bullet J_2} \\
    - \Bigg( 2 \frac{\Lambda_{1,2}^{L \{K\}\{ P J_1\}} }{\square_{IP}}\square_{LJ_2} + \frac{\Lambda_{1,1}^{L \{K\}\{ P\}} }{\square_{IP}}\square_{LJ} \Xi_{2,0}^{J \{J_1 J_2\} } + \frac{\Lambda_{1,1}^{L \{K\}\{ P\}} }{\square_{IP}}\left(V_{-++}\right)_{LJ_1 J_2}\Bigg){Z}_c^{\bullet J_1}{Z}_c^{\bullet J_2}\\
    - \Bigg(\frac{\Lambda_{1,1}^{L \{K\}\{ P\}} }{\square_{IP}} \square_{LJ} \Xi_{1,1}^{J \{J_1 \}\{J_2 \} } + 2 \frac{\Lambda_{2,1}^{L \{K J_2\}\{ P \}} }{\square_{IP}}\square_{LJ_1} + \frac{\Lambda_{1,1}^{L \{K\}\{ P\}} }{\square_{IP}} 2\left(V_{--+}\right)_{LJ_2 J_1 }\Bigg) {Z}_c^{\bullet J_1}{Z}_c^{\star J_2} \\
    - \Bigg( \frac{\Xi_{1,1}^{L \{P\}\{ K\} } }{\square_{IP}} \square_{LJ} \Lambda_{1,1}^{J \{J_1\}\{  J_2\} } + \frac{\Xi_{1,1}^{L \{P\}\{ K\} } }{\square_{IP}} 2\left(V_{-++}\right)_{J_1 L J_2} + 2\frac{\Xi_{2,1}^{L \{P J_2\}\{ K  \} } }{\square_{IP}} \square_{LJ_1}\Bigg) {Z}_c^{\star J_1} {Z}_c^{\bullet J_2} \\
    -\Bigg( \frac{\Xi_{1,1}^{L \{P\}\{ K\} } }{\square_{IP}}\square_{LJ} \Lambda_{2,0}^{J \{J_1 J_2\} } +\frac{\Xi_{1,1}^{L \{P\}\{ K\} } }{\square_{IP}}\left(V_{--+}\right)_{J_1 J_2 L} + 2\frac{\Xi_{1,2}^{L \{P\}\{ K J_1 \} } }{\square_{IP}}\square_{LJ_2} \Bigg){Z}_c^{\star J_1} {Z}_c^{\star J_2}\Bigg\}\,.
\end{multline}
The above expression is of the type $\left(\mathbb{1} +y\right)$. One can therefore compute the inverse using the following series expansion
\begin{equation}
     \left(\frac{-1}{\square} \mathrm{M}_{\star\bullet}\right)^{-1} = \left(\mathbb{1} +y\right)^{-1}= \sum_{k=0}^{\infty} (-1)^{k} y^k\,.
    \label{eq:inv_exp}
\end{equation}
\begin{multline}
   \left. \left(\frac{-1}{\square} \right)_{IP} \left(\mathrm{M}_{\bullet\bullet}\right)_{PK} \right|_{\mathrm{2nd}} = -2\,\frac{\Xi_{2,0}^{L \{PK\}  }}{\square_{IP}}\square_{LJ}Z_c^{\star J} \\ 
   -2\,\square_{LJ_1}   \frac{\Lambda_{1,2}^{L \{J_2 \}\{ PK\} }}{\square_{IP}} Z_c^{\bullet J_1}{Z}_c^{\star J_2} + 2\,  \,\frac{\mathcal{U}^{ \{J_1 J_2\}\{ PK\} }_{-\,- + \, +}}{\square_{IP}} \,
   Z_c^{\star J_1} Z_c^{\star J_2} \\
    -\Bigg(2\,\frac{\Xi_{2,0}^{L \{PK\}  }}{\square_{IP}}\square_{LJ}\Lambda_{1,1}^{J \{J_1 \}\{ J_{2}\} }  +4\,\frac{\Xi_{2,0}^{L \{PK\}  }}{\square_{IP}}\left(V_{-++}\right)_{J_1LJ_2} +6\,\square_{LJ_1} \frac{\Xi_{3,0}^{L \{PK J_2 \}}}{\square_{IP}}\Bigg) Z_c^{\star J_1}{Z}_c^{\bullet J_2}
    \\ -\Bigg(2\,\frac{\Xi_{2,0}^{L \{PK\}  }}{\square_{IP}}\square_{LJ}\Lambda_{2,0}^{J \{J_1 J_{2}\} }   
     +2\,\frac{\Xi_{2,0}^{L \{PK\}  }}{\square_{IP}}\left(V_{--+}\right)_{J_1 J_2 L} +2\,\square_{LJ_1}\frac{\Xi_{2,1}^{L \{PK\} \{J_2 \}}}{\square_{IP}}\Bigg) Z_c^{\star J_1}{Z}_c^{\star J_2} 
\end{multline}

For the final matrix in the second term of Eq.~\eqref{eq:matr_in}, we have
\begin{equation}
   \left(\frac{-1}{\square} \right)_{IP} \left(\mathrm{M}_{\star\star}\right)_{PK}= 
   \left(\frac{-1}{\square} \right)_{IP}\left\{ 
   \frac{\delta^2 S[Z_c]}{\delta Z^{\star P}\delta Z^{\star K}}
   + 
   J_{\star L}\frac{\delta^2 A^{\star L}[Z_c]}{\delta Z^{\star P}\delta Z^{\star K}} + J_{\bullet L}\frac{\delta^2 A^{\bullet L}[Z_c]}{\delta Z^{\star P}\delta Z^{\star K}} 
   \right\} \,,
   \label{eq:MSS}
\end{equation}
where the differentiation of $S[Z_c]$ with respect to $Z^{\star P} $ and $  Z^{\star K}$ reads
\begin{equation}
    \frac{\delta^2 S[Z]}{\delta Z^{\star I}\delta Z^{\star K}}
    =  -\sum_{n=4}^{\infty} 
     \, \sum_{m=2}^{n-2}
    \,m(m-1)\,\, \mathcal{U}^{ \{J_1 \dots J_m\}\{ J_{m+1}\dots J_n\} }_{\underbrace{-\,\cdots\,-}_{m}\underbrace{+ \,\cdots\, +}_{n-m}} \,\delta_{J_{1}}^I \delta_{J_{2}}^K\,
   \prod_{i=3}^{m}Z_c^{\star J_i}
   \prod_{k=m+1}^{n}Z_c^{\bullet J_k} \, .
   \label{eq:S[Z]--}
\end{equation}
Up to the second order in fields, we have
\begin{equation}
    \left.\frac{-1}{\square}_{IP}\frac{\delta^2 S[Z]}{\delta Z^{\star I}\delta Z^{\star K}} \right|_{\mathrm{2nd}}
    =  2\, \frac{\mathcal{U}^{ \{P K\}\{ J_{1} J_2\} }_{-\,-+ \, +} }{\square_{IP}}
   Z_c^{\bullet J_1} Z_c^{\bullet J_2} \, .
   \label{eq:S[Z]--2}
\end{equation}
For the remaining two terms, using the following
\begin{equation}
\frac{\delta^2 A^{\star L}[Z_c]}{\delta Z^{\star I}\delta Z^{\star K}} = 
\sum_{n=2}^{\infty} 
     \, \sum_{i=2}^{n}\,i (i-1)\, \Lambda_{i,n-i}^{L \{J_1 \dots J_i\}\{ J_{i+1}\dots J_n\} } \delta_{J_1}^I \delta_{J_2}^K\,\prod_{k=3}^{i}{Z}_c^{\star J_k} \prod_{l=i+1}^{n}{Z}_c^{\bullet J_l}\,,
    \label{eq:A_star_ss}
\end{equation}
along with the expression for the source $J_{\star L}$ following Eq.\eqref{eq:J_currZ} and Eq.~\eqref{eq:S+-Z}, up to second order in fields we get
\begin{multline}
    \left. \frac{\delta S_{\mathrm{YM}}[A[Z_c]]}{\delta A^{\star L}}\frac{\delta^2 A^{\star L}[Z_c]}{\delta Z^{\star P}\delta Z^{\star K}}\left(\frac{1}{\square} \right)_{IP}\,\,\right|_{\mathrm{2nd}} = \Bigg\{ - 2\, \frac{\Lambda_{2,0}^{L \{PK\} }}{\square_{IP}}\square_{LJ}Z_c^{\bullet J} \\
    - \Bigg(2\, \frac{\Lambda_{2,0}^{L \{PK\} }}{\square_{IP}}\left(V_{-++}\right)_{LJ_1J_2} + 2\, \frac{\Lambda_{2,0}^{L \{PK\} }}{\square_{IP}}\square_{LJ}\Xi_{2,0}^{J \{J_1 J_{2}\} } + 2\, \square_{LJ_1}\frac{\Lambda_{2,1}^{L \{PK\}\{J_2\} }}{\square_{IP}} \Bigg)Z_c^{\bullet J_1}{Z}_c^{\bullet J_2}\\
- \Bigg( 2\, \frac{\Lambda_{2,0}^{L \{PK\} }}{\square_{IP}}\square_{LJ}\Xi_{1,1}^{J \{J_1 \}\{ J_{2}\} } +4\, \frac{\Lambda_{2,0}^{L \{PK\} }}{\square_{IP}}\left(V_{--+}\right)_{LJ_2 J_1} +6\, \square_{LJ_1}\frac{\Lambda_{3,0}^{L \{PK J_2\} }}{\square_{IP}} \Bigg)Z_c^{\bullet J_1}{Z}_c^{\star J_2}\Bigg\}\,,
\label{eq:JssT}
\end{multline}
and using
\begin{equation}
\frac{\delta^2 A^{\bullet L}[Z_c]}{\delta Z^{\star I}\delta Z^{\star K}} = 
\sum_{n=3}^{\infty} 
     \, \sum_{i=1}^{n-2}\, (n-i) (n-i-1) \Xi_{i,n-i}^{L \{J_1 \dots J_i\}\{ J_{i+1}\dots J_n\} } \delta_{J_{i+1}}^I \delta_{J_{i+2}}^K \prod_{k=1}^{i}{Z}_c^{\bullet J_k} \prod_{l=i+3}^{n}{Z}_c^{\star J_l}\,,
    \label{eq:A_bul_ss}
\end{equation}
along with the expression for the source $J_{\bullet L}$ following Eq.\eqref{eq:J_currZ} and Eq.~\eqref{eq:S+-Z2}, up to second order in fields we get
\begin{equation}
    \left. \frac{\delta S_{\mathrm{YM}}[A[Z_c]]}{\delta A^{\bullet L}}\frac{\delta^2 A^{\bullet L}[Z_c]}{\delta Z^{\star P}\delta Z^{\star K}}\left(\frac{1}{\square} \right)_{IP}\,\,\right|_{\mathrm{2nd}} = \Bigg\{-2\,\square_{LJ_1}  \frac{\Xi_{1,2}^{L \{J_2 \}\{ PK\} }}{\square_{IP}} Z_c^{\star J_1} {Z}_c^{\bullet J_2}  \Bigg\}\,.
    \label{eq:JssT2}
\end{equation}

Substituting Eqs.~\eqref{eq:S[Z]--2}, \eqref{eq:JssT} and \eqref{eq:JssT2} in Eq.~\eqref{eq:MSS} we get
\begin{multline}
     \left. \left(\frac{-1}{\square} \right)_{IP} \left(\mathrm{M}_{\star\star}\right)_{PK} \right|_{\mathrm{2nd}} = - 2\, \frac{\Lambda_{2,0}^{L \{PK\} }}{\square_{IP}}\square_{LJ}Z_c^{\bullet J} \\
     -2\,\square_{LJ_1}  \frac{\Xi_{1,2}^{L \{J_2 \}\{ PK\} }}{\square_{IP}} Z_c^{\star J_1} {Z}_c^{\bullet J_2} + 2\, \frac{\mathcal{U}^{ \{P K\}\{ J_{1} J_2\} }_{-\,-+ \, +} }{\square_{IP}}
   Z_c^{\bullet J_1} Z_c^{\bullet J_2}\\
    - \Bigg(2\, \frac{\Lambda_{2,0}^{L \{PK\} }}{\square_{IP}}\left(V_{-++}\right)_{LJ_1J_2} + 2\, \frac{\Lambda_{2,0}^{L \{PK\} }}{\square_{IP}}\square_{LJ}\Xi_{2,0}^{J \{J_1 J_{2}\} } + 2\, \square_{LJ_1}\frac{\Lambda_{2,1}^{L \{PK\}\{J_2\} }}{\square_{IP}} \Bigg)Z_c^{\bullet J_1}{Z}_c^{\bullet J_2}\\
- \Bigg( 2\, \frac{\Lambda_{2,0}^{L \{PK\} }}{\square_{IP}}\square_{LJ}\Xi_{1,1}^{J \{J_1 \}\{ J_{2}\} } +4\, \frac{\Lambda_{2,0}^{L \{PK\} }}{\square_{IP}}\left(V_{--+}\right)_{LJ_2 J_1} +6\, \square_{LJ_1}\frac{\Lambda_{3,0}^{L \{PK J_2\} }}{\square_{IP}} \Bigg)Z_c^{\bullet J_1}{Z}_c^{\star J_2}
\end{multline}

Putting everything together, we have
\begin{multline}
     \left. \Tr\ln\left(\mathrm{M}_{\bullet\star}
    - \mathrm{M}_{\bullet\bullet}\mathrm{M}^{-1}_{\star\bullet}\mathrm{M}_{\star\star}\right) \right|_{\mathrm{2nd}} = \Tr \Bigg[ \Bigg\{ -\frac{\Lambda_{1,1}^{L \{K\}\{ P\}} }{\square_{IP}} \square_{LJ}{Z}_c^{\bullet J}  -\square_{LJ}\frac{\Xi_{1,1}^{L \{P\}\{ K\} } }{\square_{IP}}Z_c^{\star J}  \\
    - \Bigg( 2 \frac{\Lambda_{1,2}^{L \{K\}\{ P J_1\}} }{\square_{IP}}\square_{LJ_2} + \frac{\Lambda_{1,1}^{L \{K\}\{ P\}} }{\square_{IP}}\square_{LJ} \Xi_{2,0}^{J \{J_1 J_2\} } + \frac{\Lambda_{1,1}^{L \{K\}\{ P\}} }{\square_{IP}}\left(V_{-++}\right)_{LJ_1 J_2}\Bigg){Z}_c^{\bullet J_1}{Z}_c^{\bullet J_2}\\
    - \Bigg(\frac{\Lambda_{1,1}^{L \{K\}\{ P\}} }{\square_{IP}} \square_{LJ} \Xi_{1,1}^{J \{J_1 \}\{J_2 \} } + 2 \frac{\Lambda_{2,1}^{L \{K J_2\}\{ P \}} }{\square_{IP}}\square_{LJ_1} + \frac{\Lambda_{1,1}^{L \{K\}\{ P\}} }{\square_{IP}} 2\left(V_{--+}\right)_{LJ_2 J_1 }\Bigg) {Z}_c^{\bullet J_1}{Z}_c^{\star J_2} \\
    - \Bigg( \frac{\Xi_{1,1}^{L \{P\}\{ K\} } }{\square_{IP}} \square_{LJ} \Lambda_{1,1}^{J \{J_1\}\{  J_2\} } + \frac{\Xi_{1,1}^{L \{P\}\{ K\} } }{\square_{IP}} 2\left(V_{-++}\right)_{J_1 L J_2} + 2\frac{\Xi_{2,1}^{L \{P J_2\}\{ K  \} } }{\square_{IP}} \square_{LJ_1}\Bigg) {Z}_c^{\star J_1} {Z}_c^{\bullet J_2} \\
    -\Bigg( \frac{\Xi_{1,1}^{L \{P\}\{ K\} } }{\square_{IP}}\square_{LJ} \Lambda_{2,0}^{J \{J_1 J_2\} } +\frac{\Xi_{1,1}^{L \{P\}\{ K\} } }{\square_{IP}}\left(V_{--+}\right)_{J_1 J_2 L} + 2\frac{\Xi_{1,2}^{L \{P\}\{ K J_1 \} } }{\square_{IP}}\square_{LJ_2} \Bigg){Z}_c^{\star J_1} {Z}_c^{\star J_2} \\
    +  4  \,\frac{\mathcal{U}^{ \{K  J_1\}\{P J_2\} }_{-\,- + \, +}}{\square_{IP}} \,
   Z_c^{\star J_1}
   Z_c^{\bullet J_2} -4\,\frac{\Xi_{2,0}^{L_1 \{PK_1\}  }}{\square_{IP}}\square_{L_1 J_1} \, \frac{\Lambda_{2,0}^{L_2 \{P_1K\} }}{\square_{K_1P_1}}\square_{L_2 J_2}Z_c^{\star J_1}Z_c^{\bullet J_2} \Bigg\}\\
    -\frac{1}{2} \Bigg\{ \frac{\Lambda_{1,1}^{L_1 \{K_1\}\{ P\}} }{\square_{IP}} \square_{L_1 J_1} \frac{\Lambda_{1,1}^{L_2 \{K\}\{ P_1\}} }{\square_{K_1 P_1}} \square_{L_2 J_2}{Z}_c^{\bullet J_1}{Z}_c^{\bullet J_2} + \frac{\Lambda_{1,1}^{L_1 \{K_1\}\{ P\}} }{\square_{IP}} \square_{L_1 J_1 } \square_{L_2 J_2}\frac{\Xi_{1,1}^{L_2 \{P_1\}\{ K\} } }{\square_{K_1 P_1}}{Z}_c^{\bullet J_1} Z_c^{\star J_2} \\
    +\square_{L_1 J_1}\frac{\Xi_{1,1}^{L_1 \{P\}\{ K_1\} } }{\square_{IP}}\frac{\Lambda_{1,1}^{L_2 \{K\}\{ P_1\}} }{\square_{K_1P_1}} \square_{L_2 J_2 }Z_c^{\star J_1}{Z}_c^{\bullet J_2} + \square_{L_1 J_1}\frac{\Xi_{1,1}^{L_1 \{P\}\{ K_1\} } }{\square_{IP}}\square_{L_2 J_2}\frac{\Xi_{1,1}^{L_2 \{P_1\}\{ K\} } }{\square_{K_1P_1}}Z_c^{\star J_1}Z_c^{\star J_2} \Bigg\}\Bigg] \, .
\end{multline}
\section{Equivalence of one-loop effective actions}
\label{sec:AppB}

In \cite{Kakkad_2022}, we developed loop corrections to the MHV action \cite{Mansfield2006} via the one-loop effective action approach. This approach was slightly different than the one we employed in the current work (\emph{cf.} Section \ref{sec:GAOL}).

In this appendix, employing the approach of \cite{Kakkad_2022}, we first develop loop corrections to the Z-field action. Then, we prove that the one-loop effective action obtained in the previous step is equivalent to Eq.~\eqref{eq:gam_zn} modulo a volume divergent field-independent factor which does not contribute to amplitudes and can be absorbed into the overall normalization. This will demonstrate the fact that the two ways of developing one-loop effective action, indeed, result in the same quantity. 
\subsection{Rederiving the one-loop effective action}

In \cite{Kakkad_2022}, we employed a simplified approach to developing loop corrections to the MHV action. We started with the Yang-Mills partition function, expanded it around the classical solution up to the second order in fields, and integrated out the field fluctuations to obtain the one-loop effective Yang-Mills partition function. This procedure is symbolically shown below (for further details see \cite{Kakkad_2022, kakkad2023scattering})
\begin{align}
    \mathcal{Z}_{\mathrm{YM}}[J]&=\int[dA]\, e^{i\left(S_{\mathrm{YM}}[A] + \int\!d^4x\, \Tr \hat{J}_j(x) \hat{A}^j(x)\right) } \,,\nonumber\\
&\xdownarrow{0.6cm} \nonumber\\  
    \mathcal{Z}_{\mathrm{YM}}[J]&\approx 
    \exp\Bigg\{ iS_{\mathrm{YM}}[A_c] 
    + i\int\!d^4x\, \Tr \hat{J}_i(x) \hat{A}_c^i(x) \nonumber\\
    - \frac{1}{2} \Tr\ln \Bigg[ \frac{\delta^2 S_{\mathrm{YM}}[A_c]}
    {\delta A^{\star I}\delta A^{\bullet K}} &\, \frac{\delta^2 S_{\mathrm{YM}}[A_c]}
    {\delta A^{\star K}\delta A^{\bullet J}}  - \frac{\delta^2 S_{\mathrm{YM}}[A_c]}
    {\delta A^{\star I}\delta A^{\bullet K}} \, \frac{\delta^2 S_{\mathrm{YM}}[A_c]}
    {\delta A^{\star K}\delta A^{\star L}} \left( \frac{\delta^2 S_{\mathrm{YM}}[A_c]}
    {\delta A^{\bullet L}\delta A^{\star M}} \right)^{-1} \frac{\delta^2 S_{\mathrm{YM}}[A_c]}
    {\delta A^{\bullet M}\delta A^{\bullet J}}\Bigg]
    \Bigg\} \,.
    \label{eq:gen_YMold}
\end{align}
After that, we derived the one-loop effective Yang-Mills action $\Gamma_{\mathrm{YM}}[A_c]$ via the Legendre transform of the generating functional for the connected Green's function
\begin{multline}
   \Gamma_{\mathrm{YM}}[A_c] = S_{\mathrm{YM}}[A_c] \\
    + \frac{i}{2} \Tr\ln \left[ \frac{\delta^2 S_{\mathrm{YM}}[A_c]}
    {\delta A^{\star I}\delta A^{\bullet K}} \, \frac{\delta^2 S_{\mathrm{YM}}[A_c]}
    {\delta A^{\star K}\delta A^{\bullet J}} - \frac{\delta^2 S_{\mathrm{YM}}[A_c]}
    {\delta A^{\star I}\delta A^{\bullet K}} \, \frac{\delta^2 S_{\mathrm{YM}}[A_c]}
    {\delta A^{\star K}\delta A^{\star L}} \left( \frac{\delta^2 S_{\mathrm{YM}}[A_c]}
    {\delta A^{\bullet L}\delta A^{\star M}} \right)^{-1} \frac{\delta^2 S_{\mathrm{YM}}[A_c]}
    {\delta A^{\bullet M}\delta A^{\bullet J}}\right]\,,
    \label{eq:OLEA_YM1}
\end{multline}
where
\begin{equation}
   \frac{\delta^2 S_{\mathrm{YM}}[A_c]}
    {\delta A^{\star I}\delta A^{\bullet J}} 
    = -\square_{IJ}-2\left(V_{-++}\right)_{IJK}A_{c}^{\bullet K} -2\left(V_{--+}\right)_{KIJ}A_{c}^{\star K} -4\left(V_{--++}\right)_{LIJK}A_{c}^{\star L} A_{c}^{\bullet K}
    \label{eq:S+-ym}
    \,,
\end{equation}
\begin{equation}
   \frac{\delta^2 S_{\mathrm{YM}}[A_c]}
    {\delta A^{\bullet I}\delta A^{\bullet J}} 
    = -2\left(V_{-++}\right)_{KIJ}A_{c}^{\star K} -2\left(V_{--++}\right)_{KLIJ}A_{c}^{\star K} A_{c}^{\star L}
    \label{eq:S++ym}
    \,,
\end{equation}
\begin{equation}
   \frac{\delta^2 S_{\mathrm{YM}}[A_c]}
    {\delta A^{\star I}\delta A^{\star J}} 
    = -2\left(V_{--+}\right)_{IJK}A_{c}^{\bullet K} -2\left(V_{--++}\right)_{IJKL}A_{c}^{\bullet K} A_{c}^{\bullet L}
    \label{eq:S--ym}
    \,.
\end{equation}
Above, $V_{-++}$, $V_{--+}$ and $V_{--++}$ represent the two triple gluon vertices and the four-point vertex in the light-cone Yang-Mills action.

Finally, we applied Mansfield's transformation to Eq.~\eqref{eq:OLEA_YM1} to obtain the MHV action plus one-loop corrections. This idea can be straightforwardly extended to develop loop corrections to the Z-field field action. The only change would be that instead of Mansfield's transformation, we apply the transformation that derives the Z-field action from the Yang-Mills action to Eq.~\eqref{eq:OLEA_YM1}. By doing this we get \cite{kakkad2023scattering}
\begin{multline}
   \Gamma[Z_c] = S[Z_c] 
    + \frac{i}{2} \Tr\ln \left[ \frac{\delta^2 S_{\mathrm{YM}}[A[Z_c]]}
    {\delta A^{\star I}\delta A^{\bullet K}} \, \frac{\delta^2 S_{\mathrm{YM}}[A[Z_c]]}
    {\delta A^{\star K}\delta A^{\bullet J}} \right.\\
    \left.- \frac{\delta^2 S_{\mathrm{YM}}[A[Z_c]]}
    {\delta A^{\star I}\delta A^{\bullet K}} \, \frac{\delta^2 S_{\mathrm{YM}}[A[Z_c]]}
    {\delta A^{\star K}\delta A^{\star L}} \left( \frac{\delta^2 S_{\mathrm{YM}}[A[Z_c]]}
    {\delta A^{\bullet L}\delta A^{\star M}} \right)^{-1} \frac{\delta^2 S_{\mathrm{YM}}[A[Z_c]]}
    {\delta A^{\bullet M}\delta A^{\bullet J}}\right]\,.
    \label{eq:OLEA_Zac}
\end{multline}

Substituting Eqs.~\eqref{eq:S+-ym}-\eqref{eq:S--ym} and factoring out the inverse propagator, the log term in Eq.~\eqref{eq:OLEA_Zac} reads
\begin{multline}
     \Tr\ln \Bigg[  \Bigg\{ \Big( \delta_{IK} + \left(2\square^{-1}V_{-++}\right)_{IKP}A^{\bullet P}[Z_{c}] +\left(2\square^{-1}V_{--+}\right)_{PIK}A^{\star P}[Z_{c}] +\left(4\square^{-1}V_{--++}\right)_{PIKQ}A^{\star P}[Z_{c}] A^{\bullet Q}[Z_{c}]\Big) \\
    \times \Big( \delta_{KJ} + \left(2\square^{-1}V_{-++}\right)_{KJR}A^{\bullet R}[Z_{c}] +\left(2\square^{-1}V_{--+}\right)_{RKJ}A^{\star R}[Z_{c}] +\left(4\square^{-1}V_{--++}\right)_{RKJS}A^{\star R}[Z_{c}] A^{\bullet S}[Z_{c}]\Big) \Bigg\}\\
     - \Bigg\{ \Big(\delta_{IK} + \left(2\square^{-1}V_{-++}\right)_{IKP}A^{\bullet P}[Z_{c}] +\left(2\square^{-1}V_{--+}\right)_{PIK}A^{\star P}[Z_{c}] +\left(4\square^{-1}V_{--++}\right)_{PIKQ}A^{\star P}[Z_{c}] A^{\bullet Q}[Z_{c}]\Big) \\
     \times \Big(\left(2\square^{-1}V_{--+}\right)_{KLR}A^{\bullet R}[Z_{c}] +\left(2\square^{-1}V_{--++}\right)_{KLRS}A^{\bullet R}[Z_{c}] A^{\bullet S}[Z_{c}]\Big)\\
     \times \Big( \delta_{ML} + \left(2\square^{-1}V_{-++}\right)_{MLT}A^{\bullet T}[Z_{c}] +\left(2\square^{-1}V_{--+}\right)_{TML}A^{\star T}[Z_{c}] +\left(4\square^{-1}V_{--++}\right)_{TMLU}A^{\star T}[Z_{c}] A^{\bullet U}[Z_{c}]\Big)^{-1}\\ 
     \times \Big(\left(2\square^{-1}V_{-++}\right)_{VMJ}A^{\star V}[Z_{c}] +\left(2\square^{-1}V_{--++}\right)_{WVMJ}A^{\star W}[Z_{c}] A^{\star V}[Z_{c}]\Big)\Bigg\}\Bigg]
    \label{eq:Partition_log}
    \,.
\end{multline}
The above expression highlights the major drawback of this simplified approach. Notice, the vertices participating in the loop formation in the log term above are the Yang-Mills vertices. That is, the Z-field vertices are not explicit in the loop. To avoid this and to make the Z-field vertices explicit in the loop we derived Eq.~\eqref{eq:gam_zn} following the approach discussed in Section \ref{sec:GAOL}.

Except for the above stated drawback, there are no other issues with the one-loop Z-field action Eq.~\eqref{eq:OLEA_Zac}. It is fully operational and has no missing loop contributions. We demonstrate this via a simple example in the following section.
\subsection{One-loop \texorpdfstring{$(- - - -)$}{4amp}  amplitudes}

\begin{figure}
    \centering
    \includegraphics[width=15cm]{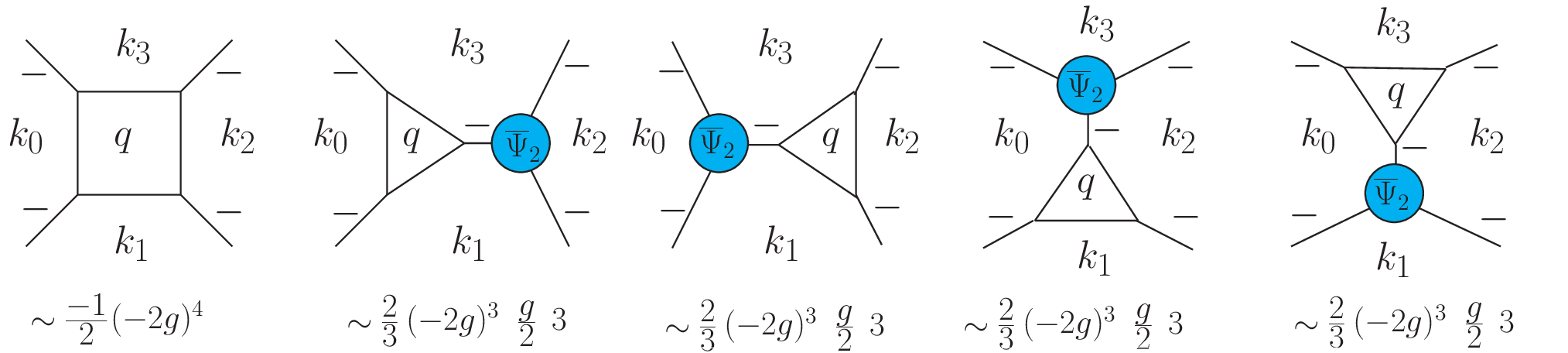}
    \caption{\small
   The contributions to the one-loop $(----)$ amplitude originating from the log term in Eq.~\eqref{eq:OLEA_YM1}. $\overline{\Psi}_2$ is the kernel Eq.~\eqref{eq:psiBar_kernel}. $q$ is the loop region momenta whereas the $k_i$ represents the region momenta outside the loop. These are related to the line momenta as $p=k_{\mathrm{right}} - k_{\mathrm{left}}$. Finally, the factor under each contribution is the symmetry factor associated with it. The origin of each of these is explained in the text.
   }
    \label{fig:4Mall}
\end{figure}

In this section, we compute the $(- - - -)$ one-loop color ordered leading trace amplitude using the Chakrabarti, Qiu, and Thorn (CQT)  regularization scheme \cite{CQT1,CQT2}. The contributions to this amplitude originate solely from the log term in the one-loop Z-field action Eq.~\eqref{eq:OLEA_Zac}. These are shown in Figure~\ref{fig:4Mall}. We name the external legs as $1,2,3,4$ starting from bottom left in anticlockwise fashion. Notice that there are no contributions involving the $(--)$ bubble. This is because the $(--)$ bubble is non-zero in the CQT scheme and is explicitly canceled via a counterterm. Also, as visible from Figure~\ref{fig:4Mall}, CQT employs the usage of region momenta: $q$ for the region enclosed in the loop, and $k_i$ for the exterior regions. In terms of these, the line momenta is given as $p=k_{\mathrm{right}} - k_{\mathrm{left}}$, where "right" and "left" are defined with respect to the orientation of the line momenta.

Finally, notice each diagram in Figure~\ref{fig:4Mall} has an associated overall symmetry factor. The first factor ($-1/2$ for the box topology and $2/3$ for the triangle topology) originates from the expansion of log term. The factor of $(-2g)^n$ originates from the $n$ $V_{--+}$ vertices in the 1PI one-loop sub-diagram of the contribution. Lastly, the factor of $g/2$ for the triangle contributions is associated with the kernel, and since the triangle $\Delta^{- - - }$ is symmetric in the three external legs, we get an additional factor of $3$.

In order to compute the diagrams in Figure~\ref{fig:4Mall}, the essential ingredient is the 1PI triangle $\Delta^{- - - }$. The latter is computed in the CQT scheme by first expressing the loop integral in terms of the region momenta and then introducing an exponential cut-off $e^{-\delta \mathbf{q}^2}$, where $\mathbf{q}^2 = 2q^{\bullet}q^{\star}$, and $\delta > 0$. It is put to zero at the end. Assuming two external legs of the $\Delta^{- - - }$ on-shell, the results reads (for details see \cite{CQT1,CQT2}. We computed $\Delta^{+ + +}$ in a similar fashion in \cite{Kakkad_2022})
\begin{equation}
    \left.\Delta^{- - -}\right|_{p_{i}^{2}=0, p_{j}^{2}=0}  =-\frac{g^{3}}{6 \pi^{2}} \frac{\left({\widetilde v}^{\star}_{ij}p_j^+\right)^{3}}{p_{i}^{+} p_{j}^{+} p_{k}^{+} p_k^{2}}\,.
\end{equation}
Above, $p_i$ and $p_j$ are on-shell and $p_k$ is off-shell. Substituting the kernel ${\widetilde \Lambda}_{2,0}^{a \left \{b_l b_m \right \} }(\mathbf{p_k}; \left \{\mathbf{p}_{l}, \mathbf{p}_{m} \right \})$ to the off-shell leg, we can obtain the expression for the triangle contributions in Figure~\ref{fig:4Mall}. We denote these as $\Delta^{- - - -}_{ij}$, where $i$ and $j$ are the external on-shell legs attached to the triangle $\Delta^{- - - }_{i j k}$. By doing the substitution, the three triangle contributions in the middle in Figure~\ref{fig:4Mall} read (in the same order)
\begin{equation}
    \Delta_{41}^{- - - -}= \frac{-g^{4}}{12 \pi^{2}} \frac{\left({\widetilde v}^{\star}_{41}p_1^+\right)^{3}p_{2}^{+}}{p_{1}^{+} p_{2}^{+} p_{3}^{+} p_{4}^{+} p_{14}^{2}{\widetilde v}_{23}}\,,
\end{equation}
\begin{equation}
   \Delta_{23}^{- - - -}= \frac{-g^{4}}{12 \pi^{2}} \frac{ \left({\widetilde v}^{\star}_{23}p_3^+\right)^{3}p_{4}^{+}}{p_{1}^{+} p_{2}^{+} p_{3}^{+} p_{4}^{+} p_{14}^{2}{\widetilde v}_{41}}\,.
\end{equation} 
\begin{equation}
   \Delta_{12}^{- - - -}= \frac{-g^{4}}{12 \pi^{2}} \frac{\left({\widetilde v}^{\star}_{12}p_2^+\right)^{3}p_{3}^{+}}{p_{1}^{+} p_{2}^{+} p_{3}^{+}p_{4}^{+} p_{34}^{2}{\widetilde v}_{34}}\,,
\end{equation}
As for the last triangle contribution (i.e. $\Delta_{34}^{+ + + +}$), it gets canceled by a similar contribution originating from the box reduction \cite{CQT1}. Essentially, the box contribution $\square^{- - - -}$ can be reduced into $-\, \Delta_{34}^{+ + + +}$ plus additional terms. Therefore summing these two diagrams, the $\Delta_{34}^{+ + + +}$ explicitly cancels out and we get
\begin{multline}
  \square^{- - - -} \,+\, \Delta_{34}^{- - - -} = \frac{-g^{4}}{12 \pi^{2}} \frac{p_{1}^{+}}{p_{1}^{+} p_{2}^{+} p_{3}^{+} p_{4}^{+} {\widetilde v}_{12} p_{14}^{2}}\\
  \left[{\widetilde v}^{\star}_{41}p_{1}^{+} {\widetilde v}^{\star}_{23}p_{3}^{+}\left({\widetilde v}^{\star}_{41}p_{1}^{+}+{\widetilde v}^{\star}_{23}p_{3}^{+}\right)+{\widetilde v}^{\star}_{34}p_{4}^{+}\left({\widetilde v}^{\star 2}_{41}p_{1}^{+ 2}+{\widetilde v}^{\star 2}_{23}p_{3}^{+ 2}\right)\right]\,.
    \label{eq:box+tri4minus}
\end{multline}

The $(- - - -)$ one-loop amplitude is the sum of all the above contributions 
\begin{equation}
      \mathcal{A}_{\mathrm{one-loop}}^{- - - -} =   \square^{- - - -} \,+\, \Delta_{34}^{- - - -} \,+\, \Delta_{12}^{- - - -} \,+\, \Delta_{41}^{- - - -} \,+\, \Delta_{23}^{- - - -} \, ,
\end{equation}
which, after a bit of algebra, simplifies to
\begin{equation}
      \mathcal{A}_{\mathrm{one-loop}}^{- - - -} = \frac{g^{4}}{24 \pi^{2}} \frac{{\widetilde v}^{\star}_{21} {\widetilde v}^{\star}_{43}}{{\widetilde v}_{21} {\widetilde v}_{43}}     \, .
       \label{eq:4_minus_one_loop}
\end{equation}
The above result agrees with the known result.

In fact, in \cite{kakkad2023scattering}, we further computed  $(- - - +)$, $(- - + +)$, $(- + + +)$ and $(+ + + +)$ one-loop color ordered leading trace amplitude using the one-loop Z-field action Eq.~\eqref{eq:OLEA_Zac} and found agreement with the known results. This provided a concrete validation of the completeness of the one-loop Z-field action Eq.~\eqref{eq:OLEA_Zac}.
\subsection{Equivalence of one-loop actions}

Although there are no missing loop contributions in the one-loop Z-field action Eq.~\eqref{eq:OLEA_Zac}, it does not have the Z-field vertices explicit in the loop. We therefore derived Eq.~\eqref{eq:gam_zn}
\begin{equation}
    \Gamma[Z_c] = S[Z_c]  +i\, \frac{1}{2} \Tr\ln \left( 
     \mathrm{M}^{\mathrm{Z}}_{IK}
     \right)\,,
     \label{eq:G_zac}
\end{equation}
 in Section \ref{sec:GAOL} such that the Z-field interaction vertices are explicit in the loop. 

 In this section, we prove that the two one-loop Z-field actions given in  Eq.~\eqref{eq:OLEA_Zac} and  Eq.~\eqref{eq:G_zac} are indeed equivalent.

 Recall, the matrix $\mathrm{M}^{\mathrm{Z}}_{IK}$ in the log term in Eq.~\eqref{eq:G_zac} has the following explicit form
\begin{multline}
\mathrm{M}^{\mathrm{Z}}_{IK}  =  \left(\begin{matrix}
     \frac{\delta^2 S[Z_c]}{\delta Z^{\bullet I}\delta Z^{\star K}} &\frac{\delta^2 S[Z_c]}{\delta Z^{\bullet I}\delta Z^{\bullet K}}\\ \\
\frac{\delta^2 S[Z_c]}{\delta Z^{\star I}\delta Z^{\star K}}
      &\frac{\delta^2 S[Z_c]}{\delta Z^{\star I}\delta Z^{\bullet K}} 
\end{matrix}\right) \\
- \left(\begin{matrix}
    \frac{\delta S_{\mathrm{YM}}[A[Z_c]]}{\delta A^{\star L}}\frac{\delta^2 A^{\star L}[Z_c]}{\delta Z^{\bullet I}\delta Z^{\star K}} + \frac{\delta S_{\mathrm{YM}}[A[Z_c]]}{\delta A^{\bullet L}}\frac{\delta^2 A^{\bullet L}[Z_c]}{\delta Z^{\bullet I}\delta Z^{\star K}} 
     & \frac{\delta S_{\mathrm{YM}}[A[Z_c]]}{\delta A^{\star L}}\frac{\delta^2 A^{\star L}[Z_c]}{\delta Z^{\bullet I}\delta Z^{\bullet K}} +  \frac{\delta S_{\mathrm{YM}}[A[Z_c]]}{\delta A^{\bullet L}}\frac{\delta^2 A^{\bullet L}[Z_c]}{\delta Z^{\bullet I}\delta Z^{\bullet K}}\\ \\
 \frac{\delta S_{\mathrm{YM}}[A[Z_c]]}{\delta A^{\star L}}\frac{\delta^2 A^{\star L}[B_c]}{\delta Z^{\star I}\delta Z^{\star K}} + \frac{\delta S_{\mathrm{YM}}[A[Z_c]]}{\delta A^{\bullet L}}\frac{\delta^2 A^{\bullet L}[B_c]}{\delta Z^{\star I}\delta Z^{\star K}}
      &\frac{\delta S_{\mathrm{YM}}[A[Z_c]]}{\delta A^{\star L}}\frac{\delta^2 A^{\star L}[B_c]}{\delta Z^{\star I}\delta Z^{\bullet K}} + \frac{\delta S_{\mathrm{YM}}[A[Z_c]]}{\delta A^{\bullet L}}\frac{\delta^2 A^{\bullet L}[B_c]}{\delta Z^{\star I}\delta Z^{\bullet K}}
\end{matrix}\right)
\label{eq:mz1}
\end{multline}
We know that the Z-field action can be obtained from the light-cone Yang-Mills action via the substitution of the solution $(A^{\bullet}[Z_c], A^{\star}[Z_c])$ of field transformation to the latter
\begin{equation}
    S[Z_c] = S_{\mathrm{YM}}[A^{\bullet}[Z_c], A^{\star}[Z_c]]\,.
\end{equation}
Differentiating the L.H.S with respect to $Z^{\bullet I} $, $ Z^{\star K}$ and using the chain rule on the R.H.S, we can write
\begin{multline}
    \frac{\delta^2 S[Z_c]}{\delta Z^{\bullet I}\delta Z^{\star K}} = \frac{\delta S_{\mathrm{YM}}[A[Z_c]]}{\delta A^{\bullet R}}\frac{\delta^2 A^{\bullet R}[Z_c]}{\delta Z^{\bullet I}\delta Z^{\star K}}+\frac{\delta S_{\mathrm{YM}}[A[Z_c]]}{\delta A^{\star R}}\frac{\delta^2 A^{\star R}[Z_c]}{\delta Z^{\bullet I}\delta Z^{\star K}}\\
    +\frac{\delta A^{\star S}[Z_c]}{\delta Z^{\bullet I}}\frac{\delta^2 S_{\mathrm{YM}}[B[Z_c]]}{\delta A^{\star S}\delta A^{\bullet R}}\frac{\delta A^{\bullet R}[Z_c]}{\delta Z^{\star K}}  
    +\frac{\delta A^{\bullet S}[Z_c]}{\delta Z^{\bullet I}}\frac{\delta^2 S_{\mathrm{YM}}[A[Z_c]]}{\delta A^{\bullet S}\delta A^{\bullet R}}\frac{\delta A^{\bullet R}[Z_c]}{\delta Z^{\star K}}\\    
    +\frac{\delta A^{\star S}[Z_c]}{\delta Z^{\bullet I}}\frac{\delta^2 S_{\mathrm{YM}}[A[Z_c]]}{\delta A^{\star S}\delta A^{\star R}}\frac{\delta A^{\star R}[Z_c]}{\delta Z^{\star K}}    
    +\frac{\delta A^{\bullet S}[Z_c]}{\delta Z^{\bullet I}}\frac{\delta^2 S_{\mathrm{YM}}[A[Z_c]]}{\delta A^{\bullet S}\delta A^{\star R}}\frac{\delta A^{\star R}[Z_c]}{\delta Z^{\star K}}\,.
    \label{eq:SZ_BS}
\end{multline}
Using this, the component $\left(\mathrm{M}^{\mathrm{Z}}_{IK}\right)_{11}$ of the matrix $\mathrm{M}^{\mathrm{Z}}_{IK}$ can be simplified as follows
\begin{align}
    \left(\mathrm{M}^{\mathrm{Z}}_{IK}\right)_{11}  =&\, \frac{\delta^2 S[Z_c]}{\delta Z^{\bullet I}\delta Z^{\star K}} - \frac{\delta S_{\mathrm{YM}}[A[Z_c]]}{\delta A^{\star L}}\frac{\delta^2 A^{\star L}[Z_c]}{\delta Z^{\bullet I}\delta Z^{\star K}} - \frac{\delta S_{\mathrm{YM}}[A[Z_c]]}{\delta A^{\bullet L}}\frac{\delta^2 A^{\bullet L}[Z_c]}{\delta Z^{\bullet I}\delta Z^{\star K}}\,, \nonumber\\
    =&\,\frac{\delta A^{\star S}[Z_c]}{\delta Z^{\bullet I}}\frac{\delta^2 S_{\mathrm{YM}}[B[Z_c]]}{\delta A^{\star S}\delta A^{\bullet R}}\frac{\delta A^{\bullet R}[Z_c]}{\delta Z^{\star K}}  
    +\frac{\delta A^{\bullet S}[Z_c]}{\delta Z^{\bullet I}}\frac{\delta^2 S_{\mathrm{YM}}[A[Z_c]]}{\delta A^{\bullet S}\delta A^{\bullet R}}\frac{\delta A^{\bullet R}[Z_c]}{\delta Z^{\star K}}\\    \nonumber
   &\, +\frac{\delta A^{\star S}[Z_c]}{\delta Z^{\bullet I}}\frac{\delta^2 S_{\mathrm{YM}}[A[Z_c]]}{\delta A^{\star S}\delta A^{\star R}}\frac{\delta A^{\star R}[Z_c]}{\delta Z^{\star K}}    
    +\frac{\delta A^{\bullet S}[Z_c]}{\delta Z^{\bullet I}}\frac{\delta^2 S_{\mathrm{YM}}[A[Z_c]]}{\delta A^{\bullet S}\delta A^{\star R}}\frac{\delta A^{\star R}[Z_c]}{\delta Z^{\star K}}\,.
\end{align}
The first expression above represents the $\left(\mathrm{M}^{\mathrm{Z}}_{IK}\right)_{11}$ component from Eq.~\eqref{eq:mz1}. The second expression is obtained by substituting Eq.~\eqref{eq:SZ_BS} to the first expression. Notice, the two terms originating from the source matrix (the second matrix in Eq.~\eqref{eq:mz1}) explicitly cancel out.

Repeating the above steps for the remaining components i.e. $\left(\mathrm{M}^{\mathrm{Z}}_{IK}\right)_{12}$, $\left(\mathrm{M}^{\mathrm{Z}}_{IK}\right)_{21}$, and $\left(\mathrm{M}^{\mathrm{Z}}_{IK}\right)_{22}$, one can see that in each case the two terms originating from the source matrix cancels out. As a result, we get
\begin{equation}
\mathrm{M}_{IK}  =   \left(\begin{matrix}
     \substack{\frac{\delta A^{\star S}[Z_c]}{\delta Z^{\bullet I}}\frac{\delta^2 S_{\mathrm{YM}}[B[Z_c]]}{\delta A^{\star S}\delta A^{\bullet R}}\frac{\delta A^{\bullet R}[Z_c]}{\delta Z^{\star K}} \\ 
    +\frac{\delta A^{\bullet S}[Z_c]}{\delta Z^{\bullet I}}\frac{\delta^2 S_{\mathrm{YM}}[A[Z_c]]}{\delta A^{\bullet S}\delta A^{\bullet R}}\frac{\delta A^{\bullet R}[Z_c]}{\delta Z^{\star K}}\\    
   +\frac{\delta A^{\star S}[Z_c]}{\delta Z^{\bullet I}}\frac{\delta^2 S_{\mathrm{YM}}[A[Z_c]]}{\delta A^{\star S}\delta A^{\star R}}\frac{\delta A^{\star R}[Z_c]}{\delta Z^{\star K}}    \\
    +\frac{\delta A^{\bullet S}[Z_c]}{\delta Z^{\bullet I}}\frac{\delta^2 S_{\mathrm{YM}}[A[Z_c]]}{\delta A^{\bullet S}\delta A^{\star R}}\frac{\delta A^{\star R}[Z_c]}{\delta Z^{\star K}}} & \substack{\frac{\delta A^{\star S}[Z_c]}{\delta Z^{\bullet I}}\frac{\delta^2 S_{\mathrm{YM}}[B[Z_c]]}{\delta A^{\star S}\delta A^{\bullet R}}\frac{\delta A^{\bullet R}[Z_c]}{\delta Z^{\bullet K}}  \\
    +\frac{\delta A^{\bullet S}[Z_c]}{\delta Z^{\bullet I}}\frac{\delta^2 S_{\mathrm{YM}}[A[Z_c]]}{\delta A^{\bullet S}\delta A^{\bullet R}}\frac{\delta A^{\bullet R}[Z_c]}{\delta Z^{\bullet K}}\\    
    +\frac{\delta A^{\star S}[Z_c]}{\delta Z^{\bullet I}}\frac{\delta^2 S_{\mathrm{YM}}[A[Z_c]]}{\delta A^{\star S}\delta A^{\star R}}\frac{\delta A^{\star R}[Z_c]}{\delta Z^{\bullet K}}    \\
    +\frac{\delta A^{\bullet S}[Z_c]}{\delta Z^{\bullet I}}\frac{\delta^2 S_{\mathrm{YM}}[A[Z_c]]}{\delta A^{\bullet S}\delta A^{\star R}}\frac{\delta A^{\star R}[Z_c]}{\delta Z^{\bullet K}}}\\ \\
\substack{\frac{\delta A^{\star S}[Z_c]}{\delta Z^{\star I}}\frac{\delta^2 S_{\mathrm{YM}}[B[Z_c]]}{\delta A^{\star S}\delta A^{\bullet R}}\frac{\delta A^{\bullet R}[Z_c]}{\delta Z^{\star K}}\\  
    +\frac{\delta A^{\bullet S}[Z_c]}{\delta Z^{\star I}}\frac{\delta^2 S_{\mathrm{YM}}[A[Z_c]]}{\delta A^{\bullet S}\delta A^{\bullet R}}\frac{\delta A^{\bullet R}[Z_c]}{\delta Z^{\star K}}\\    
    +\frac{\delta A^{\star S}[Z_c]}{\delta Z^{\star I}}\frac{\delta^2 S_{\mathrm{YM}}[A[Z_c]]}{\delta A^{\star S}\delta A^{\star R}}\frac{\delta A^{\star R}[Z_c]}{\delta Z^{\star K}}    \\
    +\frac{\delta A^{\bullet S}[Z_c]}{\delta Z^{\star I}}\frac{\delta^2 S_{\mathrm{YM}}[A[Z_c]]}{\delta A^{\bullet S}\delta A^{\star R}}\frac{\delta A^{\star R}[Z_c]}{\delta Z^{\star K}}}
      & \substack{\frac{\delta A^{\star S}[Z_c]}{\delta Z^{\star I}}\frac{\delta^2 S_{\mathrm{YM}}[B[Z_c]]}{\delta A^{\star S}\delta A^{\bullet R}}\frac{\delta A^{\bullet R}[Z_c]}{\delta Z^{\bullet K}}  \\
    +\frac{\delta A^{\bullet S}[Z_c]}{\delta Z^{\star I}}\frac{\delta^2 S_{\mathrm{YM}}[A[Z_c]]}{\delta A^{\bullet S}\delta A^{\bullet R}}\frac{\delta A^{\bullet R}[Z_c]}{\delta Z^{\bullet K}}\\    
    +\frac{\delta A^{\star S}[Z_c]}{\delta Z^{\star I}}\frac{\delta^2 S_{\mathrm{YM}}[A[Z_c]]}{\delta A^{\star S}\delta A^{\star R}}\frac{\delta A^{\star R}[Z_c]}{\delta Z^{\bullet K}}    \\
    +\frac{\delta A^{\bullet S}[Z_c]}{\delta Z^{\star I}}\frac{\delta^2 S_{\mathrm{YM}}[A[Z_c]]}{\delta A^{\bullet S}\delta A^{\star R}}\frac{\delta A^{\star R}[Z_c]}{\delta Z^{\bullet K}}}
\end{matrix} \right)\,.
\end{equation}
The complicated looking above matrix can be reduced to the product of the following three matrices
\begin{equation}
\mathrm{M}^{\mathrm{Z}}_{IK}  = \left(\begin{matrix}
     \frac{\delta A^{\bullet S}[Z_c]}
    {\delta Z^{\bullet I}} 
     & \frac{\delta A^{\star S}[Z_c]}
    {\delta Z^{\bullet I}} \\ \\
 \frac{\delta A^{\bullet S}[Z_c]}
    {\delta Z^{\star I}}
     &\frac{\delta A^{\star S}[Z_c]}
    {\delta Z^{\star I}} 
\end{matrix}\right) \left(\begin{matrix}
     \frac{\delta^2 S_{\mathrm{YM}}[A[Z_c]]}{\delta A^{\bullet S}\delta A^{\star R}} &\frac{\delta^2 S_{\mathrm{YM}}[A[Z_c]]}{\delta A^{\bullet S}\delta A^{\bullet R}}\\ \\
\frac{\delta^2 S_{\mathrm{YM}}[A[Z_c]]}{\delta A^{\star S}\delta A^{\star R}}
      &\frac{\delta^2 S_{\mathrm{YM}}[A[Z_c]]}{\delta A^{\star S}\delta A^{\bullet R}} 
\end{matrix}\right) \left(\begin{matrix}
     \frac{\delta A^{\star R}[Z_c]}
    {\delta Z^{\star K}} 
     & \frac{\delta A^{\star R}[Z_c]}
    {\delta Z^{\bullet K}}  \\ \\
 \frac{\delta A^{\bullet R}[Z_c]}
    {\delta Z^{\star K}}
     &\frac{\delta A^{\bullet R}[Z_c]}
    {\delta Z^{\bullet K}} 
\end{matrix}\right)\,.
\end{equation}
Therefore, the determinant reads
\begin{equation}
\det\mathrm{M}^{\mathrm{Z}}_{IK}  = \begin{vmatrix}
     \frac{\delta A^{\bullet S}[Z_c]}
    {\delta Z^{\bullet I}} 
     & \frac{\delta A^{\star S}[Z_c]}
    {\delta Z^{\bullet I}} \\ \\
 \frac{\delta A^{\bullet S}[Z_c]}
    {\delta Z^{\star I}}
     &\frac{\delta A^{\star S}[Z_c]}
    {\delta Z^{\star I}} 
\end{vmatrix} \begin{vmatrix}
     \frac{\delta^2 S_{\mathrm{YM}}[A[Z_c]]}{\delta A^{\bullet S}\delta A^{\star R}} &\frac{\delta^2 S_{\mathrm{YM}}[A[Z_c]]}{\delta A^{\bullet S}\delta A^{\bullet R}}\\ \\
\frac{\delta^2 S_{\mathrm{YM}}[A[Z_c]]}{\delta A^{\star S}\delta A^{\star R}}
      &\frac{\delta^2 S_{\mathrm{YM}}[A[Z_c]]}{\delta A^{\star S}\delta A^{\bullet R}} 
\end{vmatrix} \begin{vmatrix}
     \frac{\delta A^{\star R}[Z_c]}
    {\delta Z^{\star K}} 
     & \frac{\delta A^{\star R}[Z_c]}
    {\delta Z^{\bullet K}}  \\ \\
 \frac{\delta A^{\bullet R}[Z_c]}
    {\delta Z^{\star K}}
     &\frac{\delta A^{\bullet R}[Z_c]}
    {\delta Z^{\bullet K}} 
\end{vmatrix}\,.
\label{eq:det_rels}
\end{equation}

The determinant of the first and the third matrix is related to the Jacobian of the field transformation that derives the Z-field action from the Yang-Mills action via a volume divergent field-independent factor as shown below (for the sake of simplicity we restore the position coordinates and hats over the fields for color)
\begin{equation}
  \mathcal{J}_{\mathrm{A}\rightarrow\mathrm{Z_c}} \equiv \begin{vmatrix}
     \frac{\delta {\hat A}^{\bullet}[Z_c] (x^+;\mathbf{x})}
    {\delta {\hat Z}^{\bullet} (x^+;\mathbf{y})} 
     & \frac{\delta {\hat A}^{\star}[Z_c](x^+;\mathbf{x})}
    {\delta {\hat Z}^{\bullet}(x^+;\mathbf{y})}\\ \\
 \frac{\delta {\hat A}^{\bullet}[Z_c] (x^+;\mathbf{x})}
    {\delta {\hat Z}^{\star} (x^+;\mathbf{y})}
     &\frac{\delta {\hat A}^{\star}[Z_c](x^+;\mathbf{x})}
    {\delta {\hat Z}^{\star}(x^+;\mathbf{y})} 
\end{vmatrix}=\det \delta(y^+-x^+) \begin{vmatrix}
     \frac{\delta {\hat A}^{\bullet}[Z_c](x)}
    {\delta {\hat Z}^{\bullet}(y)} 
     & \frac{\delta {\hat A}^{\star}[Z_c](x)}
    {\delta {\hat Z}^{\bullet}(y)}\\ \\
 \frac{\delta {\hat A}^{\bullet}[Z_c](x)}
    {\delta {\hat Z}^{\star}(y)}
     &\frac{\delta {\hat A}^{\star}[Z_c](x)}
    {\delta {\hat Z}^{\star}(y)} 
\end{vmatrix}  \,.
\end{equation}
The determinant on L.H.S. is the Jacobian $\mathcal{J}_{\mathrm{A}\rightarrow\mathrm{Z_c}} $ for the transformation defined over the constant light-cone time $x^+$ hypersurface. It is field independent. The factor $\det \delta(y^+-x^+)$ is volume divergent and field independent. Owing to the latter, it does not contribute to amplitude computation and can therefore be ignored. Thus, we can rewrite Eq.~\eqref{eq:det_rels} (in collective index notation) as
\begin{equation}
\det\mathrm{M}^{\mathrm{Z}}_{IK}  =  \begin{vmatrix}
     \frac{\delta^2 S_{\mathrm{YM}}[A[Z_c]]}{\delta A^{\bullet I}\delta A^{\star K}} &\frac{\delta^2 S_{\mathrm{YM}}[A[Z_c]]}{\delta A^{\bullet I}\delta A^{\bullet K}}\\ \\
\frac{\delta^2 S_{\mathrm{YM}}[A[Z_c]]}{\delta A^{\star I}\delta A^{\star K}}
      &\frac{\delta^2 S_{\mathrm{YM}}[A[Z_c]]}{\delta A^{\star I}\delta A^{\bullet K}} 
\end{vmatrix} \,.
\label{eq:det_mik}
\end{equation}
The determinant in the above expression reads \cite{Kakkad_2022}
\begin{multline}
    \det\mathrm{M}^{\mathrm{Z}}_{IK} = \det\left[ \frac{\delta^2 S_{\mathrm{YM}}[A[Z_c]]}
    {\delta A^{\star I}\delta A^{\bullet K}} \, \frac{\delta^2 S_{\mathrm{YM}}[A[Z_c]]}
    {\delta A^{\star K}\delta A^{\bullet J}} \right.\\
    \left.- \frac{\delta^2 S_{\mathrm{YM}}[A[Z_c]]}
    {\delta A^{\star I}\delta A^{\bullet K}} \, \frac{\delta^2 S_{\mathrm{YM}}[A[Z_c]]}
    {\delta A^{\star K}\delta A^{\star L}} \left( \frac{\delta^2 S_{\mathrm{YM}}[A[Z_c]]}
    {\delta A^{\bullet L}\delta A^{\star M}} \right)^{-1} \frac{\delta^2 S_{\mathrm{YM}}[A[Z_c]]}
    {\delta A^{\bullet M}\delta A^{\bullet J}}\right]\,.
\end{multline}
Substituting the above to Eq.~\eqref{eq:G_zac}, we get
\begin{multline}
   \Gamma[Z_c] = S[Z_c] 
    + \frac{i}{2} \Tr\ln \left[ \frac{\delta^2 S_{\mathrm{YM}}[A[Z_c]]}
    {\delta A^{\star I}\delta A^{\bullet K}} \, \frac{\delta^2 S_{\mathrm{YM}}[A[Z_c]]}
    {\delta A^{\star K}\delta A^{\bullet J}} \right.\\
    \left.- \frac{\delta^2 S_{\mathrm{YM}}[A[Z_c]]}
    {\delta A^{\star I}\delta A^{\bullet K}} \, \frac{\delta^2 S_{\mathrm{YM}}[A[Z_c]]}
    {\delta A^{\star K}\delta A^{\star L}} \left( \frac{\delta^2 S_{\mathrm{YM}}[A[Z_c]]}
    {\delta A^{\bullet L}\delta A^{\star M}} \right)^{-1} \frac{\delta^2 S_{\mathrm{YM}}[A[Z_c]]}
    {\delta A^{\bullet M}\delta A^{\bullet J}}\right]\,.
    \label{eq:OLEA_Zac2}
\end{multline}
which is the same as the one-loop Z-field action we obtained via the simplified approach in Eq.~\eqref{eq:OLEA_Zac}.

In proving the equivalence, we used the fact that the Jacobian $\mathcal{J}_{\mathrm{A}\rightarrow\mathrm{Z_c}} $ is field independent. This can also be verified in a different way which utilizes the fact that the Z-field field can be derived from the light-cone Yang-Mills action in two different ways \cite{Kakkad:2021uhv}. First is the direct approach for which the Jacobian is $\mathcal{J}_{\mathrm{A}\rightarrow\mathrm{Z_c}} $. The second is via two consecutive canonical transformations. The first transformation maps the self-dual part of the Yang-Mills action to a free term in the MHV action (Mansfield's transformation \cite{Mansfield2006})
\begin{equation}
\mathcal{L}_{-+}[A^{\bullet},A^{\star}]+\mathcal{L}_{-++}[A^{\bullet},A^{\star}]
\,\, \longrightarrow \,\,
\mathcal{L}_{-+}[B^{\bullet},B^{\star}]
\,,
\label{eq:AtoBtransform}
\end{equation}
The Jacobian for this transformation is
\begin{equation}
    \mathcal{J}_{\mathrm{A}\rightarrow\mathrm{B}}  =  \begin{vmatrix}
     \frac{\delta {\hat A}^{\bullet} (x^+;\mathbf{x})}
    {\delta {\hat B}^{\bullet} (x^+;\mathbf{y})} 
     &\mathbb{0} \\ \\
\frac{\delta {\hat A}^{\star}(x^+;\mathbf{x})}
    {\delta {\hat B}^{\bullet}(x^+;\mathbf{y})} 
     &\frac{\delta {\hat A}^{\star}(x^+;\mathbf{x})}
    {\delta {\hat B}^{\star}(x^+;\mathbf{y})} 
    \label{eq:MT_jac}
\end{vmatrix} \,.
\end{equation}
and is field independent. The second transformation maps the anti-self-dual part of the MHV action to the kinetic term in the Z-field action
\begin{equation}
\mathcal{L}_{-+}[B^{\bullet},B^{\star}]+\mathcal{L}_{--+}[B^{\bullet},B^{\star}]
\,\, \longrightarrow \,\,
\mathcal{L}_{-+}[Z^{\bullet},Z^{\star}]
\,,
\label{eq:BtoZtransform1}
\end{equation}
The Jacobian for this transformation is
\begin{equation}
    \mathcal{J}_{\mathrm{B}\rightarrow\mathrm{Z}}  =  \begin{vmatrix}
     \frac{\delta {\hat B}^{\star} (x^+;\mathbf{x})}
    {\delta {\hat Z}^{\star} (x^+;\mathbf{y})} 
     &\mathbb{0} \\ \\
\frac{\delta {\hat B}^{\bullet}(x^+;\mathbf{x})}
    {\delta {\hat Z}^{\star}(x^+;\mathbf{y})} 
     &\frac{\delta {\hat B}^{\bullet}(x^+;\mathbf{x})}
    {\delta {\hat Z}^{\bullet}(x^+;\mathbf{y})} 
\end{vmatrix} \,.
\label{eq:Z_jac}
\end{equation}
and is also field independent.

Using chain rule, the Jacobian $\mathcal{J}_{\mathrm{A}\rightarrow\mathrm{Z}} $ can be rewritten as
\begin{align}
    \mathcal{J}_{\mathrm{A}\rightarrow\mathrm{Z}} &= \begin{vmatrix}
     \frac{\delta {\hat A}^{\bullet}[Z] (x^+;\mathbf{x})}
    {\delta {\hat Z}^{\bullet} (x^+;\mathbf{y})} 
     & \frac{\delta {\hat A}^{\star}[Z](x^+;\mathbf{x})}
    {\delta {\hat Z}^{\bullet}(x^+;\mathbf{y})}\\ \\
 \frac{\delta {\hat A}^{\bullet}[Z] (x^+;\mathbf{x})}
    {\delta {\hat Z}^{\star} (x^+;\mathbf{y})}
     &\frac{\delta {\hat A}^{\star}[Z](x^+;\mathbf{x})}
    {\delta {\hat Z}^{\star}(x^+;\mathbf{y})} 
\end{vmatrix}\,, \nonumber \\
&=\begin{vmatrix}
     \frac{\delta {\hat A}^{\bullet}[B] (x^+;\mathbf{x})}
    {\delta {\hat B}^{\bullet} (x^+;\mathbf{q})} \frac{\delta {\hat B}^{\bullet}[Z] (x^+;\mathbf{q})}
    {\delta {\hat Z}^{\bullet} (x^+;\mathbf{y})} 
     & \frac{\delta {\hat A}^{\star}[B](x^+;\mathbf{x})}
    {\delta {\hat B}^{\bullet}(x^+;\mathbf{q})} \frac{\delta {\hat B}^{\bullet}[Z] (x^+;\mathbf{q})}
    {\delta {\hat Z}^{\bullet} (x^+;\mathbf{y})} \\ \\
 \frac{\delta {\hat A}^{\bullet}[B] (x^+;\mathbf{x})}{\delta {\hat B}^{\bullet} (x^+;\mathbf{q})} \frac{\delta {\hat B}^{\bullet}[Z] (x^+;\mathbf{q})}
    {\delta {\hat Z}^{\star} (x^+;\mathbf{y})}
     &\frac{\delta {\hat A}^{\star}[B](x^+;\mathbf{x})}{\delta {\hat B}^{\bullet} (x^+;\mathbf{q})} \frac{\delta {\hat B}^{\bullet}[Z] (x^+;\mathbf{q})}
    {\delta {\hat Z}^{\star}(x^+;\mathbf{y})} + \frac{\delta {\hat A}^{\star}[B](x^+;\mathbf{x})}{\delta {\hat B}^{\star} (x^+;\mathbf{q})} \frac{\delta {\hat B}^{\star}[Z] (x^+;\mathbf{q})}
    {\delta {\hat Z}^{\star}(x^+;\mathbf{y})} 
\end{vmatrix}\,, \nonumber \\
&= \begin{vmatrix}
    \frac{\delta {\hat B}^{\bullet}(x^+;\mathbf{q})}
    {\delta {\hat Z}^{\bullet}(x^+;\mathbf{y})} 
     &\mathbb{0} \\ \\
\frac{\delta {\hat B}^{\bullet}(x^+;\mathbf{q})}
    {\delta {\hat Z}^{\star}(x^+;\mathbf{y})} 
     & \frac{\delta {\hat B}^{\star} (x^+;\mathbf{q})}
    {\delta {\hat Z}^{\star} (x^+;\mathbf{y})} 
\end{vmatrix} \,\, \begin{vmatrix}
     \frac{\delta {\hat A}^{\bullet} (x^+;\mathbf{x})}
    {\delta {\hat B}^{\bullet} (x^+;\mathbf{q})} 
     &\frac{\delta {\hat A}^{\star}(x^+;\mathbf{x})}
    {\delta {\hat B}^{\bullet}(x^+;\mathbf{q})}  \\ \\
\mathbb{0}
     &\frac{\delta {\hat A}^{\star}(x^+;\mathbf{x})}
    {\delta {\hat B}^{\star}(x^+;\mathbf{q})} 
\end{vmatrix} \,, \nonumber\\
&= \mathcal{J}_{\mathrm{B}\rightarrow\mathrm{Z}} \,\,\mathcal{J}_{\mathrm{A}\rightarrow\mathrm{B}} \,.
\end{align}
The R.H.S. is the product of the Jacobians Eqs.~\eqref{eq:MT_jac} and \eqref{eq:Z_jac}, each of which is field independent.
\end{document}